\documentclass{aa}  

\usepackage{natbib}
\bibpunct{(}{)}{;}{a}{}{,} 
\usepackage{graphicx}
\usepackage{txfonts}
\usepackage[colorlinks=true,linkcolor=blue,citecolor=blue, urlcolor=blue]{hyperref}

\newcommand{\teff}{$T_{\rm eff}$}
\newcommand{\eexc}{$E_{\rm exc}$}
\def\vt{$\xi_{\rm t}$}
\def\kms{$\rm km~s^{-1}$}
\def\ione{\,{\sc i}}
\def\ii{\,{\sc ii}}
\newcommand{\eps}{\log\varepsilon}
\newcommand{\gkai}[1]{}

\begin{document}

\title{HR-GO I: Comprehensive NLTE abundance analysis of the \\ Cetus stream\thanks{Tables 2 and 4 are only available in electronic form at the CDS via anonymous ftp to cdsarc.u-strasbg.fr (130.79.128.5) or via \url{http://cdsweb.u-strasbg.fr/cgi-bin/qcat?J/A+A/}.}}
\titlerunning{The NLTE abundances of the Cetus stream}
\authorrunning{Sitnova, Yuan, Matsuno et al.}

\author{T. M. Sitnova\inst{1}, Z. Yuan \gkai{袁珍} \inst{2,3,4}, T. Matsuno\inst{5}, L. I. Mashonkina\inst{1}, S. A. Alexeeva\inst{6}, E. Holmbeck\inst{7},  F. Sestito\inst{8}, L.~Lombardo\inst{9}, P. Banerjee\inst{10}, N. F. Martin\inst{4}, F. Jiang\inst{11,7}
}

   \institute{Institute of Astronomy, Russian Academy of Sciences, Pyatnitskaya 48, 119017, Moscow, Russia,
              \email{sitamih@gmail.com}\
         \and
         School of Astronomy and Space Science, Nanjing University, Nanjing 210093, China, \email{zhen.yuan@nju.edu.cn}   
         \and
         Key Laboratory of Modern Astronomy and Astrophysics (Nanjing University), Ministry of Education, Nanjing 210093, China 
         \and
             Universit\'e de Strasbourg, CNRS, Observatoire astronomique de Strasbourg, UMR 7550, F-67000 Strasbourg, France
          \and
          Astronomisches Rechen-Institut, Zentrum f\"ur Astronomie der Universit\"at Heidelberg, M\"onchhofstra{\ss}e 12-14, 69120 Heidelberg, Germany
          \and
          CAS Key Laboratory of Optical Astronomy, National Astronomical Observatories, Chinese Academy of Sciences, Beijing, 100101, China
          \and
The Observatories of the Carnegie Institution for Science, 813 Santa Barbara Street, Pasadena, CA 91101, USA
          \and
          Department of Physics and Astronomy, University of Victoria, PO Box 3055, STN CSC, Victoria BC V8W 3P6, Canada  
          \and
          Goethe University Frankfurt, Institute for Applied Physics (IAP), Max-von-Laue-Str. 12, 60438, Frankfurt am Main
          \and  
          Department of Physics, Indian Institute of Technology Palakkad, Kerala 678558, India
        \and
          Kavli Institute for Astronomy and Astrophysics, Peking University, Beijing 100871, China
           }

   \date{Received ; accepted }

 
\abstract
{Dwarf galaxy streams encode vast amounts of information essential to understanding early galaxy formation and nucleosynthesis channels. Due to the variation in the timescales of star formation history in their progenitors, stellar streams serve as `snapshots' that record different stages of galactic chemical evolution.}
{This study focusses on the Cetus stream, stripped from a low-mass dwarf galaxy. We aim to uncover its chemical evolution history as well as the different channels of its element production from detailed elemental abundances.}
{We carried out a comprehensive analysis of the chemical composition of 22 member stars based on their high-resolution spectra. We derived abundances for up to 28 chemical species from C to Dy and, for 20 of them, we account for the departures from local thermodynamic equilibrium (NLTE effects).}
{We confirm that the Cetus stream has a mean metallicity of [Fe/H] = $-2.11$ $\pm$ 0.21. 
All observed Cetus stars are $\alpha$ enhanced with [$\alpha$/Fe] $\simeq$ 0.3. The absence of the $\alpha$-`knee' implies that star formation stopped before iron production in type Ia supernovae (SNe~Ia) became substantial.
Neutron capture element abundances suggest that both the rapid (r-) and the main slow (s-) processes contributed to their origin. The decrease in [Eu/Ba] from a typical r-process value of [Eu/Ba] = 0.7 to 0.3 with increasing [Ba/H] indicates a distinct contribution of the r- and s-processes to the chemical composition of different Cetus stars. For barium, the r-process contribution varies from 100\%  to 20\% in different sample stars, with an average value of 50\%.}
{Our abundance analysis indicates that the star formation in the Cetus progenitor ceased after the onset of the main s-process in low- to intermediate-mass asymptotic giant branch stars but before SNe~Ia played an important role. A distinct evolution scenario is revealed by comparing the abundances in the Ursa Minor dwarf spheroidal galaxy, showing the diversity in ---and uniqueness of--- the chemical evolution of low-mass dwarf galaxies.}


   \maketitle
%
\section{Introduction}

Ancient stars, which contain very small amounts of metals ---typically less than one percent of the solar value---, encode vast amounts of information about the very early Universe. The studies of these old and low-metallicity stars belong to the field of Galactic archaeology \citep{1979Msngr..16....7S}. In the {\it Gaia} era \citep{2021A&A...649A...1G}, a library of stellar streams is being built, which are mainly populated by old stars \citep[see e.g.][]{ibata2021, li2022, martin2022, 2024ApJ...967...89I}. These streams are the debris of ancient stellar systems, such as the globular clusters and dwarf galaxies that merged with the Milky Way (MW) during its long assembly history.

Among them, dwarf galaxy streams have unique value. Because of the relatively short star formation histories of their progenitors, the elemental abundances of the member stars are very sensitive to the initial mass function, star-forming activities, and chemical enrichment from different nucleosynthesis channels. Therefore, they serve as pristine laboratories for studying these physical processes that shaped the Universe we see today, and the elemental abundances of their member stars are the key to decoding such information. With modern large telescopes accompanied by high-resolution (HR) spectrographs, we are able to determine detailed abundances using high-quality stellar spectra. In this context, stellar streams from now-disrupted dwarf galaxies offer a great advantage over intact dwarf galaxies: streams are much closer to than the surviving dwarf galaxies with similar progenitor masses. This is the natural outcome of hierarchical formation. Consequently, we have larger samples of bright stars from stellar streams for which high-quality HR spectra are attainable. Moreover, the observation efficiency for individual stars is greatly enhanced as the brightness increases.

Several recent works demonstrated the power of this approach and revealed the nature of the progenitors from the chemical composition of their stellar streams and substructures. Some of those streams arise from globular clusters, such as Typhon \citep[][]{2023MNRAS.519.4467J}, ATLAS, Aliqa Uma, Phoenix \citep[][]{2020AJ....160..181J}, GD-1 \citep[][]{2022MNRAS.515.5802B}, C-19 \citep[][]{2022Natur.601...45M}, and $\omega$ Cen stream \citep[][]{2021ApJ...912...52G}, while others are from dwarf galaxies: Chenab, Elqui, Indus, Jhelum \citep[][]{2020AJ....160..181J}, Helmi/S2 \citep[][]{2010ApJ...711..573R,2021ApJ...913L..28L,2021ApJ...912...52G,2022A&A...665A..46M}, {\it Gaia}~Sausage/Enceladus \citep[][]{2021ApJ...908L...8A,2024A&A...684A..37C}, Sequoia \citep[][]{2021ApJ...908L...8A,2022A&A...661A.103M,2024A&A...684A..37C}, Wukong/LMS-1 \citep[][]{2024MNRAS.530.2512L}, Orphan-Chenab \citep[][]{2023ApJ...948..123H}, and Specter, which formed from an ultrafaint dwarf (UFD) galaxy \citep[][]{2022ApJ...940..127C};
for some streams, their progenitor nature is debated: for example, Sylgr \citep[][]{roederer2019}, Nyx \citep[][]{2023ApJ...955..129W}, Antaeus, and ED-2 \citep[][]{2024A&A...684A..37C}.

This study is conducted within a novel {H}igh-{R}esolution spectroscopic program on the {G}alactic {O}rigins of elements (HR-GO) initiated by Z. Yuan. A detailed introduction to HR-GO will be given in a forthcoming paper (Yuan et al., in preparation). This program focusses on stars in stellar streams and substructures that are stripped from accreted dwarf galaxies. The goal is to understand the production channels of elements in dwarf galaxies as well as their evolution histories by observing stars in their debris. In the analysis procedure, we aim to obtain accurate stellar atmosphere parameters and detailed chemical compositions, accounting for the departures from local thermodynamic equilibrium (i.e. NLTE effects). This paper, as the first in this series of studies, reveals the detailed abundances of the Cetus stream for the first time.

The Cetus stream was discovered by \citet{newberg2009} from ancient star tracers, blue straggler and blue horizontal branch stars in the Sloan Digital Sky Survey \citep[SDSS,][]{2000AJ....120.1579Y}. After {\it Gaia} DR2, using stellar orbits, \citet{yuan2019} identified about 150 members from the K giant sample selected by \citet{liu2014} from the Large Sky Area Multi-Object Fiber Spectroscopic Telescope survey \citep[LAMOST,][]{2012RAA....12.1197C,2012RAA....12..735D,2012RAA....12..723Z}. To understand the disruption history in the MW potential, \citet{chang2020} performed a series of N-body simulations and predicted a vast extension of the stream towards the south, surprisingly coincident with the Palca stream \citep{shipp2018}. Later on, \citet{thomas2022} revealed that the Palca and Cetus streams are two parts of one stream and named it the Cetus-Palca stream. At about the same time, \citet{yuan2022} found that the Cetus stream system has two wraps with different orbital phases, one of which contains the Palca stream and  was thus named the Cetus-Palca wrap, while the other contains the Cetus
stream, and was named the Cetus-New wrap. 

All of the above studies give us a coherent picture of the Cetus progenitor system. It is mainly populated by very metal-poor (VMP, [Fe/H]\footnote{We use a standard designation, [X/Y] = $\log($N$_{\rm X}$/N$_{\rm Y}$)$_{*} - \log($N$_{\rm X}$/N$_{\rm Y}$)$_{\odot}$, where N$_{\rm X}$ and N$_{\rm Y}$ are total number densities of elements X and Y, respectively.} $<$ $-$2)  stars and has an average metallicity of $-2.1$ with an intrinsic dispersion of 0.2 dex in both hemispheres, based on low-resolution spectroscopic and narrow-band photometric surveys \citep{newberg2009, yam2013, yuan2019, yuan2022}. According to the stellar mass--metallicity relationship of the MW satellite dwarf galaxies \citep{kirby2013}, the Cetus progenitor is compatible with a low-mass dwarf galaxy with a stellar mass of M$_{\ast}\gtrsim$10$^{6}$M$_{\odot}$. By summing the fluxes of all identified members, the Cetus progenitor is estimated to have a stellar mass of M$_{\ast}\gtrsim$10$^{5.6}$M$_{\odot}$ \citep{yuan2022}, which is similar to the Ursa Minor (UMi) dwarf spheroidal (dSph) galaxy (M$_{\ast}\sim$10$^{5.7}$M$_{\odot}$) and Sextans dSph with M$_{\ast}$ $\sim$10$^{5.8}$M$_{\odot}$ \citep{kirby2013}.

To reconstruct the history of chemical element formation in a given system, high-quality spectra are required, enabling the measurement of the detailed chemical composition, including $\alpha$-elements, iron-peak elements, and neutron capture elements. The orbits of stars in the Cetus stream span a distance range from 10 to 40 kpc \citep{chang2020}, making its giant stars perfect targets for high-resolution spectroscopic studies of chemical abundances. For comparison, the UMi dSph is located at a distance of 69 kpc \citep{1999AJ....118..366M}. For UMi stars with V magnitudes from 16.7 to 17.2, \citet{2004PASJ...56.1041S} took exposures of 150 to 210 minutes to obtain spectra with signal-to-noise ratios of 50 to 60 in the red wavelength region using the High Dispersion Spectrograph at the Subaru telescope. With the same instrument, exposures of only 30 minutes are needed to obtain spectra of the same quality for the Cetus stars with V $\simeq$~15.

The paper is constructed as follows. We describe our target stars and the observations in Sect.~\ref{sample_obs}. Section~\ref{st_par} presents a determination of stellar atmosphere parameters. The abundance determination method is described in Sect.~\ref{abund}. In Sect.~\ref{comparison_sample}, we present the comparison sample stars. The element abundances are given in Sect.~\ref{results} along with a discussion of our findings. Our conclusions are presented in Sect.~\ref{conclusions}.

\section{Stellar sample and observations}\label{sample_obs}
\begin{figure*}
  \includegraphics[width=\linewidth]{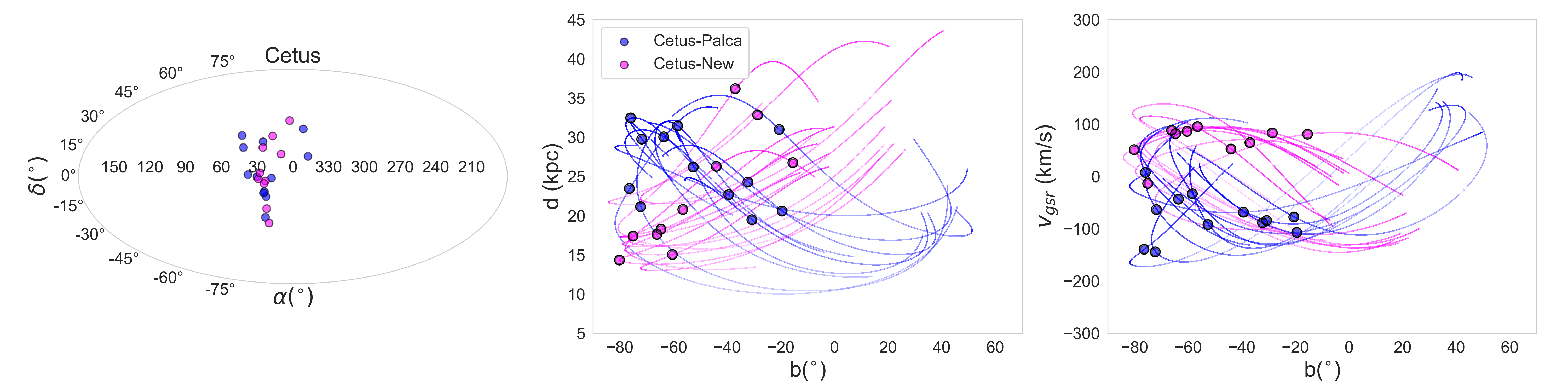}  
   \caption{Cetus stream members in the HR-GO program are shown in the ($\alpha$, $\delta$) and ($b$, $d$, $v_{\rm gsr}$) space. The members belonging to the Cetus-Palca and Cetus-New wraps are denoted by blue and magenta circles, respectively. These two wraps have different orbital phases revealed by the integrated orbits of their member stars in the ($b$, $d$) and ($b$, $v_{\rm gsr}$) spaces.}
   \label{fig:cetus} 
\end{figure*}

Combining the member lists in \citet{yuan2019} and \citet{yuan2022}, we select all the Cetus stars brighter than G = 16, yielding 22 stars in total for high-resolution spectroscopic follow up. According to \citet{yuan2022}, the main Cetus stream can be decomposed into two wraps (Cetus-Palca and Cetus-New), which have different orbital phases. In the HR target list, there are 12 stars from the Cetus-Palca wrap and 10 from the Cetus-New wrap denoted in Fig.~\ref{fig:cetus} as blue and magenta circles, respectively. Although the on-sky projections of these two wraps overlap, they can be distinguished in the ($b$, $d$) and ($b$, $v_{\rm gsr}$) spaces where $b$ is the galactic latitude, $d$ is the heliocentric distance, and $v_{\rm gsr}$ is the Galactic rest-frame radial velocity. These are the two signature spaces that best reveal the properties of the Cetus stream, as shown in \citet{newberg2009}, \citet{yam2013}, \citet{yuan2019}, and \citet{yuan2022}. The distances used in this work are derived from $\log$~g, which are determined iteratively from {\it Gaia} parallax, photometry, and HR spectra, as described in Sect.\ref{st_par}. For each HR target, we trace its orbit backward and forward for about one-quarter of its period time and clearly see that stars in these two wraps follow distinct tracks, indicating that they are at different orbital phases. The orbital tracks of the Cetus-Palca wrap (solid blue line) exhibit a decreasing trend in $d$ as $b$ increases, and the Cetus-New wrap (solid magenta line) shows an opposite trend. Similarly, we can see that stars belonging to these two wraps follow different tracks in the ($b$, $v_{\rm gsr}$) space.

The high-resolution spectra of the 22 selected Cetus stars were taken in three observational programs with different telescopes and spectrographs and were processed with the corresponding pipelines:

\begin{enumerate}
    \item The UV-visual echelle spectrograph (UVES) at the UT2 Kueyen Telescope on 23--24 October 2021 via program 0108.B-0431(A) (PI: Z.~Yuan). The UVES spectra are derived with the 1.0" slit width, which provides a resolving power of R = $\lambda/\Delta \lambda$ $\sim$ 40~000. The wavelength coverage is 3750 -- 5000, 5700 -- 7500, and 7660 -- 9450 \AA\ \citep{2000SPIE.4008..534D}. The observed spectra are reduced with the ESO pipeline\footnote{https://www.eso.org/sci/software/pipelines/}.
    \item The Magellan Inamori Kyocera Echelle (MIKE) spectrograph at the Magellan Telescope on 19 August 2022 (PI: F.~Jiang). The MIKE spectra are taken with the 0.7" slit width, which yields  R $\sim$ 28~000  and $\sim$ 35~000 in the red and blue wavelength regions, respectively \citep{2003SPIE.4841.1694B}. The wavelength coverage is 3300 -- 9600 \AA. The data reduction is carried out with the MIKE Carnegie Python pipeline \citep{mike}. The barycentric velocity correction was performed using the online interface\footnote{https://astroutils.astronomy.osu.edu/exofast/barycorr.html}, developed by \citet{2014PASP..126..838W}.
    \item The High Dispersion Spectrograph (HDS) at the Subaru Telescope on 21 September 2022 via program S22B-0094N (PI: Z.~Yuan). The HDS spectra are derived using the standard StdYd setup, which provides a wavelength coverage of 4000 -- 5340 and 5450 -- 6800 \AA, with  R =  45~000 \citep{2002PASJ...54..855N}. The data were reduced using the IRAF\footnote{IRAF is distributed by the National Optical Astronomy Observatory, which is operated by the Association of Universities for Research in Astronomy (AURA) under a cooperative agreement with the National Science Foundation} script hdsql\footnote{\tiny{http://www.subarutelescope.org/Observing/Instruments/HDS/hdsql-e.html}} , which includes CCD linearity correction, scattered light subtraction, aperture extraction, flat-fielding, wavelength calibration, and barycentric velocity correction. 
\end{enumerate}

We list the coordinates, {\it Gaia} IDs, and characteristics of all target stars in Table~\ref{table_par}. Observation information is also provided, including the telescope and spectrograph, the exposure time, and the resulting signal-to-noise ratio in the blue and red wavelength regions (S/N$_{\rm B}$ and S/N$_{\rm R}$), corresponding to wavelengths around $\lambda \sim$ 4500 \AA\ and 6000 \AA, respectively. We provide the S/N per pixel, which is calculated as the ratio between the mean normalised flux (F$_{\rm mean}$), which nearly equals unity, and its standard deviation $\sigma = \sqrt{ \Sigma (F_{\rm mean} - F_i )^2 /(N - 1)}$, where F$_{\rm i}$ and N are the flux at a given wavelength and the number of data points in the selected spectral region, respectively. It is worth noting that spectra obtained with different instruments are characterised by a different number of pixels per resolving element. For our UVES, MIKE, and HDS spectra, a wavelength interval of 1 \AA\ falls on 55, 50, and 70 pixels in the blue range, and on 40, 20, and 55 pixels in the red range, respectively.

\begin{figure}
\centering
\includegraphics[trim={0 37 0 0},width=\hsize]{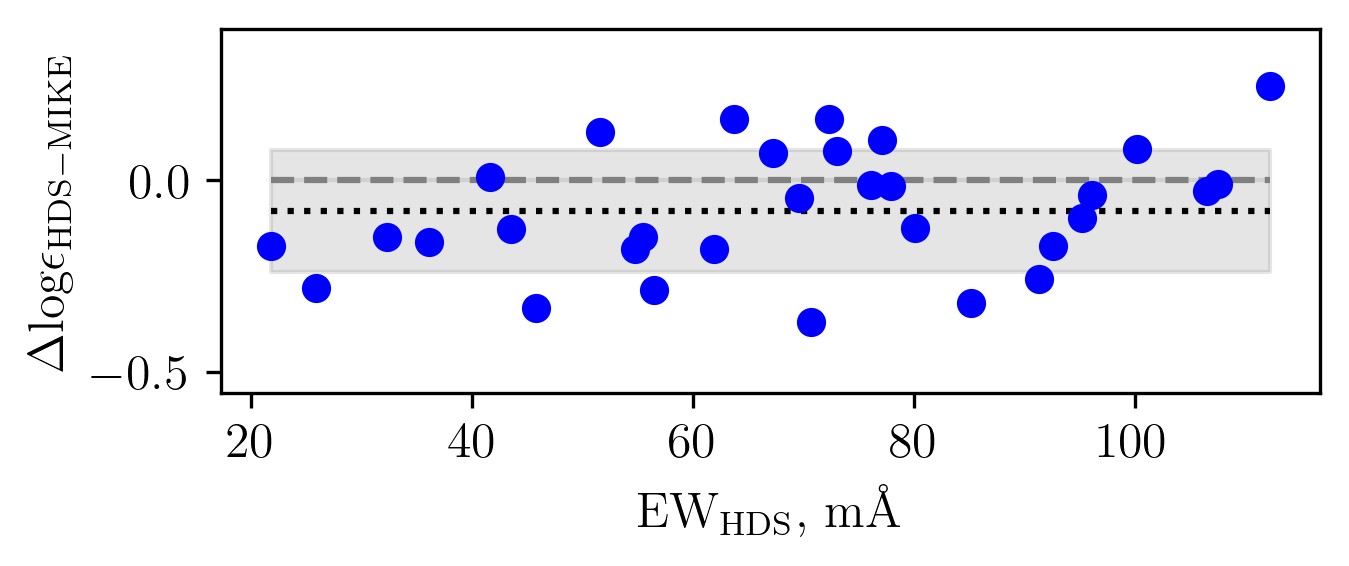}
\includegraphics[trim={0 18 0 0},width=\hsize]{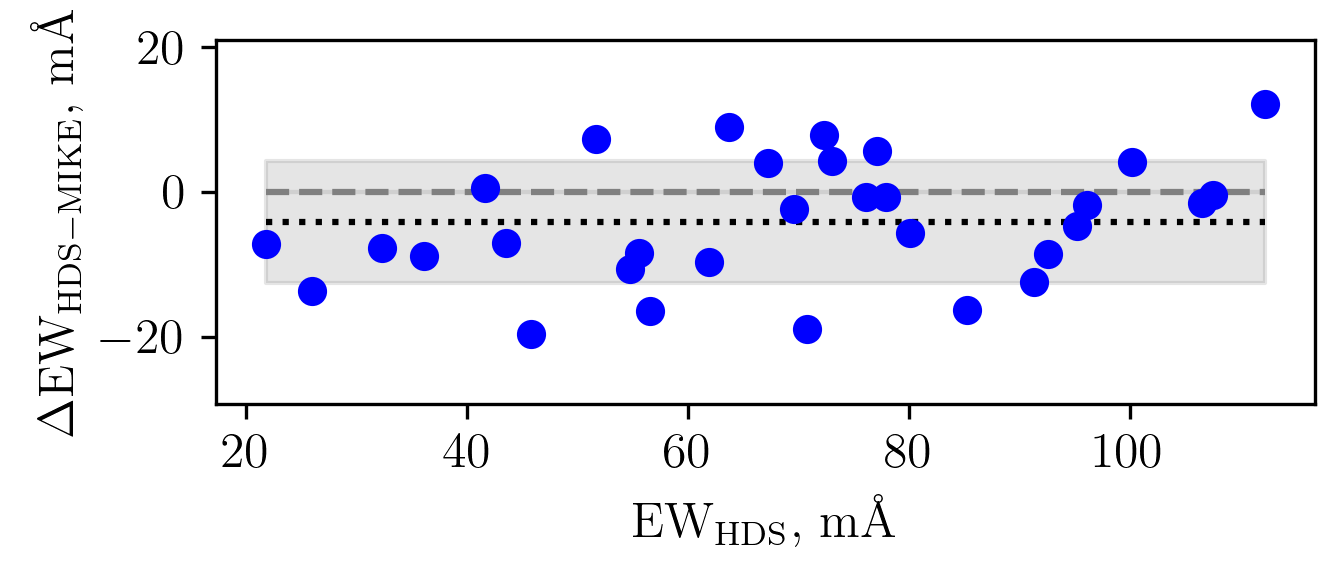}
\caption{Abundance difference between iron lines measured in {\it Gaia} DR3 2505061738639700608 spectra taken with the HDS and MIKE instruments (top panel) and the corresponding abundance differences in EWs (bottom panel) as a function of EW. The dotted lines and shaded areas in both panels represent the mean difference and 1~$\sigma$ dispersion.}
\label{mike_hds}
\end{figure}

To ensure the absence of systematic differences in our abundance analysis arising from the use of spectral observations obtained with different instruments, we observed star {\it Gaia} DR3 2505061738639700608 using both MIKE and HDS. Figure~\ref{mike_hds} presents a comparison of the equivalent widths (EWs) and abundances from iron lines measured from the two spectra. We find consistent EWs, with an average difference of $\Delta$ EW$_{\rm HDS - MIKE}$ = $-4$ $\pm$ 8 m\AA, which translates to an average abundance difference of $-0.08$ $\pm$ 0.16~dex. For the majority of spectral lines in this star, we use its MIKE spectrum.

With our high-resolution spectra, we measure radial velocities (v$_{\rm r}$, Table~\ref{table_par}) with an uncertainty of 1~\kms. For the majority of the stars of our sample, our measurements are consistent with those provided in the {\it Gaia} DR3 catalogue. The exception is {\it Gaia} DR3~2507540140928136832 , for which we find v$_{\rm r}$ = $-68.0$ $\pm$ 1~\kms, while {\it Gaia} provides $-88.68$ $\pm$ 9.98~\kms\ and a renormalised unit weight error of RUWE = 1.035. We assume this star could be a binary.

\begin{table*}
\caption{Stellar sample, stellar parameters, and characteristics of the observed spectra} 
\setlength{\tabcolsep}{1.0mm}            
\label{table_par}      
\centering          
\begin{tabular}{l r r c l l l r r l l l l l r}   
\hline      
\tiny{{\it Gaia} ID}  &  \tiny{RA}  &  \tiny{DEC} &  \tiny{G} & \tiny{E(B--V)} & \tiny{S/T$^1$} & \tiny{t$_{\rm exp}$}, & \tiny{S/N$_{\rm B}$}    & \tiny{S/N$_{\rm R}$}   & \teff ,  & log g & \tiny{[Fe/H]}        & \vt ,       &  d$_{\rm spec}$,  & \tiny{v$_{\rm r}$}, \\
   & \tiny{deg.} & \tiny{deg.} & \tiny{mag.} & \tiny{mag.}  &   & \tiny{min.}    &      &            & \tiny{K}        & \tiny{$\rm cm~s^{-2}$} &                & \tiny{\kms} & \tiny{kpc} & \tiny{\kms}      \\
\hline                    
\multicolumn{15}{c}{Cetus-New}\\ 
\hline 
 \tiny{2483903595868274048} &   23.194442 &   --3.811474 & 13.99 & 0.032 & U & 30 &  29 & 46 & 4440 & 0.94 & --2.10\tiny{$\pm$0.05} & 1.7 & 18.2\tiny{$\pm$2.8} &  43.5  \\
 \tiny{2795091499930032768} &   10.415858 &    18.780757 & 14.28 & 0.045 & H & 30 &  20 & 53 & 4400 & 0.70 & --2.53\tiny{$\pm$0.15} & 1.8 & 26.3\tiny{$\pm$4.0} & --83.5   \\
 \tiny{2479610243479857792 } &  24.259206 &   --5.924668 & 14.41 & 0.035 & U & 30 &  26 & 41 & 4560 & 1.20 & --1.99\tiny{$\pm$0.08} & 1.7 & 17.6\tiny{$\pm$2.6} &  58.0  \\
 \tiny{2512532782711324288} &   27.890108 &     3.048892 & 14.91 & 0.027 & H & 30 &  24 & 77 & 4625 & 1.30 & --1.98\tiny{$\pm$0.06} & 1.7 & 20.8\tiny{$\pm$3.1} &  44.5  \\
 \tiny{320073130541422080  } &  19.339750 &    33.959461 & 15.20 & 0.040 & H & 30 &  14 & 21 & 4620 & 1.00 & --2.20\tiny{$\pm$0.11} & 2.1 & 32.8\tiny{$\pm$4.9} & --65.5   \\
 \tiny{2505061738639700608} &   29.090208 &   --1.975403 & 15.22 & 0.024 & M & 25 &  37 & 53 & 4930 & 1.85 & --1.85\tiny{$\pm$0.10} & 1.8 & 15.0\tiny{$\pm$2.2} &  53.5  \\
 \tiny{                   } &         &         &       &       & H & 30 &  27 & 62 &      &      &                        &     &                       &      \\
 \tiny{5036037656380706560} &   23.899406 &  --26.841977 & 15.23 & 0.012 & M & 30 &  32 & 35 & 4905 & 1.90 & --1.50\tiny{$\pm$0.08} & 1.9 & 14.3\tiny{$\pm$2.1} &   81.9 \\
 \tiny{392208686827199744 } &    3.583298 &    46.930416 & 15.24 & 0.093 & H & 30 &  15 & 53 & 4540 & 1.10 & --2.10\tiny{$\pm$0.11} & 1.9 & 26.8\tiny{$\pm$4.0} & --113.5  \\
 \tiny{291112234783352320  } &  26.944332 &    24.114584 & 15.29 & 0.102 & H & 30 &  12 & 42 & 4285 & 0.71 & --1.94\tiny{$\pm$0.14} & 1.9 & 36.2\tiny{$\pm$5.5} &  --46.5  \\
 \tiny{5008704278351937280} &   23.906688 &  --38.859709 & 15.67 & 0.011 & M & 50 &  39 & 51 & 4940 & 1.92 & --2.04\tiny{$\pm$0.13} & 1.5 & 17.4\tiny{$\pm$2.6} &   54.5 \\
\hline 
\multicolumn{15}{c}{Cetus-Palca}\\ 
\hline 
 \tiny{1918128584760819968} &  349.622498 &    39.983749 & 14.25 & 0.109 & H & 30 & 19 & 70 & 4420 & 0.85 & --2.18\tiny{$\pm$0.11} & 1.8 & 20.6\tiny{$\pm$3.1} & --315.0  \\ 
 \tiny{2454009833214133376} &   25.256498 &  --13.688868 & 14.70 & 0.016 & M & 30 & 36 & 52 & 4600 & 1.20 & --1.99\tiny{$\pm$0.09} & 1.9 & 21.1\tiny{$\pm$3.2} & --144.1  \\ 
 \tiny{299322872223728000 } &   27.626137 &    28.865772 & 14.73 & 0.052 & H & 30 & 25 & 44 & 4500 & 1.00 & --2.12\tiny{$\pm$0.09} & 1.7 & 24.3\tiny{$\pm$3.7} & --209.0  \\ 
 \tiny{110194640179380864 } &   44.152267 &    23.833143 & 14.84 & 0.158 & H & 20 & 15 & 43 & 4610 & 1.19 & --2.22\tiny{$\pm$0.08} & 1.7 & 19.5\tiny{$\pm$2.9} & --152.0  \\ 
 \tiny{137407896564837376 } &   48.409805 &    33.542347 & 14.91 & 0.165 & H & 30 & 18 & 59 & 4325 & 0.65 & --2.17\tiny{$\pm$0.11} & 1.7 & 31.0\tiny{$\pm$4.7} & --161.0  \\ 
 \tiny{2457328468544765184} &   24.417424 &  --12.552211 & 14.97 & 0.018 & M & 30 & 32 & 46 & 4545 & 0.98 & --2.04\tiny{$\pm$0.09} & 1.8 & 29.8\tiny{$\pm$4.5} &  --81.9 \\ 
 \tiny{2507540140928136832 } &  30.152781 &   --0.309415 & 15.10 & 0.024 & H & 30 & 15 & 53 & 4500 & 0.95 & --2.35\tiny{$\pm$0.06} & 1.8 & 31.4\tiny{$\pm$4.7} &  --68.0 \\ 
 \tiny{2532901132536856320} &   18.053215 &   --1.234548 & 15.15 & 0.053 & M & 30 & 29 & 51 & 4455 & 0.96 & --1.90\tiny{$\pm$0.17} & 1.8 & 30.0\tiny{$\pm$4.5} & --106.3  \\ 
 \tiny{2502895425855079424} &   37.896091 &     1.442148 & 15.41 & 0.023 & M & 25 & 28 & 37 & 4685 & 1.33 & --2.48\tiny{$\pm$0.06} & 1.7 & 26.2\tiny{$\pm$3.9} & --109.4  \\ 
 \tiny{2818454438393552256} &  347.123444 &    16.893238 & 15.42 & 0.126 & H & 30 & 14 & 48 & 4650 & 1.34 & --2.09\tiny{$\pm$0.12} & 1.9 & 22.7\tiny{$\pm$3.4} & --244.0  \\ 
 \tiny{5012153068370401664} &   26.448789 &  --33.961611 & 15.56 & 0.021 & M & 50 & 40 & 49 & 4790 & 1.54 & --2.32\tiny{$\pm$0.11} & 1.6 & 23.4\tiny{$\pm$3.5} &  --81.2 \\ 
 \tiny{2450913123138677376} &   23.160807 &  --16.835150 & 15.85 & 0.018 & M & 50 & 33 & 53 & 4675 & 1.32 & --2.21\tiny{$\pm$0.13} & 2.0 & 32.5\tiny{$\pm$4.8} &    4.7 \\ 
\hline      
\multicolumn{15}{l}{1 -- Spectrograph/Telescope: U -- UVES/VLT, M -- MIKE/Magellan, H -- HDS/Subaru.}\\ 
\end{tabular}
\end{table*}

\section{Stellar parameters}\label{st_par}

\begin{figure}
\centering
\includegraphics[trim={0 20 0 0},width=\hsize]{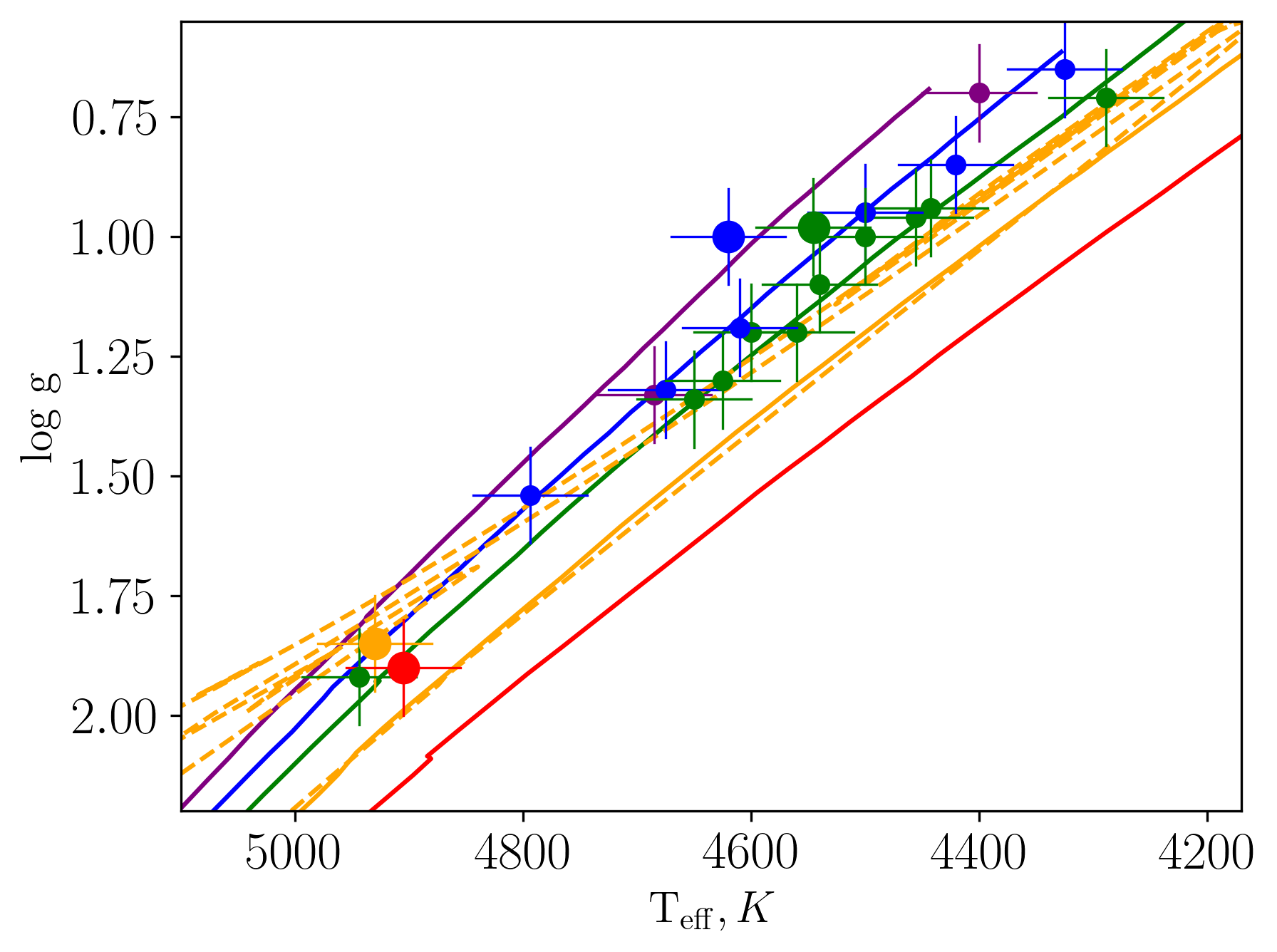}
\caption{Positions of the sample stars on giant branches of 12 Gyr  isochrones \citep[][solid lines]{2008ApJS..178...89D} with [Fe/H] from $-2.5$ (violet) to $-1.5$ (red) with a step of 0.25 dex. For comparison, we show the 12 Gyr \citet{2016ApJS..222....8D} isochrone with [Fe/H] = $-1.75,$ which includes advanced evolutionary stages (dashed orange line). Larger symbols represent  stars at more advanced evolutionary stages. The symbols have the same metallicity colour coding as the lines.}
\label{iso}
\end{figure}

Atmospheric parameters were derived through an iterative procedure. They are presented in Fig.~\ref{iso} and Table~\ref{table_par}.
First, we calculate effective temperatures (\teff ) using the {\it Gaia} ${\rm BP-G}$, ${\rm G-RP}$, ${\rm BP-RP}$ dereddened colours and the calibration of \citet{2021A&A...653A..90M}. The extinction ${\rm E(B-V)}$ is adopted from \citet{2011ApJ...737..103S} and the colours are corrected according to \citet{2018MNRAS.479L.102C}. Using different colours yields very similar effective temperatures and the uncertainty in \teff\ is therefore mainly defined by an uncertainty in the calibration  of 80~K, as given by \citet{2021A&A...653A..90M}.

To estimate our initial surface gravities (log~g), we made use of distances based on {\it Gaia} parallaxes and isochrones. Parallaxes are corrected for the zero offset according to \citet{2021A&A...649A...4L} and distances are computed from the maximum of the probability distribution function as described in \citet{2015PASP..127..994B}. With these distances, effective temperatures, bolometric corrections of \citet{2018MNRAS.479L.102C}, and assuming a mass of 0.8 solar masses, we derive surface gravities using the formula $\log g = 4.44+\log(m/m_{\odot})+0.4(M_{\rm bol} - 4.75) + 4\log($\teff$/5780)$, where $m_{\odot}$ is a solar mass and $M_{\rm bol}$ is an absolute bolometric magnitude. For our sample stars, the ratio between the parallax error and the parallax exceeds 0.2 and the uncertainty in distance results in an uncomfortably large uncertainty of 0.3~dex in log~g. Consequently, we refine the initially calculated log~g values using the 12 Gyr isochrones with [Fe/H] = $-2$ and [$\alpha$/Fe] = 0.4 \citep{2008ApJS..178...89D}, as described below. 

Starting with the first step parameters, we calculated NLTE abundances using Fe\ione\ and Fe\ii\ lines. To further refine the stellar parameters, we used the difference in NLTE abundance from Fe\ione\ and Fe\ii, along with the derived [Fe/H]. This refinement process involves fitting isochrones with [Fe/H] values ranging from $-2.5$ to $-1.5$, and is designed to keep abundances from the Fe\ione\ and Fe\ii\ lines consistent within the error bars. It is worth noting that using the same method for iron NLTE abundance determination, \citet{2017A&A...604A.129M} found consistent Fe\ione\ and Fe\ii\ abundances in VMP giants with photometric effective temperatures and accurate distance-based surface gravities.

\begin{figure}
 \includegraphics[trim={0 15 0 10},width=80mm]{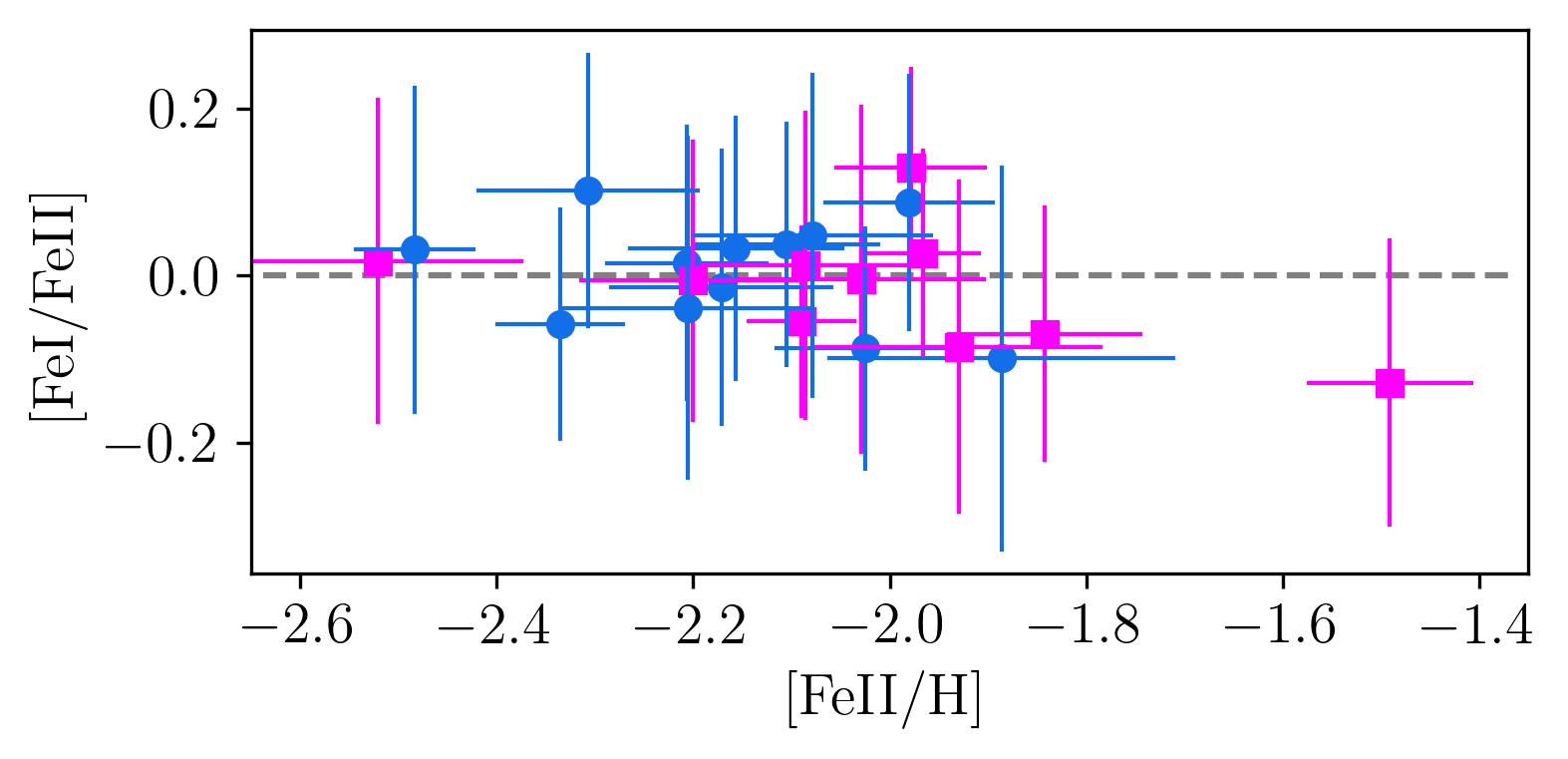}        
   \caption{NLTE abundance difference between Fe\ione\ and Fe\ii\ as a function of metallicity [Fe\ii/H]. The Cetus-New stars and the Cetus-Palca stars are shown with magenta squares and blue circles, respectively.}
   \label{fe12} 
\end{figure}

The tuning procedure is repeated until a reasonable convergence between isochrones, ionisation balance, and metallicity is achieved. 
When a given star's parameters (\teff, log~g,  and [Fe/H]) are found to be within 50~K and 0.1~dex of the corresponding isochrone, and the NLTE abundances from Fe\ione\ and Fe\ii\ lines agree within the error bars, we adopt these parameters as the final values. Based on the criteria mentioned above, we assign uncertainties of 50~K for \teff\ and 0.1~dex for log~g. Figure~\ref{fe12} shows the NLTE abundance difference between Fe\ione\ and Fe\ii\ as a function of metallicity based on the Fe\ii\ lines.  The difference never exceeds 0.13~dex in absolute value and, for the 22 sample stars, the average NLTE difference amounts to Fe\ione --Fe\ii = $-0.01$ $\pm$  0.07. The microturbulent velocity is derived from the lines of Fe\ione\ and Fe\ii\ and its uncertainty amounts to 0.2~\kms.

Among the sample stars, four do not  align with the giant branch  with the corresponding [Fe/H] and their positions are shifted towards higher \teff\ and/or lower log~g. Changing \teff\ and log~g leads to a discrepancy between abundances from Fe\ione\ and Fe\ii. 
These stars might be at more advanced evolutionary stages than the red giant branch, such as early asymptotic giant branch (AGB) stars in a phase prior to the onset of thermal pulses. The \citet{2016ApJS..222....8D} isochrone, which accounts for advanced evolutionary stages, supports this guess (see Fig.~\ref{iso}).

With our final surface gravities, we calculate the corresponding distances, referred to as  spectroscopic distances (d$_{\rm spec}$). These are listed in Table~\ref{table_par}. The distances to our sample stars range from 14 to 36 kpc. These results can be employed for improving the precision of the orbits and characteristics of the Cetus stream's progenitor. It is worth noting that our distances differ significantly from those provided by the {\it Gaia} DR3 catalogue \citep[][distances from GSP-Phot Aeneas best library using BP/RP spectra]{2021A&A...649A...1G} and the \citet{2021AJ....161..147B} catalogue, which provide unexpectedly small values for our sample stars from 2.3 to 8.5 kpc and from 5.2 to 12.4 kpc, respectively. 

\section{Abundance analysis methods}\label{abund}

\subsection{Codes and model atmospheres}

We use classical 1D model atmospheres from the \textsc {marcs} model grid \citep{marcs}, interpolated for the given \teff, log~g, and [Fe/H] of the stars.
We solve the coupled radiative transfer and statistical equilibrium equations with the \textsc {detail} code \citep{Giddings81,Butler84} using the updated opacity package as presented by  \citet{mash_fe}. For synthetic spectra calculations, we use the \textsc {synthV\_NLTE} code \citep{Tsymbal2018} attached to the \textsc {idl binmag} code  \citep{2018ascl.soft05015K}. This technique allows us to obtain the best fit to the observed line profiles with the NLTE effects taken into account via  pre-calculated departure coefficients (the ratio between NLTE and LTE atomic level populations) for a given model atmosphere. When fitting the line profiles, the abundance of the element of interest is varied together with the macroturbulent velocity (v$_{\rm mac}$) and the radial velocity (v$_{\rm r}$). 

The line list for spectral synthesis is extracted from a recent version of the Vienna Atomic Line Database \citep[VALD,][]{2019ARep...63.1010P,2015PhyS...90e4005R}, which provides isotopic and hyperfine structure components of the spectral lines for a number of studied chemical elements. We adopt the oscillator strengths recommended by VALD. The exception is Fe\ii, where we take gf-values from \citet{ru1998} corrected by +0.11~dex, following the recommendation of \citet{1999A&A...347..348G}. The same approach was adopted in the earlier studies of VMP stars \citep{2017A&A...604A.129M,2017A&A...608A..89M,2019AA...631A..43M}. For barium, we determine abundances using two different isotope ratios ---solar mixture, as provided by VALD by default, and the r-process ratio taken from \citet{1999ApJ...525..886A}. For the Ba\ii\ resonance line at 4554 \AA, using the r-process ratio leads to lower abundances compared to those derived with the solar mixture, and the abundance shifts vary from 0.06~dex to 0.20~dex in our sample stars with EW$_{\rm 4554}$ from 280 to 140  m\AA, respectively. The subordinate lines of Ba\ii\ are weaker and their abundances are nearly immune to the adopted isotope ratio. Our average barium abundances rely only on the subordinate Ba\ii\ lines. For each star, we provide abundances from individual spectral lines together with their atomic data and measured EWs; see Table~\ref{tab:individual}.

\begin{table}
\caption{NLTE and LTE abundances from individual lines and their atomic data.}
\label{tab:individual}
\setlength{\tabcolsep}{0.95mm}
\begin{tabular}{lllcrrrr} 
\hline
{\it Gaia} ID &Sp.  &  $\lambda$,  & \eexc, & log~gf & EW,  & $\eps$ & $\eps$ \\
  &  &  \AA\        &  eV    &        &  m\AA & LTE  & NLTE   \\
\hline
2483... &   CH      &   4192.56 & 0.64 & --1.28 &   19.0 & 5.37 &      \\      
2483... &   CH      &   4210.94 & 0.46 & --1.34 &   24.4 & 5.29 &      \\ 
2483... &   O\ione\ &   6300.30 & 0.00 & --9.78 &   24.1 & 7.25 & 7.25 \\ 
2483... &  Na\ione\ &   5889.95 & 0.00 &   0.11 &  249.1 & 4.05 & 3.86 \\ 
2483... &  Na\ione\ &   5895.92 & 0.00 & --0.19 &  222.2 & 4.05 & 3.81 \\ 
2483... &  Na\ione\ &   8183.25 & 2.10 &   0.24 &   63.3 & 3.99 & 3.90 \\ 
2483... &  Na\ione\ &   8194.80 & 2.10 &   0.49 &   87.4 & 3.99 & 3.85 \\ 
2483... &  Mg\ione\ &   4702.99 & 4.35 & --0.44 &  116.2 & 5.76 & 5.71 \\ 
2483... &  Mg\ione\ &   5711.09 & 4.35 & --1.72 &   41.4 & 5.78 & 5.76 \\ 
2483... &  Al\ione\ &   3961.52 & 0.01 & --0.32 &  186.0 & 3.95 & 4.05 \\ 
2483... &  Si\ione\ &   4102.94 & 1.91 &  -3.14 &  120.7 & 5.86 &      \\ 
2483... &  Si\ione\ &   5948.54 & 5.08 &  -1.23 &   25.1 & 5.84 &      \\ 
2483... &  Si\ione\ &   6155.13 & 5.62 & --0.76 &   12.5 & 5.65 &      \\ 
2483... &  Si\ione\ &   7932.35 & 5.96 & --0.47 &   11.7 & 5.68 &      \\ 
2483... &  S\ione\  &   9212.86 & 6.52 &   0.47 &   43.5 & 5.40 & 5.17 \\ 
2483... &  K\ione\  &   7698.96 & 0.00 & --0.15 &  108.2 & 3.39 & 3.01 \\ 
2483... &  Ca\ione\ &   5857.45 & 2.93 &   0.23 &   70.6 & 4.49 & 4.55 \\ 
2483... &  Ca\ione\ &   6102.72 & 1.88 & --0.79 &   91.2 & 4.45 & 4.52 \\ 
2483... &  Ca\ione\ &   6122.22 & 1.89 & --0.31 &  124.8 & 4.54 & 4.57 \\ 
2483... &  Ca\ione\ &   6169.04 & 2.52 & --0.80 &   45.6 & 4.57 & 4.73 \\ 
2483... &  Ca\ione\ &   6439.07 & 2.53 &   0.39 &  108.6 & 4.40 & 4.34 \\ 
2483... &  Ca\ione\ &   6471.66 & 2.53 & --0.69 &   44.9 & 4.43 & 4.50 \\ 
2483... &  Ca\ione\ &   6493.78 & 2.52 & --0.11 &   73.8 & 4.30 & 4.32 \\ 
2483... &  Ca\ione\ &   6499.65 & 2.52 & --0.82 &   32.5 & 4.34 & 4.43 \\ 
2483... &  Sc\ii\   &   4400.38 & 0.61 & --0.54 &  107.8 & 1.01 & 0.98 \\ 
2483... &  Sc\ii\   &   4415.54 & 0.60 & --0.68 &  111.5 & 1.16 & 1.15 \\ 
2483... &  Sc\ii\   &   6245.64 & 1.51 & --1.02 &   35.1 & 1.03 & 1.05 \\ 
2483... &  Ti\ione\ &   4840.87 & 0.90 & --0.43 &   52.6 &      &      \\ 
2483... &  Ti\ione\ &   4913.61 & 1.87 &   0.16 &   14.3 &      &      \\ 
2483... &  Ti\ione\ &   4981.73 & 0.85 &   0.57 &  100.2 &      &      \\ 
2483... &  Ti\ione\ &   6261.10 & 1.43 & --0.53 &   22.0 &      &      \\ 
2483... &  Ti\ii\   &   4764.52 & 1.24 & --2.69 &   51.5 & 3.18 & 3.18 \\ 
2483... &  Ti\ii\   &   4798.53 & 1.08 & --2.66 &   73.3 & 3.35 & 3.35 \\ 
2483... &  Ti\ii\   &   4911.19 & 3.12 & --0.64 &   21.1 & 2.85 & 2.85 \\ 
2483... &  Ti\ii\   &   6491.57 & 2.06 & --1.94 &   35.8 & 3.07 & 3.07 \\ 
\hline 
\end{tabular}\\
The table is accessible in a machine-readable format at the CDS. A portion is shown to illustrate its format and content.
\end{table}

\subsection{NLTE effects}

We take into account the departure from LTE for a number of species: O\ione\ \citep{2018AstL...44..411S},  Na\ione\ \citep{2014AstL...40..406A}, Mg\ione\ \citep{2013AA...550A..28M}, Al\ione\ \citep{2022A&A...665A..33L}, S\ione\ \citep{2005PASJ...57..751T}, K\ione\ \citep{2020AstL...46..621N}, Ca\ione\  \citep{2017AA...605A..53M}, Sc\ii\ \citep{2022AstL...48..455M}, Ti\ii\  \citep{2020AstL...46..120S}, Cr\ione\ \citep{2010A&A...522A...9B}, Mn\ione\ \citep{2019A&A...631A..80B}, Fe\ione\ \citep{mash_fe}, Co\ione\ \citep{2010MNRAS.401.1334B}, Cu\ione\ \citep{2023MNRAS.518.3796C}, Zn\ione\ \citep{2022MNRAS.515.1510S}, Sr\ii\  \citep{sr_nlte,atoms10010033}, Y\ii\ \citep{2023ApJ...957...10A}, Zr\ii\ \citep{Velichko2010_zr}, Ba\ii\ \citep{2019AstL...45..341M}, and Eu\ii\ \citep{mash_eu}.
We refer the reader to the papers listed above for the description of the model atoms and the mechanism of the NLTE effects. For those interested in the original NLTE studies conducted from the 1960s to the present, we suggest the bibliography compiled by \citet{2018abmc.conf...40S} and available on the web page \url{non-lte.com/bibliography.html}. 

For species where NLTE effects are taken into account, we present the differences between the average NLTE and LTE abundances in our Cetus stars as a function of metallicity (Fig.~5). 
For oxygen, our abundance determinations mostly rely on the O\ione\ 6300~\AA\ line, where NLTE effects are negligible; hence, we do not show the corresponding differences. In the most metal-rich star of our sample, we also use the O\ione\ 7771~\AA\ line and apply a NLTE abundance correction of $-0.20$~dex. Abundances of sodium and aluminium are strongly affected by NLTE effects. For sodium, the difference between NLTE and LTE abundances reaches down to $-0.6$~dex. For the Al\ione\ 3961 \AA\ line, NLTE corrections are positive and they take values of up to 0.2~dex. For magnesium and calcium, NLTE effects are small and they result in up to 0.1~dex lower and higher abundances, respectively. Accounting for the NLTE effects is particularly essential for S\ione\ and K\ione, where NLTE typically yields 0.4 dex lower abundances compared to LTE. Sc\ii\ and Ti\ii\ are the majority species in the investigated stellar parameter range and NLTE leads to a moderate positive shift in their abundances of up to 0.1~dex. NLTE corrections are positive for lines of neutral species of the iron-group elements and copper. However, the magnitudes of the NLTE effects are different for different elements, with a typical shift between NLTE and LTE abundances of 0.1~dex for Fe\ione\ and of as much as 1~dex for Co\ione. Abundances of n-capture elements are derived from lines of the singly ionised atoms, which are all the majority species in the atmospheres of the sample stars. The NLTE effects for these lines are moderate and they result in an abundance difference of up to 0.2~dex in absolute value.

\begin{figure*}
\label{fig_diff_nlte_lte}
 \includegraphics[trim={0 20 0 10},width=95mm]{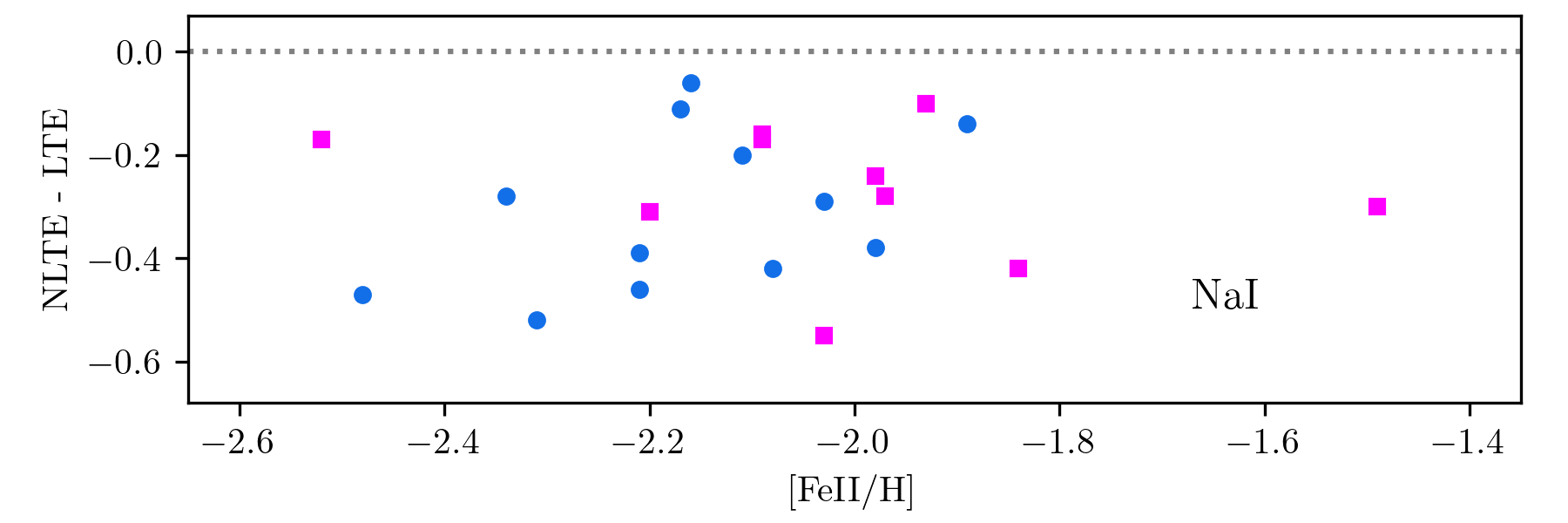}      
 \includegraphics[trim={0 20 0 10},width=95mm]{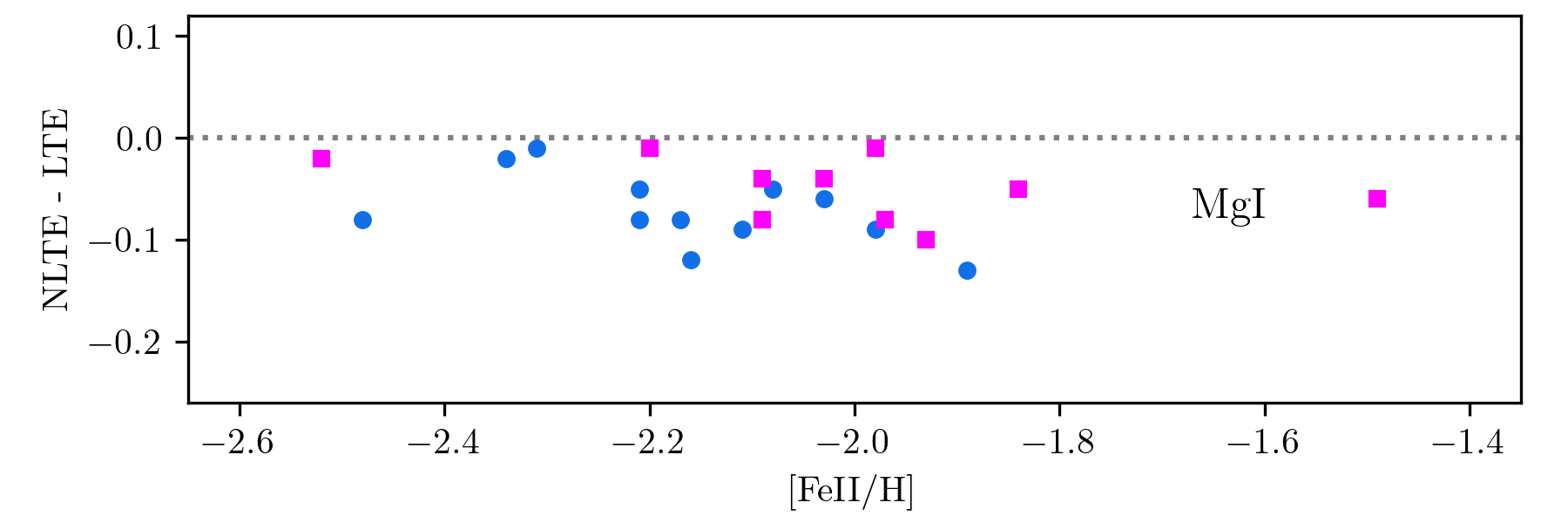}  
 \includegraphics[trim={0 20 0 10},width=95mm]{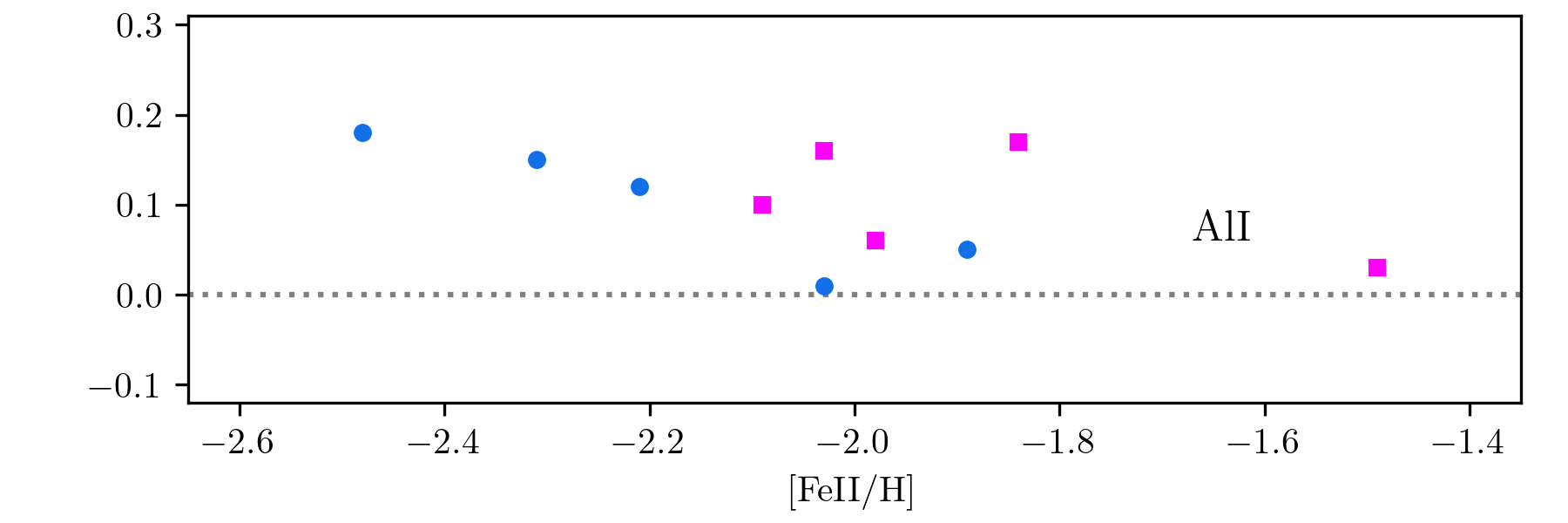}  
 \includegraphics[trim={0 20 0 10},width=95mm]{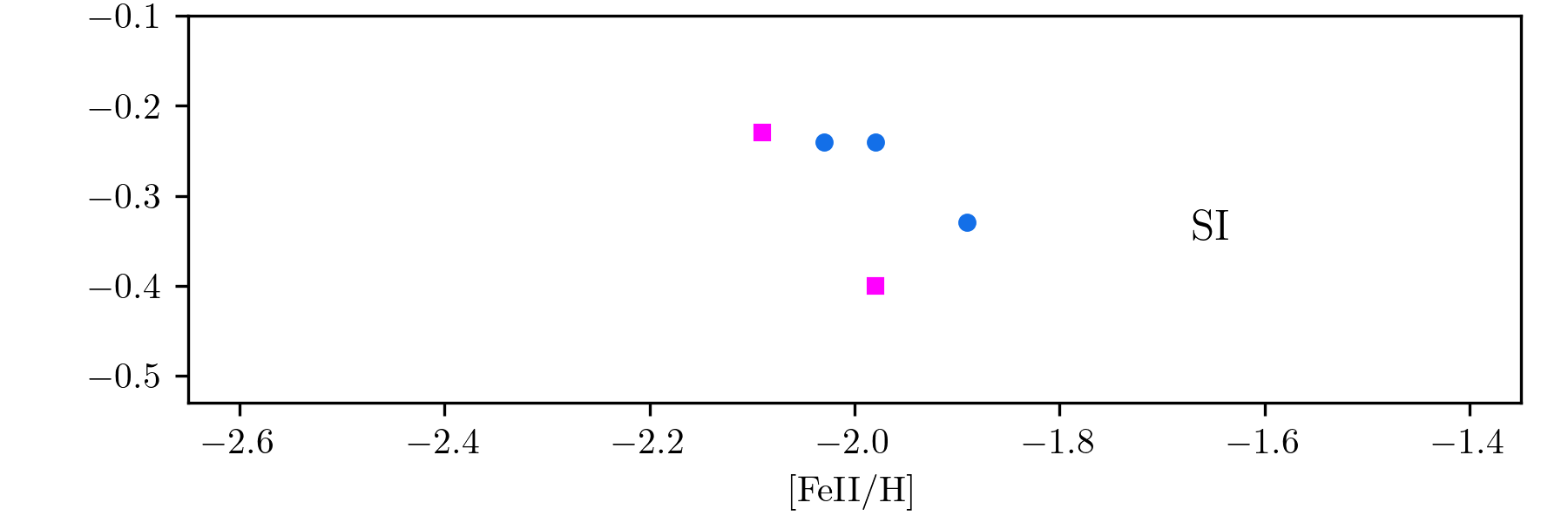}       
 \includegraphics[trim={0 20 0 10},width=95mm]{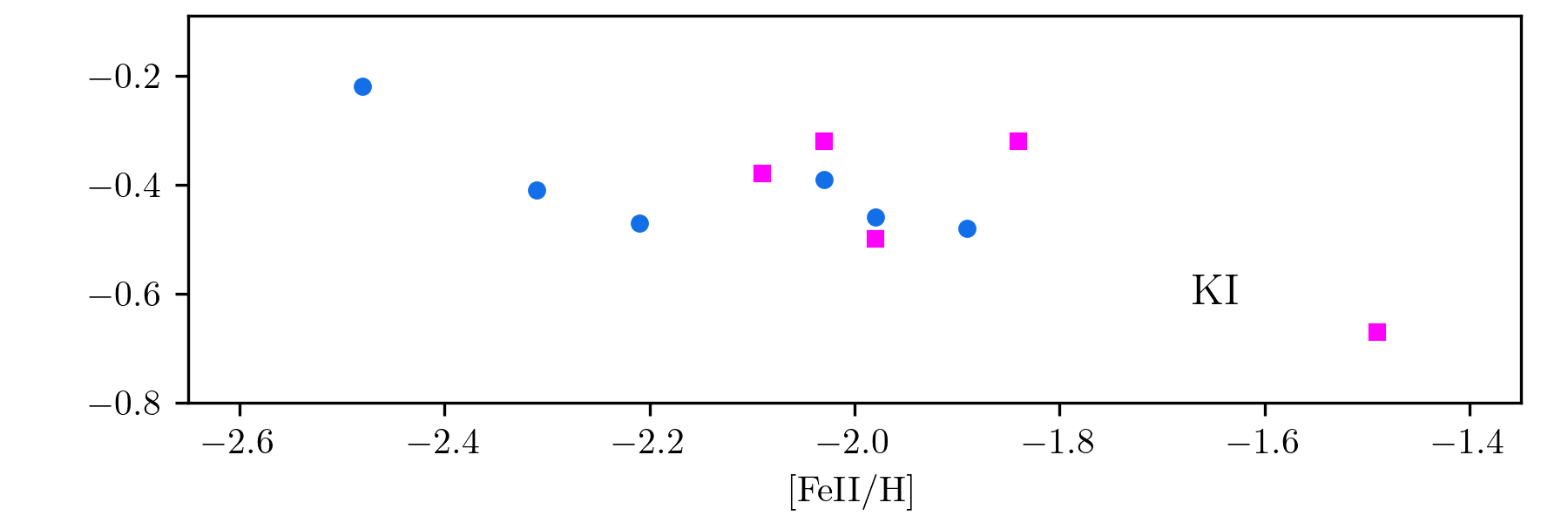}      
 \includegraphics[trim={0 20 0 10},width=95mm]{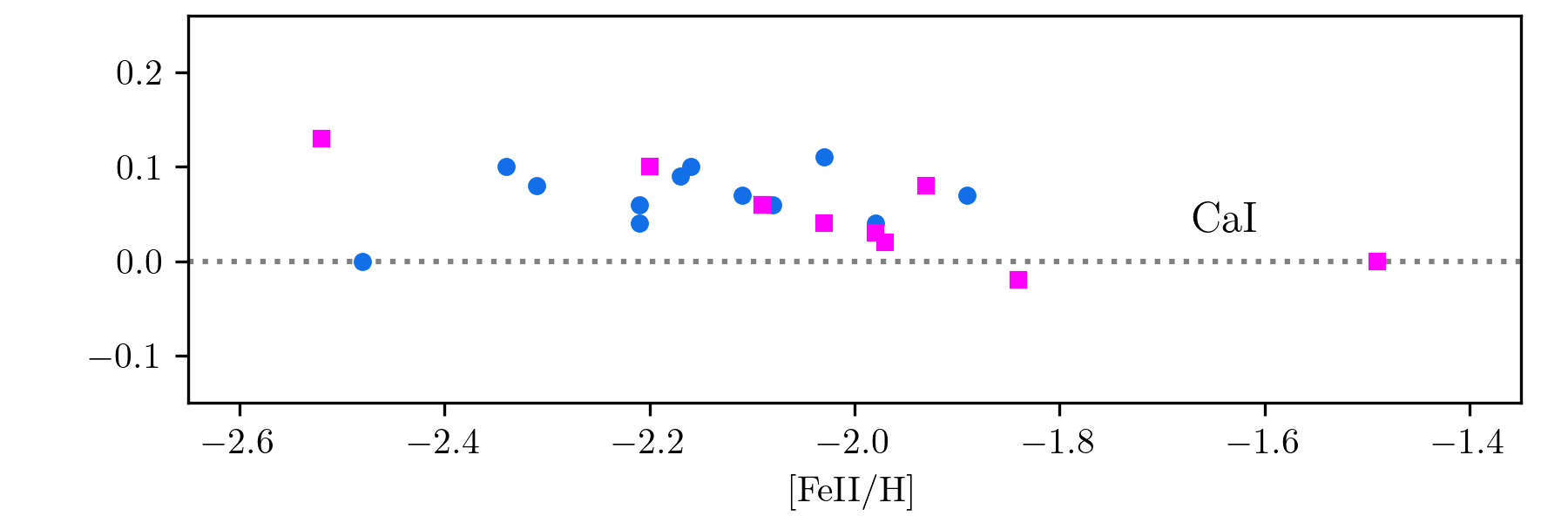}  
 \includegraphics[trim={0 20 0 10},width=95mm]{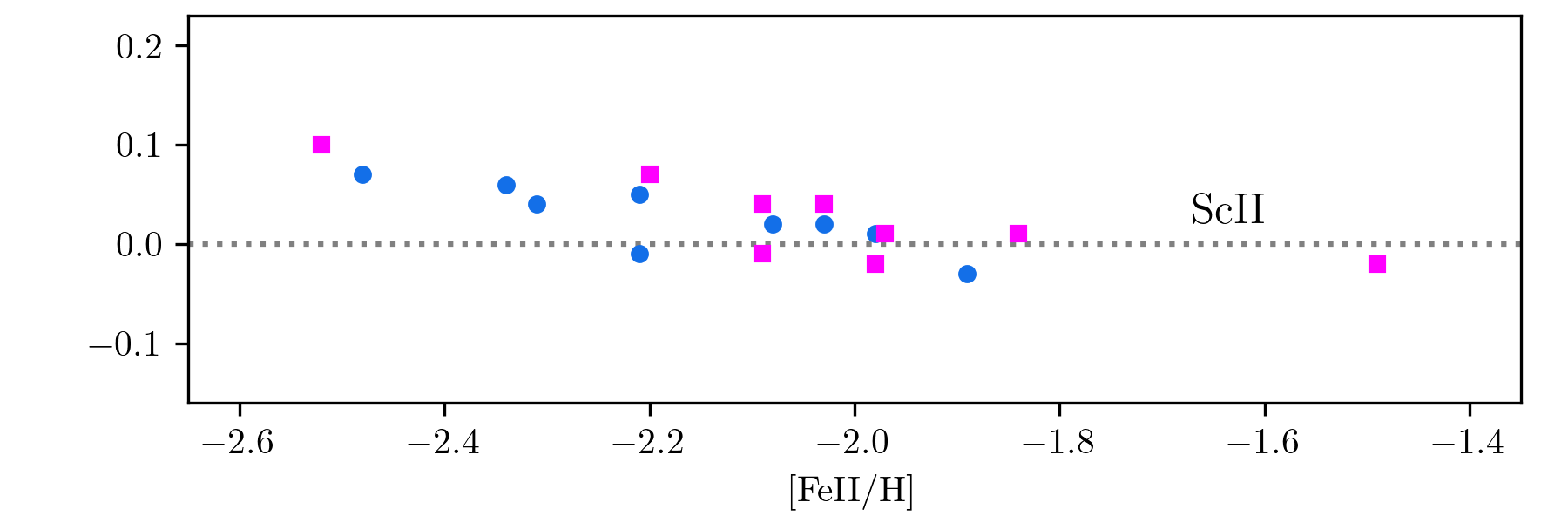}  
 \includegraphics[trim={0 20 0 10},width=95mm]{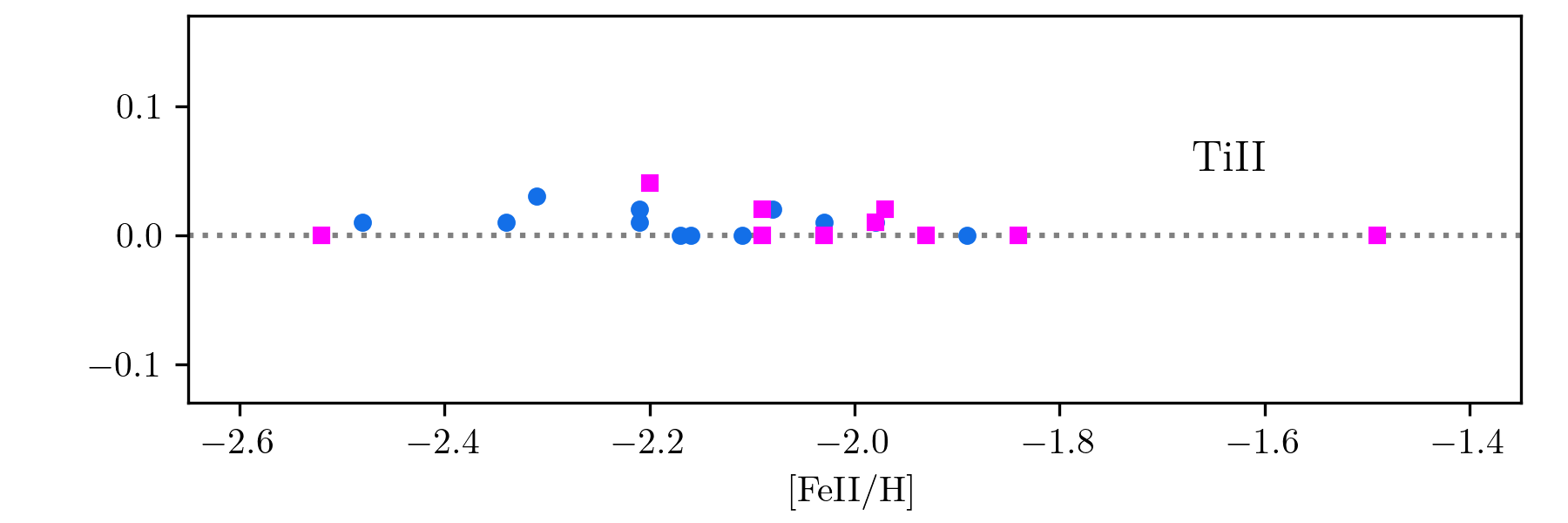} 
 \includegraphics[trim={0 20 0 10},width=95mm]{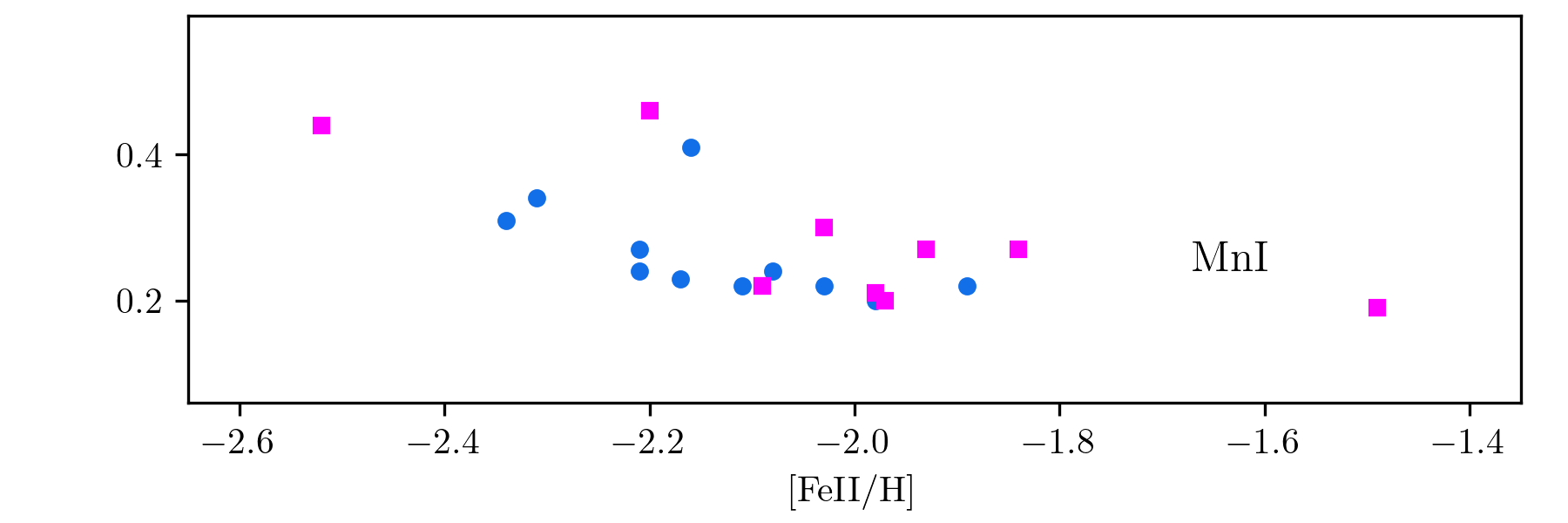}       
 \includegraphics[trim={0 20 0 10},width=95mm]{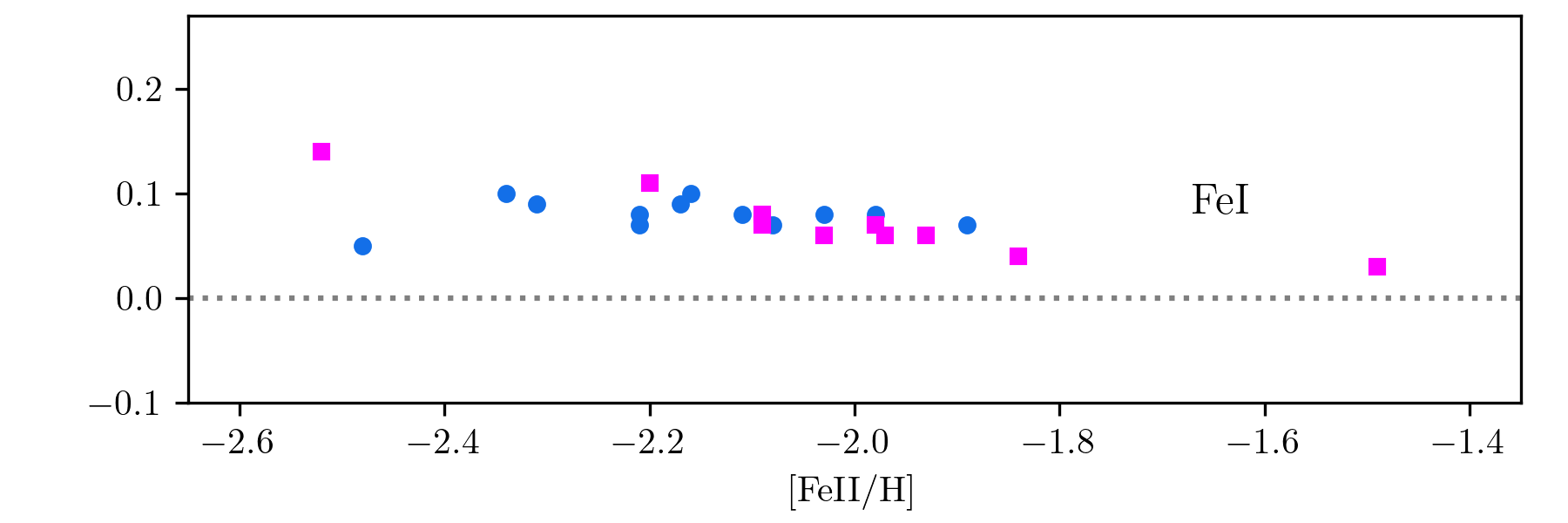}      
 \includegraphics[trim={0 20 0 10},width=95mm]{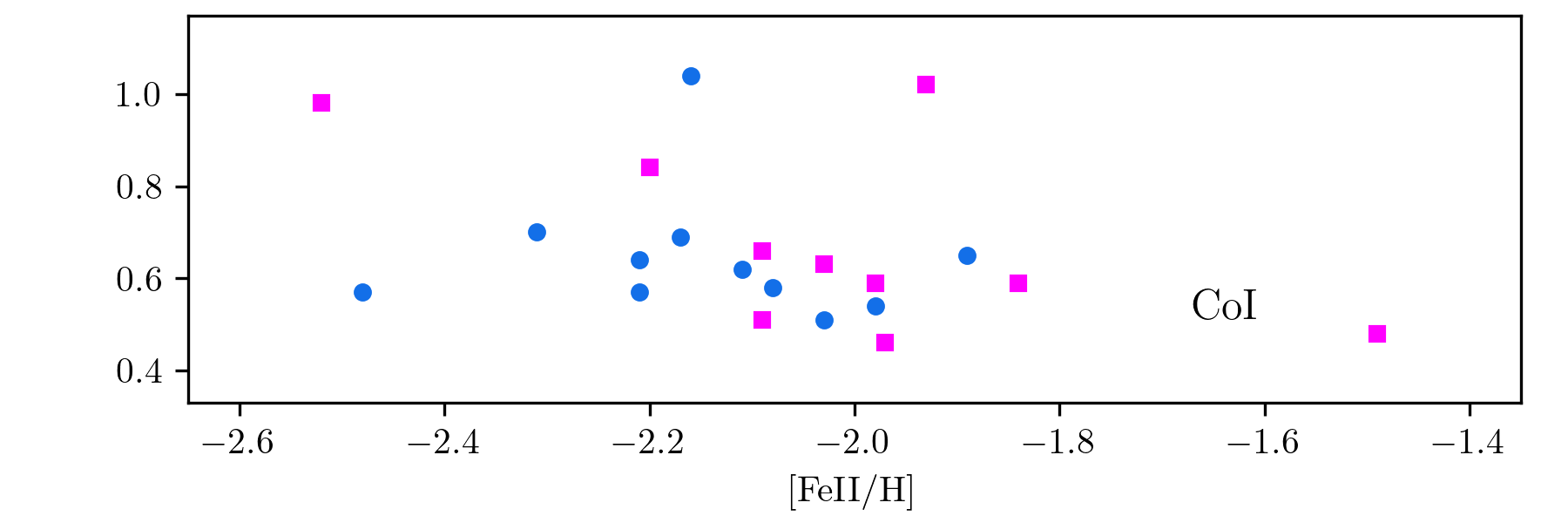}  
 \includegraphics[trim={0 20 0 10},width=95mm]{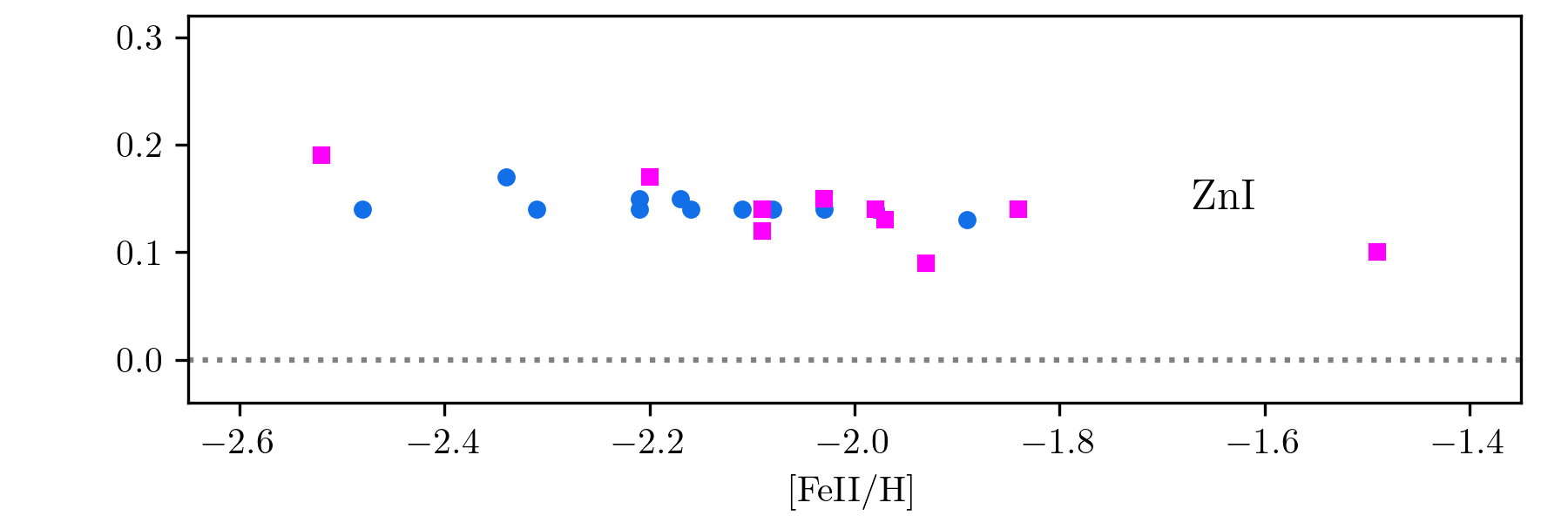}  
 \includegraphics[trim={0 20 0 10},width=95mm]{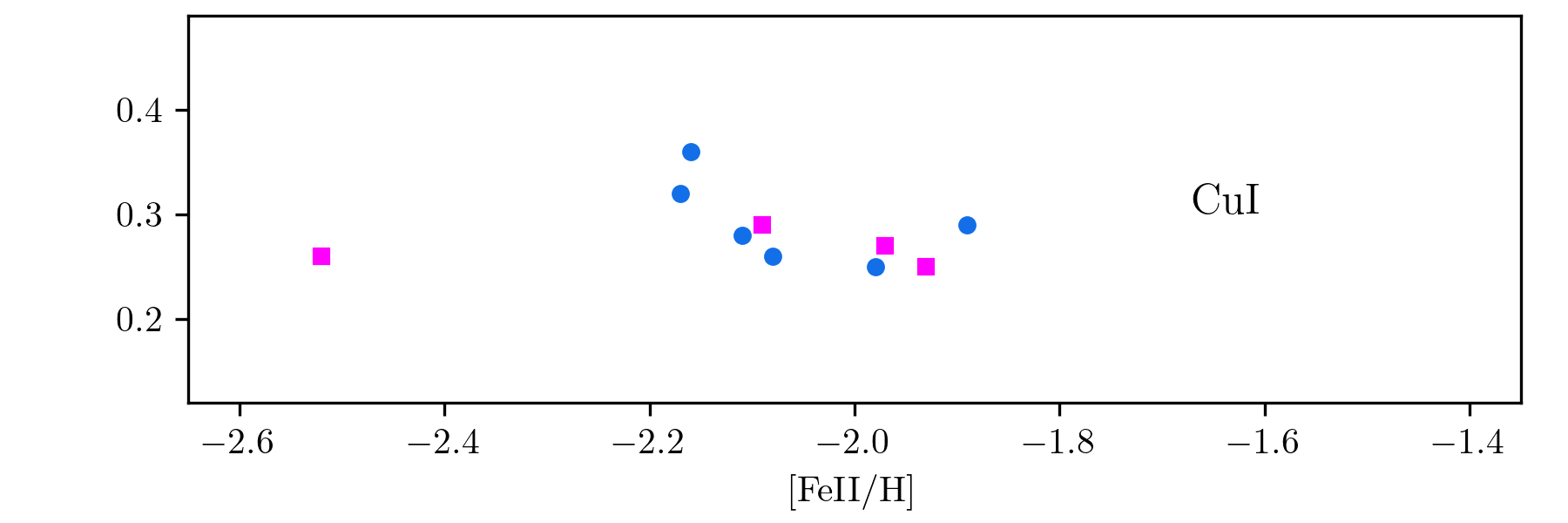} 
 \includegraphics[trim={0 20 0 10},width=95mm]{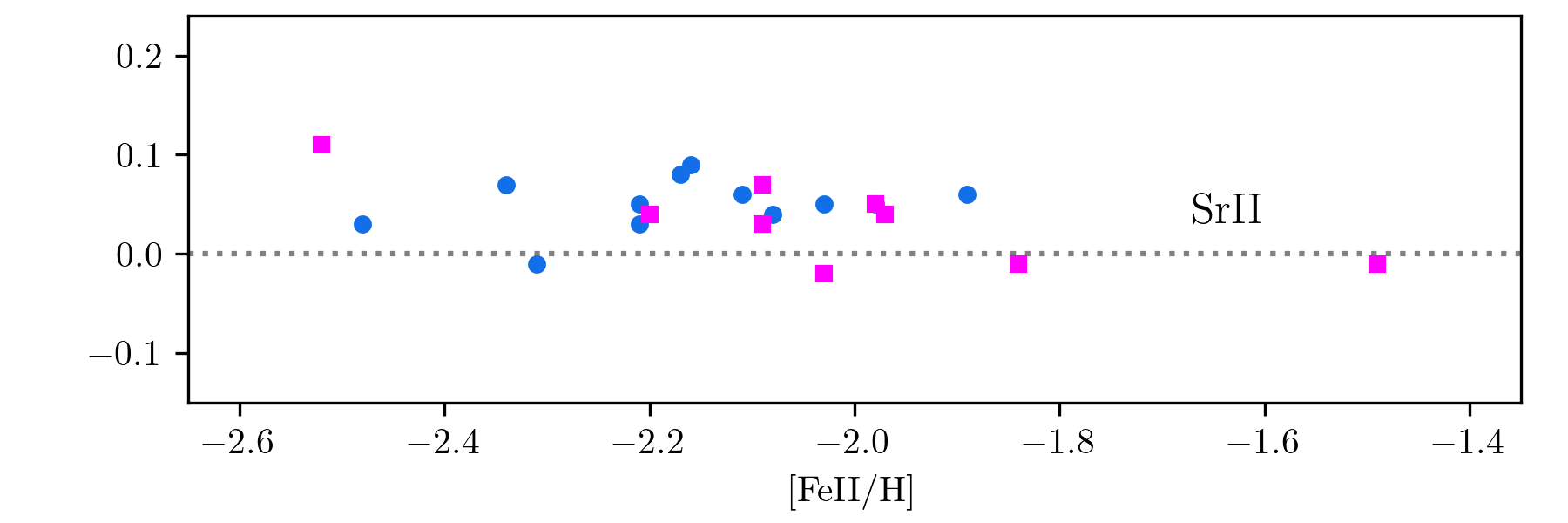}  
 \includegraphics[trim={0 20 0 10},width=95mm]{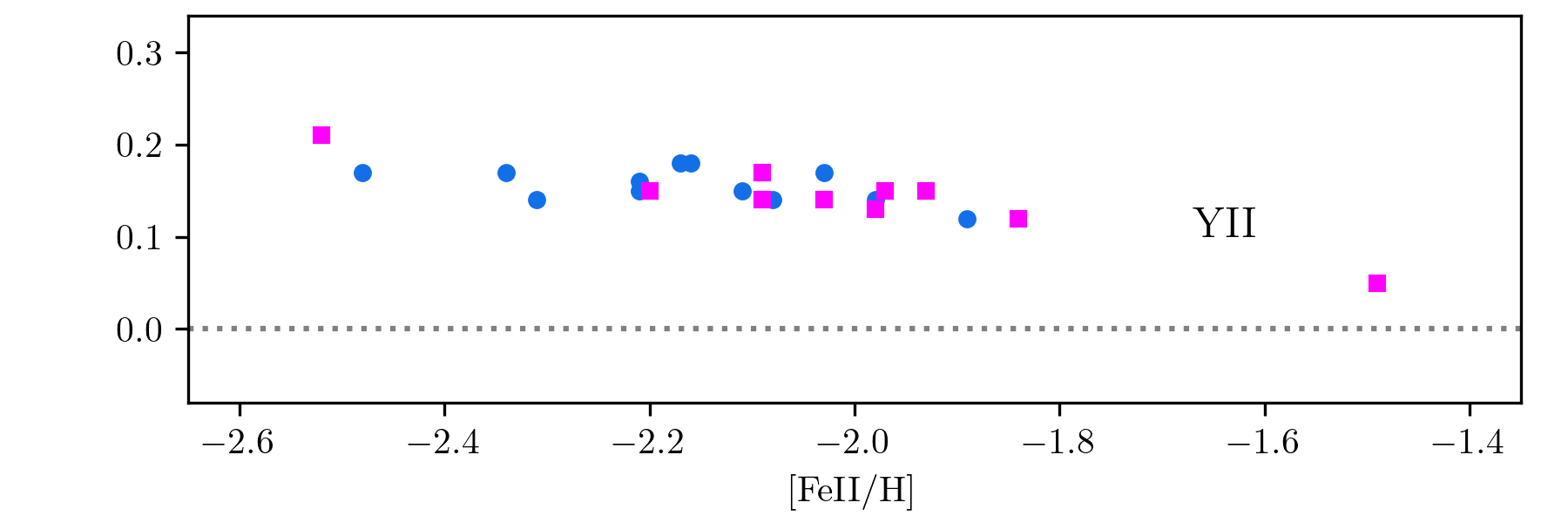} 
 \includegraphics[trim={0 20 0 10},width=95mm]{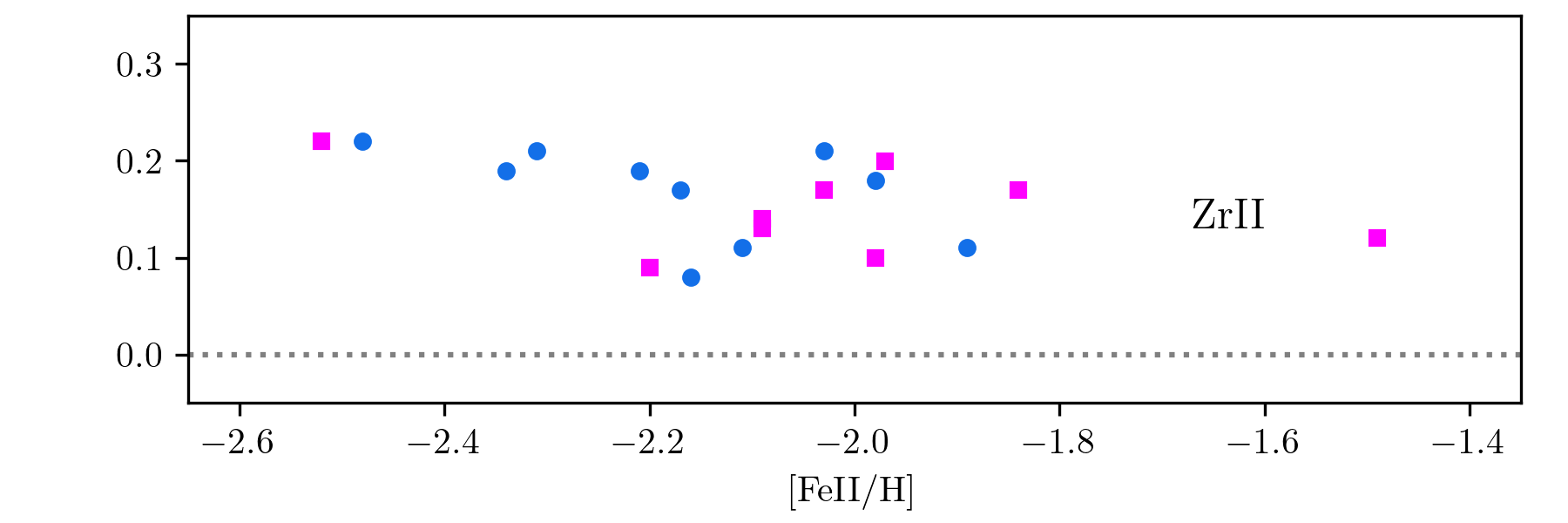}  
 \includegraphics[trim={0 10 0 10},width=95mm]{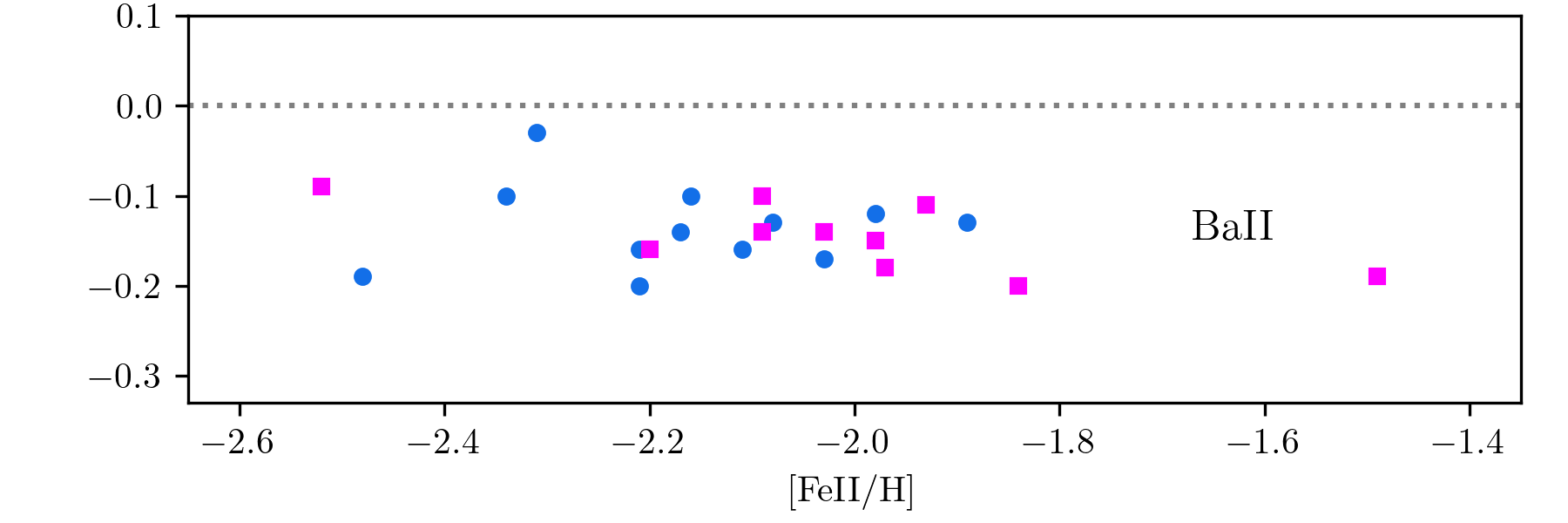} 
 \includegraphics[trim={0 10 0 10},width=95mm]{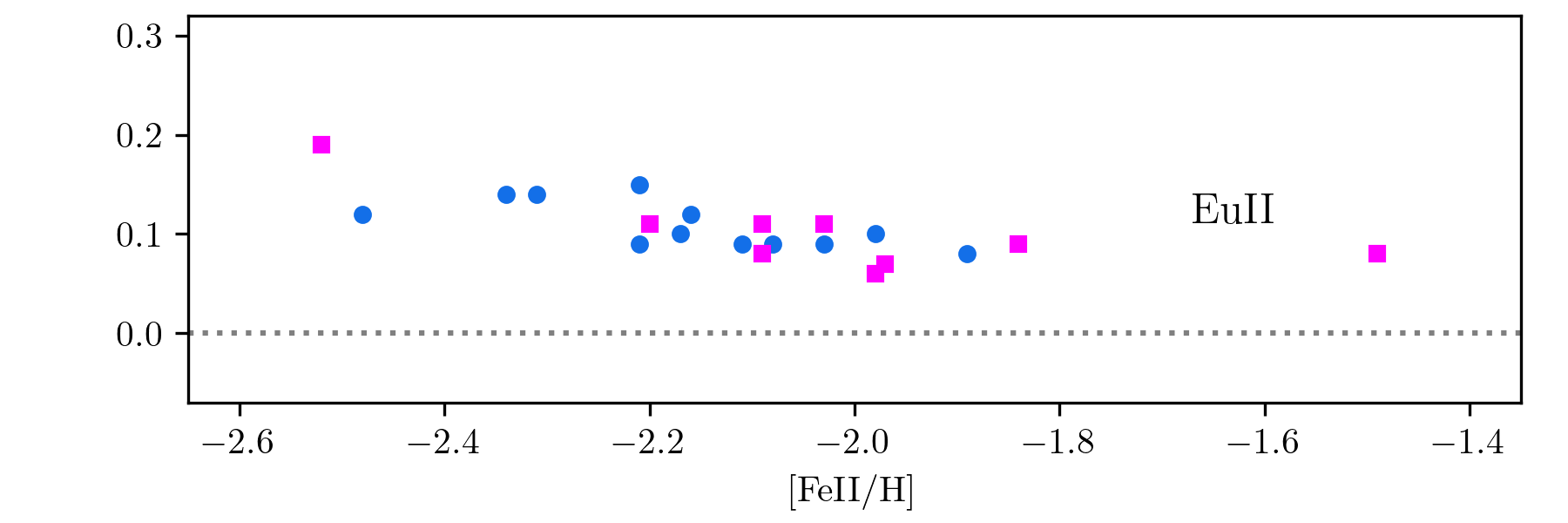}  
   \caption{Differences between the NLTE and LTE average abundance of different chemical species in the Cetus-New (magenta squares) and Cetus-Palca (blue circles) sample stars plotted as a function of metallicity.}   
\end{figure*}

For most chemical elements, we perform NLTE calculations with the specific model atmospheres, while for Al\ione, Cr\ione, Mn\ione, Co\ione, and Zn\ione, we interpolate the NLTE abundance corrections in the precalculated grids available in the literature. It is worth noting that a reasonable amount of data has been accumulated on the NLTE abundance corrections for different species, and these can be retrieved from the following user-friendly databases: \url{https://nlte.mpia.de/gui-siuAC_secEnew.php} (M.~Kovalev and M.~Bergemann), inspect-stars.com (K.~Lind), and \url{http://spectrum.inasan.ru/nLTE2/} \citep{2023MNRAS.524.3526M}. Abundances of C, Si, V, Ni, La, Ce, Nd, Sm, and Dy are determined under the LTE assumption.

Our carbon abundance is based on the LTE analysis of the molecular CH lines. \citet{2023A&A...670A..25P} report notable NLTE effects for the CH lines, with positive NLTE corrections that grow with decreasing metallicity. These authors calculate $\Delta_{\rm NLTE}$ = 0.15~dex for the model 4500/2.0/$-2.0$ with [C/Fe] = 0.0. At the same time, \citet{2015MNRAS.453.1619A} found the LTE abundance from the CH lines to be consistent with the NLTE abundance from the C\ione\ lines for cool stars with well-determined stellar atmosphere parameters.

For Si\ione, NLTE effects are minor and can be neglected even in metal-poor stars \citep{2016AstL...42..366M}.
For V\ione -\ii\ NLTE calculations are not available in the literature. For our sample of stars, where both V\ione\ and V\ii\ lines are detected, we find a systematically lower LTE abundance from V\ione\ compared to V\ii, and the difference spans from $-0.3$ to $-0.7$ dex. This result can be explained by the fact that V\ione\ is the least abundant species and may suffer from overionisation, which leads to positive NLTE abundance corrections, while V\ii\ is the most abundant species in the investigated atmosphere parameter range and minor NLTE abundance corrections are expected. 

The departures from LTE for Ni\ione\ in the solar atmosphere were investigated by \citet{1993AA...269..509B}, \citet{2013ApJ...769..103V}, \citet{2021MNRAS.508.2236B}, and \citet{2022A&A...661A.140M} and in FGK stars by \citet{eitnerni}. For Ni\ione\ lines in the visible range, \citet{eitnerni} predict positive NLTE abundance corrections, increasing towards higher \teff\ and lower log~g. For example, these latter authors found $\Delta_{\rm NLTE} \sim$ 0.2~dex in the model atmosphere with \teff/log~g/[Fe/H] = 5000/3/$-2.5$.

We plan to study the NLTE calculations for La\ii, Ce\ii, Nd\ii, Sm\ii, and Dy\ii\  in late-type stars. The above species form the majority of the species found in atmospheres of cool stars, and one might expect moderate or minor NLTE effects.

\section{Comparison samples: MW halo and UMi dSph}\label{comparison_sample}

For the purpose of comparison, we adopted a sample of the MW halo stars and a sample of stars from the UMi~dSph that were analysed using similar methods to those of this study. The choice of the UMi~dSph is motivated by its stellar mass of M$_{\rm *} = 10^{5.5}M_{\odot}$ \citep{2012AJ....144....4M}, 10$^{5.7}M_{\odot}$ \citep{kirby2013}, which is similar to the progenitor mass of the Cetus stream M$_{\rm *} = 10^{5.6}M_{\odot}$ \citep{yuan2022}. Another reason is that the UMi dSph is one of the most well-studied nearby dwarf galaxies. High-resolution optical spectra were obtained for more than a dozen UMi stars \citep{2001ApJ...548..592S,2004PASJ...56.1041S, 2007PASJ...59L..15A, 2010ApJ...719..931C, 2015MNRAS.449..761U, 2012AJ....144..168K, 2023MNRAS.525.2875S}. These latter analysis yielded valuable insights into the chemical evolution history of the UMi dSph. In this study, we do not aim to reinvestigate the UMi chemical evolution and we simply use its stellar abundances for comparison with those derived in the Cetus stars. For consistency, we reanalyse abundances for some of the UMi stars with the methods adopted in this study.

\subsection{MW halo stars}
Our MW halo comparison sample includes VMP dwarfs and giants. For dwarfs, photometric effective temperatures and distance-based surface gravities were determined by \citet{2015ApJ...808..148S}. For each star, NLTE abundances of Fe\ione\ and Fe\ii\ 
 were obtained that are consistent within
the error bars. 
For the MW halo giants, surface gravities were determined from NLTE abundances of Fe\ione\ and Fe\ii\  \citep[][hereafter MJ17a]{2017A&A...604A.129M}. These gravities were confirmed by calculations of \citet{2023MNRAS.524.3526M}  based on accurate {\it Gaia} parallaxes.

For the dwarfs, the NLTE abundances for a number of species are taken from \citet{2016ApJ...833..225Z}. In addition, we employ previously determined NLTE abundances of scandium \citep{2022AstL...48..455M}, zinc \citep{2022MNRAS.515.1510S}, and yttrium \citep{2023ApJ...957...10A} for our star sample. For the giants, the NLTE abundances are taken from \citet[][hereafter MJ17b]{2017A&A...608A..89M}. For oxygen, potassium, scandium, chromium, manganese, cobalt, copper, and zinc, MJ17b do not present abundances and so we adopted those of Sitnova et al. (in prep.). The same line lists and NLTE methods used for the Cetus stars (see Sect.~\ref{abund}) were applied to determine the abundances in the MW halo stars.

\subsection{UMi dSph}
We adopt the NLTE abundances of the UMi~dSph VMP stars from MJ17b, which are based on the high-resolution spectra collected by \citet[][hereafter CH10]{2010ApJ...719..931C}, \citet{2012AJ....144..168K}, and \citet{2015MNRAS.449..761U}. The studies of MJ17a and  MJ17b focus on the early chemical enrichment and are therefore limited to stars with [Fe/H] < $-2$. Fortunately, CH10 provide EWs for the UMi stars in the metal-poor regime ($-1$ $>$ [Fe/H] $>$ $-2$) based on their high-resolution spectra. Here, we supplement the UMi sample from MJ17b by including four UMi stars from CH10 and redetermine their abundances using our linelist and considering NLTE effects. In addition, we include a VMP UMi star recently discovered by \citet[][hereafter, SZ23]{2023MNRAS.525.2875S}. For consistency, we reanalysed this star using our methods.

For the four metal-poor stars, namely COS171, JI2, N37, and 27940, we take the photometric \teff\ and distance-based log~g from CH10. Stellar parameters of CH10 rely on V--I, V--J, and V--H colours, isochrones, distances, and analyses of the spectra. For comparison, we calculated \teff\ and log~g for COS171 using {\it Gaia} photometry, E(B-V) = 0.028, a distance of 69 kpc \citep{1999AJ....118..366M}, the bolometric corrections from \citet{2018MNRAS.479L.102C}, [Fe/H] = $-1.3$, and m = 0.8m$_{\odot}$. We find \teff = 4315 $\pm$ 80~K and log~g = 0.84 $\pm$ 0.06, which are consistent within the error bars with \teff\ = 4380~K and log~g = 0.80, as recommended by CH10. For the VMP star from SZ23, we find \teff, log~g,  [Fe/H], and \vt \ of  4615$\pm$85~K, 1.24$\pm$0.06, $-2.01$$\pm0.17$, and 1.8$\pm0.2$\kms , respectively, which are consistent within the error bars with the values found by SZ23.

For the four CH10 stars, we employ the EWs provided in the original study and our linelist to determine their abundances, while for SZ23 star, we use the observed spectrum. We note that the barium abundances of UMi stars rely only on the subordinate lines, which are not affected by the adopted barium odd/even isotope ratio. The NLTE and LTE abundance ratios are presented in Table~\ref{umi}, along with the stellar parameters for the five metal-poor UMi stars. It is worth noting that the UMi sample contains a peculiar star, COS171, which is enhanced in iron, and thus shows low [X/Fe] ratios; for example, [Mg/Fe] = $-0.63$ $\pm$ 0.13. Similar iron-rich stars are known in the literature, such as the star ET0381 discovered by \citet{2015A&A...583A..67J} in the Sculptor dSph and the three MW halo stars identified by \citet{2023AJ....166..128R}.

\begin{table}
\caption{Stellar parameters and abundance ratios of the UMi dSph stars}
\setlength{\tabcolsep}{0.80mm}
\label{umi}
\centering   
   \begin{tabular}{lllrlrr} 
   \hline 
Sp.    & $\eps_{\rm \odot}^{*}$  &  [X/H] & [X/Fe\ii] & [X/H] & [X/Fe\ii] & N \\ 
      &         & LTE   & LTE       &  NLTE & NLTE     &    \\ 
      \hline
\multicolumn{7}{c}{UMi COS171: 4380 / 0.80 / --1.28 / 1.8} \\
 Na\ione  &  6.27 & --2.74 $\pm$  0.13 &  --1.47 &  --2.77 $\pm$   0.16 &  --1.50 &    3 \\ 
 Mg\ione  &  7.52 & --1.84 $\pm$  0.16 &  --0.57 &  --1.89 $\pm$   0.13 &  --0.62 &    3 \\ 
 Ca\ione  &  6.27 & --1.63 $\pm$  0.09 &  --0.36 &  --1.49 $\pm$   0.11 &  --0.23 &    4 \\ 
 Ti\ii    &  4.90 & --1.70 $\pm$  0.06 &  --0.43 &  --1.70 $\pm$   0.06 &  --0.43 &    5 \\ 
 Cr\ione  &  5.63 & --2.07 $\pm$ 0.11  &  --0.84 &  --1.95 $\pm$ 0.07   &  --0.72 &    3 \\ 
 Fe\ione  &  7.45 & --1.50 $\pm$  0.08 &  --0.24 &  --1.46 $\pm$   0.07 &  --0.20 &   20 \\ 
 Fe\ii    &  7.45 & --1.27 $\pm$  0.05 &    0.00 &  --1.27 $\pm$   0.05 &    0.00 &    4 \\ 
 Ni\ione  &  6.20 & --1.95             &  --0.68 &                      &         &    1 \\ 
 Zn\ione  &  4.61 & --2.45 $\pm$  0.04 &  --1.18 &  --2.33 $\pm$   0.05 &  --1.06 &    2 \\ 
 Sr\ii    &  2.88 & --1.86             &  --0.59 &  --1.83              &  --0.56 &    1 \\ 
 Ba\ii    &  2.17 & --2.11 $\pm$  0.14 &  --0.84 &  --2.17 $\pm$   0.14 &  --0.91 &    3 \\ 
 Eu\ii    &  0.51 & --1.42             &  --0.15 &  --1.35              &  --0.08 &    1 \\ 
\multicolumn{7}{c}{UMi JI2: 4415 / 0.85 / --1.76 / 1.9} \\
 Na\ione  &  6.27 & --2.46 $\pm$  0.07 &  --0.71 &  --2.55 $\pm$   0.14 &  --0.80 &    4 \\ 
 Mg\ione  &  7.52 & --1.69 $\pm$  0.11 &    0.06 &  --1.76 $\pm$   0.07 &  --0.01 &    3 \\ 
 Ca\ione  &  6.27 & --1.69 $\pm$  0.02 &    0.07 &  --1.59 $\pm$   0.04 &   0.16 &    4 \\ 
 Ti\ii    &  4.90 & --1.59 $\pm$  0.13 &    0.17 &  --1.59 $\pm$   0.13 &   0.17 &    4 \\ 
 Cr\ione  &  5.63 & --2.11 $\pm$ 0.06  &  --0.40 &  --1.99 $\pm$ 0.08   &  --0.28 &    2 \\ 
 Fe\ione  &  7.45 & --1.90 $\pm$  0.08 &  --0.15 &  --1.84 $\pm$   0.08 &  --0.09 &   25 \\ 
 Fe\ii    &  7.45 & --1.75 $\pm$  0.10 &   0.00 &  --1.75 $\pm$   0.10 &   0.00 &    3 \\ 
 Ni\ione   &  6.20 & --1.91             &  --0.16 &                     &        &    1 \\ 
 Zn\ione  &  4.61 & --2.14 $\pm$  0.06 &  --0.38 &  --2.03 $\pm$   0.07 &  --0.28 &    2 \\ 
 Ba\ii    &  2.17 & --1.44 $\pm$  0.09 &   0.31 &  --1.56 $\pm$   0.08 &   0.19 &    3 \\ 
 Eu\ii    &  0.51 & --0.85 $\pm$  0.05 &   0.90 &  --0.81 $\pm$   0.04 &   0.95 &    2 \\ 
\multicolumn{7}{c}{UMi N37: 4390 / 0.80 --1.55 / 1.8}\\
 Na\ione  &  6.27 & --2.29 $\pm$  0.09 &  --0.65 &  --2.35 $\pm$   0.14 &  --0.71 &    4 \\ 
 Mg\ione  &  7.52 & --1.54 $\pm$  0.10 &   0.10 &  --1.61 $\pm$   0.04 &   0.03 &    3 \\ 
 Ca\ione  &  6.27 & --1.51 $\pm$  0.08 &   0.13 &  --1.41 $\pm$   0.11 &   0.23 &    4 \\ 
 Ti\ii    &  4.90 & --1.46 $\pm$  0.03 &   0.18 &  --1.46 $\pm$   0.03 &   0.18 &    4 \\ 
 Cr\ione  &  5.63 & --1.93 $\pm$ 0.05  &  --0.33 &  --1.77  $\pm$ 0.13  &  --0.17 &    3 \\
 Fe\ione  &  7.45 & --1.67 $\pm$  0.09 &  --0.03 &  --1.61 $\pm$   0.09 &   0.03 &   21 \\ 
 Fe\ii    &  7.45 & --1.64 $\pm$  0.09 &   0.00 &  --1.64 $\pm$   0.09 &   0.00 &    5 \\ 
 Ni\ione   &  6.20 & --1.65             &  --0.01 &                     &        &    1 \\ 
 Zn\ione  &  4.61 & --2.05 $\pm$  0.02 &  --0.41 &  --1.98 $\pm$   0.01 &  --0.33 &    2 \\ 
 Ba\ii    &  2.17 & --1.35 $\pm$  0.03 &   0.29 &  --1.45 $\pm$   0.01 &   0.19 &    3 \\ 
 Eu\ii    &  0.51 & --0.69 $\pm$  0.08 &   0.95 &  --0.65 $\pm$   0.07 &   1.00 &    2 \\ 
\multicolumn{7}{c}{UMi 27940: 4290 / 0.70 / --1.77 / 1.9}\\
 Na\ione  &  6.27 & --2.32 $\pm$  0.14 &  --0.56 &  --2.35 $\pm$   0.13 &  --0.59 &    4 \\ 
 Mg\ione  &  7.52 & --1.70 $\pm$  0.12 &   0.07 &  --1.78 $\pm$   0.05 &  --0.02 &    3 \\ 
 Ca\ione  &  6.27 & --1.77 $\pm$  0.03 &  --0.01 &  --1.69 $\pm$   0.05 &   0.08 &    3 \\ 
 Ti\ii    &  4.90 & --1.73 $\pm$  0.06 &   0.03 &  --1.73 $\pm$   0.06 &   0.03 &    4 \\ 
 Cr\ione  &  5.63 & --2.28 $\pm$ 0.11  &  --0.56 &  --2.05 $\pm$ 0.05   &  --0.33 &    3 \\ 
 Fe\ione  &  7.45 & --2.03 $\pm$  0.09 &  --0.26 &  --1.98 $\pm$   0.09 &  --0.21 &   23 \\ 
 Fe\ii    &  7.45 & --1.76 $\pm$  0.18 &   0.00 &  --1.76 $\pm$   0.18 &   0.00 &    4 \\ 
 Ni\ione  &  6.20 & --2.08             &  --0.32 &                     &        &    1 \\ 
 Zn\ione  &  4.61 & --2.24 $\pm$  0.02 &  --0.47 &  --2.13 $\pm$   0.02 &  --0.37 &    2 \\ 
 Sr\ii    &  2.88 & --2.19             &  --0.43 &  --2.13              &  --0.37 &    1 \\ 
 Ba\ii    &  2.17 & --1.82 $\pm$  0.16 &  --0.06 &  --1.91 $\pm$   0.18 &  --0.15 &    3 \\ 
 Eu\ii    &  0.51 & --1.40 $\pm$  0.01 &   0.36 &  --1.31 $\pm$   0.00 &   0.45 &    2 \\ 
\multicolumn{7}{c}{UMi T1SZ23: 4615 / 1.24 / --2.01 / 1.8}\\ 
 Na\ione   &  6.27 & --2.57 $\pm$   0.02 &  --0.56 &  --2.83 $\pm$   0.12 & --0.82 &    3 \\ 
 Mg\ione   &  7.52 & --1.21 $\pm$   0.10 &    0.80 &  --1.33 $\pm$   0.18 &   0.68 &    2 \\ 
 Si\ione   &  7.51 & --1.68 $\pm$   0.21 &    0.33 &  --1.68 $\pm$   0.21 &   0.33 &    4 \\ 
  K\ione   &  5.07 & --1.99              &    0.02 &  --2.20              & --0.19 &    1 \\ 
 Ca\ione   &  6.27 & --1.79 $\pm$   0.16 &    0.22 &  --1.77 $\pm$   0.20 &   0.24 &    9 \\ 
 Sc\ii     &  3.04 & --1.94 $\pm$   0.12 &    0.07 &  --1.93 $\pm$   0.12 &   0.08 &    3 \\ 
 Ti\ii     &  4.90 & --1.89              &    0.12 &  --1.89              &   0.12 &    1 \\ 
 Cr\ione   &  5.63 & --2.34 $\pm$   0.30 &  --0.33 &  --2.06 $\pm$   0.29 & --0.05 &    4 \\ 
 Fe\ione   &  7.45 & --2.07 $\pm$   0.22 &  --0.06 &  --2.00 $\pm$   0.22 &   0.01 &   45 \\ 
 Fe\ii     &  7.45 & --2.01 $\pm$   0.16 &    0.00 &  --2.01 $\pm$   0.16 &   0.00 &    8 \\ 
  Ni\ione  &  6.20 & --2.17 $\pm$   0.11 &  --0.16 &                      &        &    5 \\ 
 Ba\ii     &  2.17 & --2.90 $\pm$   0.09 &  --0.88 &  --2.90 $\pm$   0.10 & --0.89 &    2 \\ 
 \hline
\multicolumn{7}{l}{$^{*}$ Solar abundances are taken from \citet{2021SSRv..217...44L}.}\\
      \end{tabular}
\end{table}

\section{Results}\label{results}

\subsection{Metallicity distribution function}
\begin{figure}
\includegraphics[trim={0 0 30 10},width=80mm]{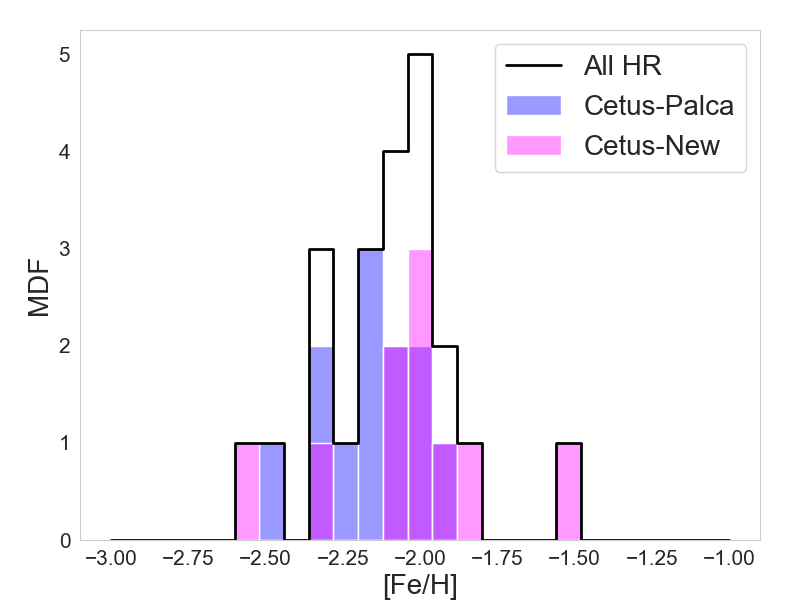}     
   \caption{ Metallicity distribution function of all 22 Cetus stars in the HR follow up gives a mean [Fe/H] = $-$2.11 with an intrinsic dispersion of 0.21 dex (black histogram). The MDFs of the Cetus-Palca (blue histogram) and Cetus-New (magenta histogram) wraps almost overlap.}
   \label{fig:mdf} 
\end{figure}

Using the derived metallicities, we confirm that the two Cetus wraps have similar metallicity distribution functions (MDFs). They are shown in Fig.~\ref{fig:mdf}. By taking into account the uncertainty in [Fe/H] for each star, the mean metallicity in the Cetus-Palca is $-2.19$ with an intrinsic dispersion of 0.12 dex. The Cetus-New wrap has a mean metallicity of $-$2.02, and a  larger dispersion of 0.25 dex. The total sample of 22 Cetus stars gives a mean [Fe/H] = $-$2.11 $\pm$ 0.21, consistent with the previous findings: [Fe/H] = $-$2.07 $\pm$ 0.12 from LAMOST K giants in \citet{liu2014} and \citet{yuan2019}; and [Fe/H] = $-$2.17 \citep{yuan2022} from 35 member stars with photometric metallicities based on Pristine survey \citep{starkenburg2017} and SkyMapper DR2 \citep{huang2022}. 

Using the two-sample Kolmogorov-Smirnov test, we examined whether the samples from Cetus-New and Cetus-Palca originate from the same distribution. We find a resulting probability p-value = 14 \%, indicating a 14 \% chance of rejecting our hypothesis that the two samples come from the same distribution. The metallicity dispersion of 0.21~dex confirms that the Cetus progenitor was a dwarf galaxy rather than a globular cluster. Globular clusters exhibit a narrower metallicity dispersion, typically less than 0.05 dex \citep[see e.g.][]{li2022}.

\subsection{Abundance trends}

Table~\ref{abund_table} presents the LTE and NLTE abundances together with their statistical uncertainties and the number of lines used. The abundance uncertainty is calculated as the dispersion of the single line measurements around the mean $\sigma = \sqrt{ \Sigma (\eps - \eps_i )^2 /(N - 1)}$, where N is the total number of lines. The abundance trends are presented in Figs.~\ref{cfe_l}--\ref{lepp_ncapture}. For our sample stars and for the comparison samples, we calculated abundance ratios using solar abundances from \citet{2021SSRv..217...44L}.

\begin{table}
\caption{NLTE and LTE abundance ratios of the Cetus stars}
\setlength{\tabcolsep}{0.90mm}
 \label{abund_table}
\centering   
   \begin{tabular}{lllrlrc}    
      \hline
\tiny{{\it Gaia} ID} & Sp. &  ~~~~~~~\tiny{[X/H]}  & \tiny{[X/Fe\ii]} & ~~~~~\tiny{[X/H]}  & \tiny{[X/Fe\ii]} & \tiny{N} \\ 
&  & ~~~~~~~~\tiny{LTE}   & \tiny{LTE}~~~ & ~~~~~\tiny{NLTE}  & \tiny{NLTE}~~ &  \\
   \hline 
2483... &   CH      &   --3.14 $\pm$ 0.04 & --1.05 &                   &        &  2 \\
2483... &   O\ione  &   --1.48            &   0.61 & --1.48            &   0.61 &  1 \\
2483... &   Na\ione &   --2.25 $\pm$ 0.03 & --0.16 & --2.41 $\pm$ 0.03 & --0.32 &  4 \\
2483... &   Mg\ione &   --1.75 $\pm$ 0.01 &   0.34 & --1.79 $\pm$ 0.02 &   0.31 &  2 \\
2483... &   Al\ione &   --2.47            & --0.38 & --2.37            &  -0.28 &  1 \\ 
2483... &   Si\ione &   --1.75 $\pm$ 0.09 &   0.34 &                   &        &  4 \\ 
2483... &   S\ione  &   --1.75            &   0.34 & --1.98            &   0.11 &  1 \\ 
2483... &   K\ione  &   --1.68            &   0.41 & --2.06            &   0.03 &  1 \\ 
2483... &   Ca\ione &   --1.83 $\pm$ 0.09 &   0.26 & --1.77 $\pm$ 0.12 &   0.32 &  8 \\ 
2483... &   Sc\ii   &   --1.97 $\pm$ 0.07 &   0.12 & --1.98 $\pm$ 0.07 &   0.11 &  3 \\ 
2483... &   Ti\ii   &   --1.79 $\pm$ 0.18 &   0.30 & --1.79 $\pm$ 0.18 &   0.30 &  4 \\ 
2483... &   V\ione  &   --2.65 $\pm$ 0.01 & --0.56 &                   &        &  2 \\ 
 \hline
      \end{tabular}
The table is accessible in a machine-readable format at the CDS. A portion is shown to illustrate its format and content. 
\end{table}

\subsubsection{Carbon}
\begin{figure}
\includegraphics[trim={0 10 0 10},width=80mm]{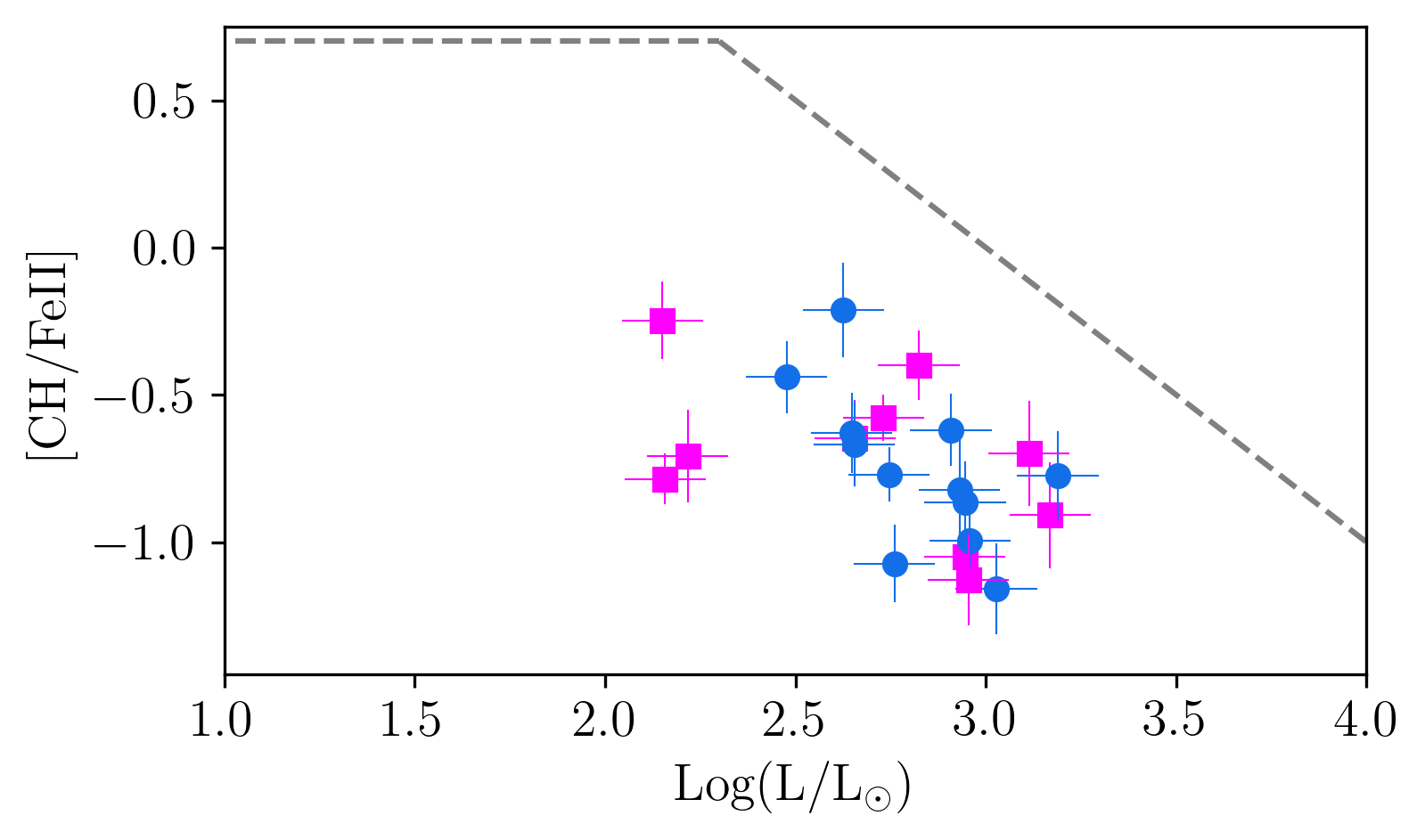}
\caption{[C/Fe] abundance ratio as a function of luminosity. The dashed line corresponds to the empirical separation between the carbon-rich and carbon-normal populations defined by \citet{2007ApJ...655..492A}. Carbon-normal stars populate the area below the dashed line. The symbols are the same as in Fig.~\ref{fe12}.}
\label{cfe_l}
\end{figure}

Our sample stars exhibit a depleted carbon abundance with respect to iron, and [C/Fe] spans from $-1.2$ to $-0.2$ depending on stellar luminosity (Fig.~\ref{cfe_l}). The sample stars are giants and therefore suffer from carbon depletion caused by the dredge up. According to the classification suggested by \citet{2007ApJ...655..492A}, the Cetus stars belong to the carbon-normal population. 

Our analysis reveals that none of the Cetus stars with measured carbon abundance exhibit carbon enhancement, which is noteworthy given that carbon-enhanced stars with [C/Fe] > 0.7 \citep[CE,][]{2007ApJ...655..492A} are numerous among VMP stars \citep{2014ApJ...797...21P,2020ApJ...889...27J}. Our result aligns with the significantly lower fraction of CE metal-poor (CEMP) stars in dSph galaxies compared to that in the MW and UFDs found by \citet{2024A&A...686A.266L}. In the metallicity range that we are concerned with in this study, nearly 20\% of the stars in the MW and  UFDs are CEMP stars, while the corresponding fraction  of dSph sources amounts to only 6\%\ \citep{2024A&A...686A.266L}. The absence of CEMP stars in Cetus supports the hypothesis that the progenitor of the Cetus stream was a dSph galaxy. Otherwise, it would be highly improbable (\textless\ 0.5\%\  chance) to select a sample of 22 stars with normal carbon abundances from an UFD galaxy with a typical fraction of CEMP stars.

\subsubsection{$\alpha$-elements and titanium}
Oxygen lines are detected in the spectra of the 12 stars with [Fe/H] > $-2.2$. These stars are enhanced in oxygen with respect to iron, with the average ratio [O/Fe] = 0.71 $\pm$ 0.10. This is similar to [O/Fe] = 0.7 found for the MW halo stars with [Fe/H] from $-2.5$ to $-1.5$ \citep{2018AstL...44..411S}. 

\begin{figure}
   \includegraphics[trim={ 0 30 25  0},width=80mm]{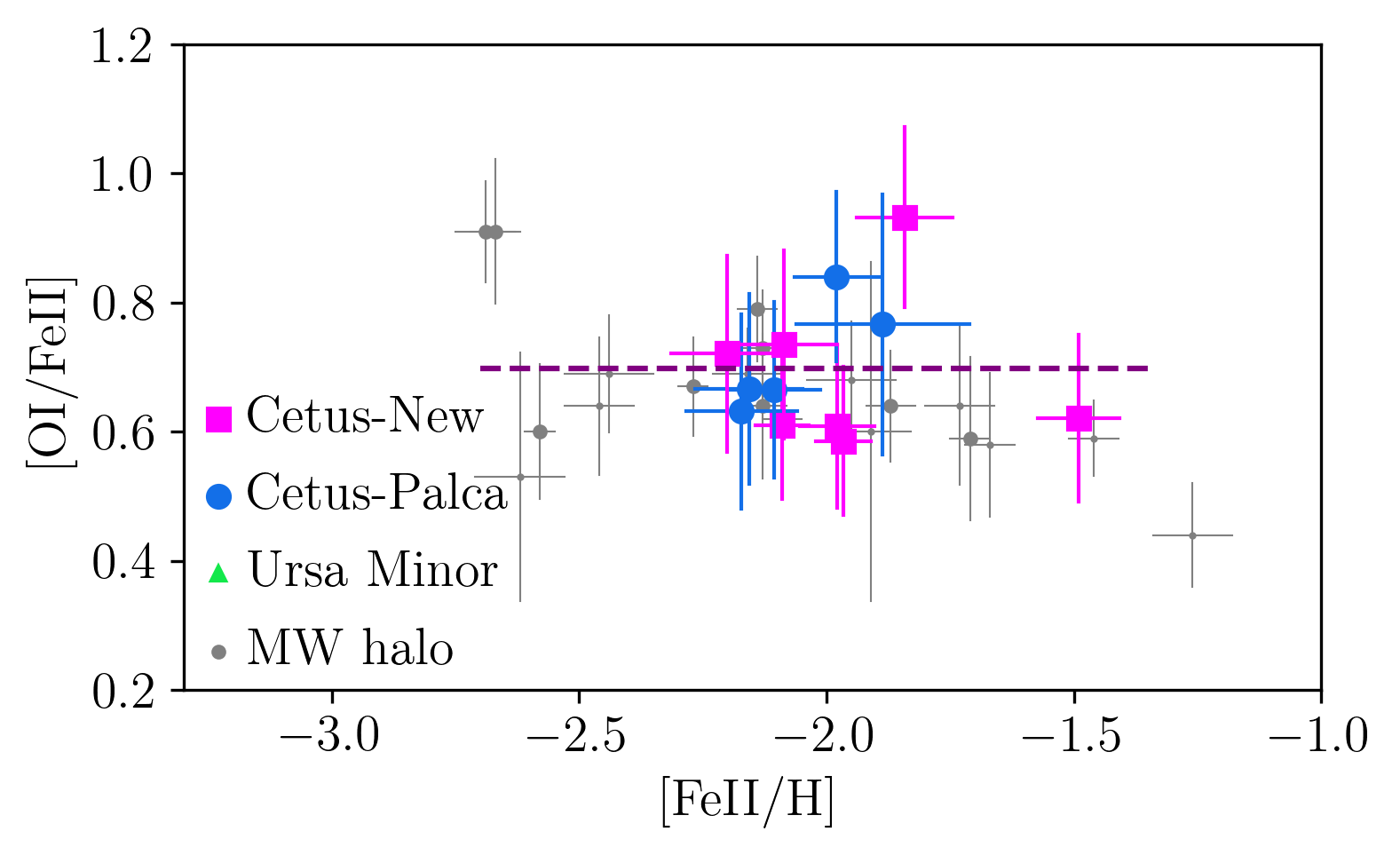}
   \includegraphics[trim={10 30 25 10},width=80mm]{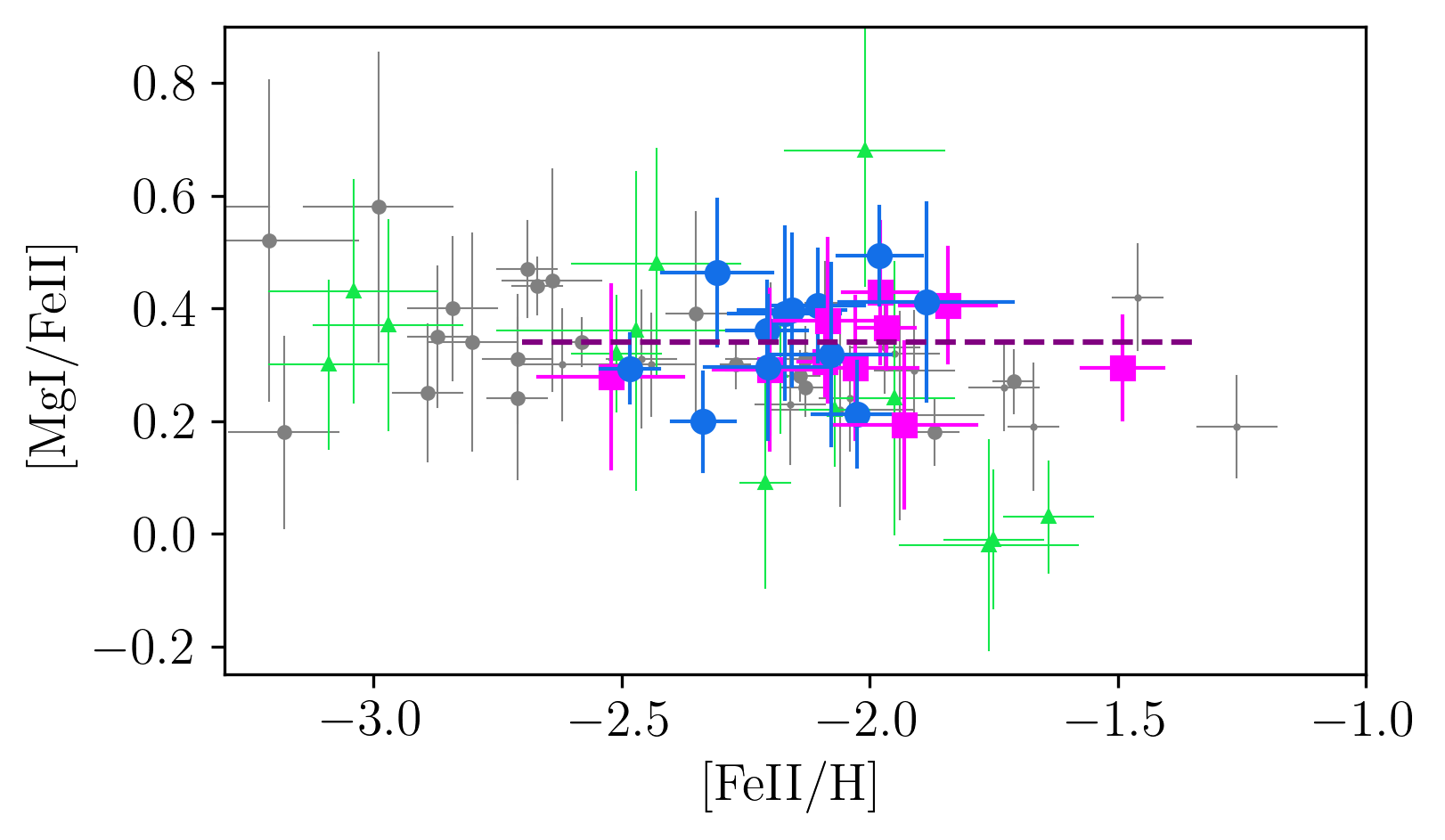}
   \includegraphics[trim={10 30 25 10},width=80mm]{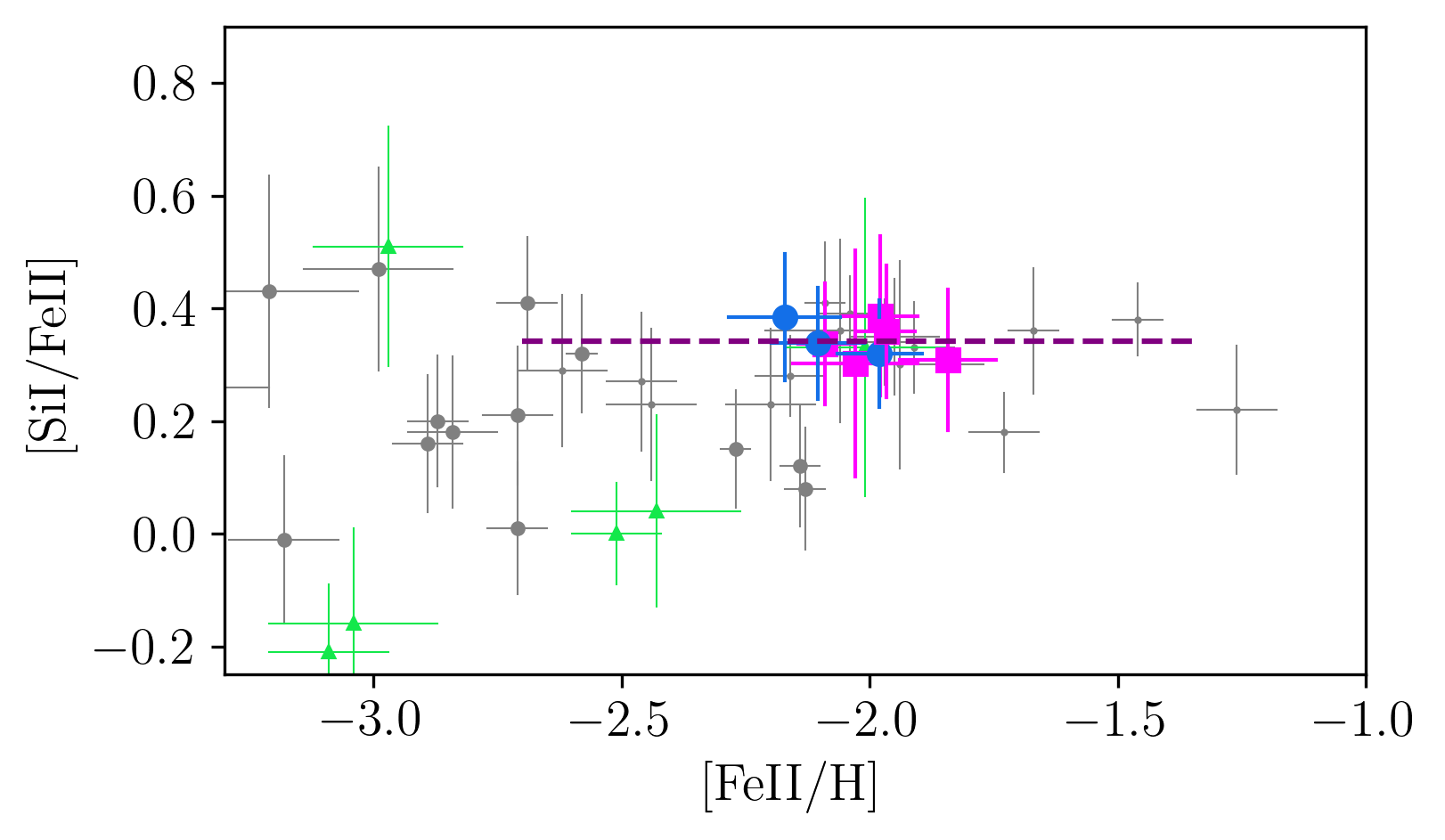}
   \includegraphics[trim={10 30 25 10},width=80mm]{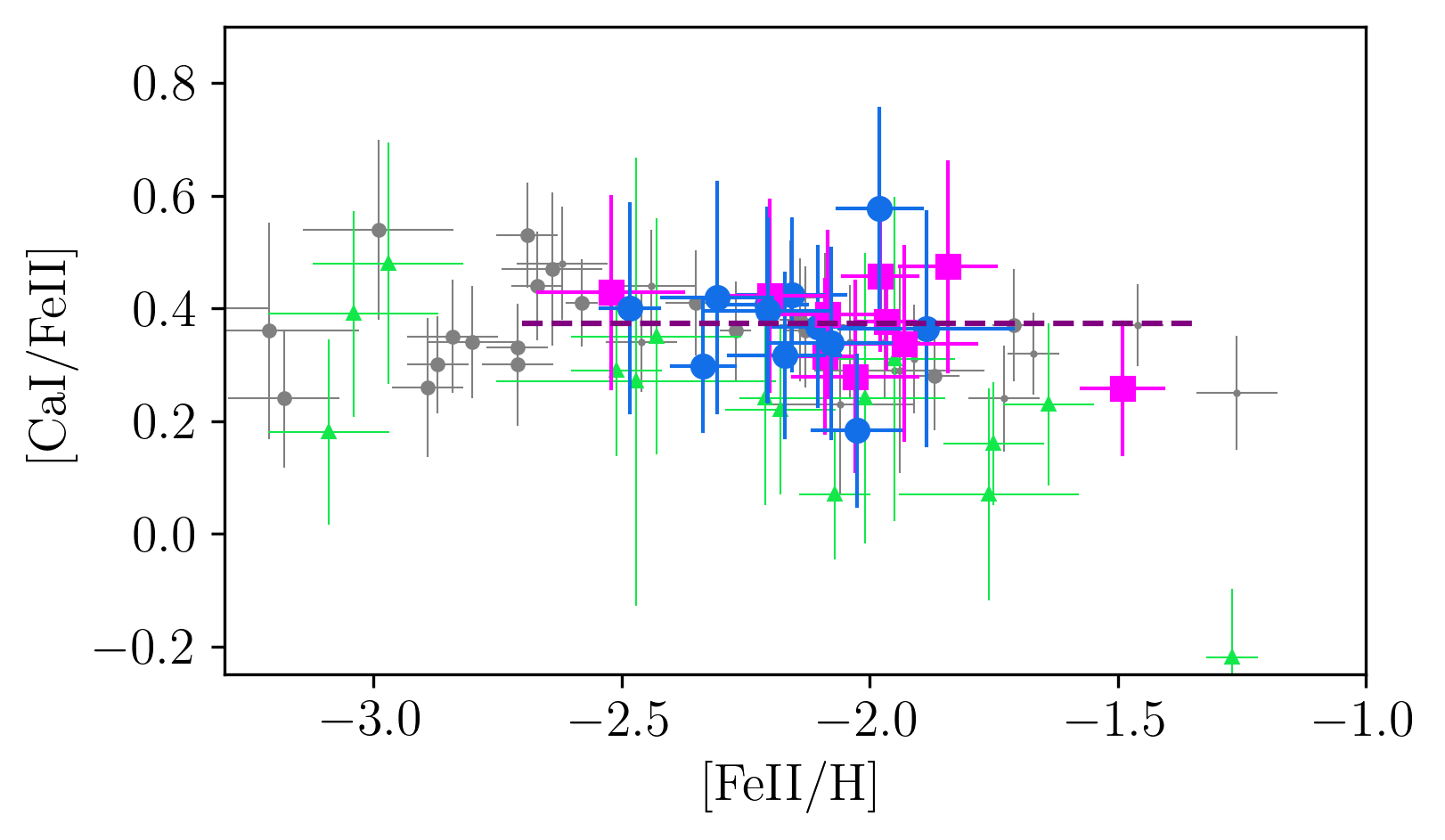}  
   \includegraphics[trim={10  0 25 10},width=80mm]{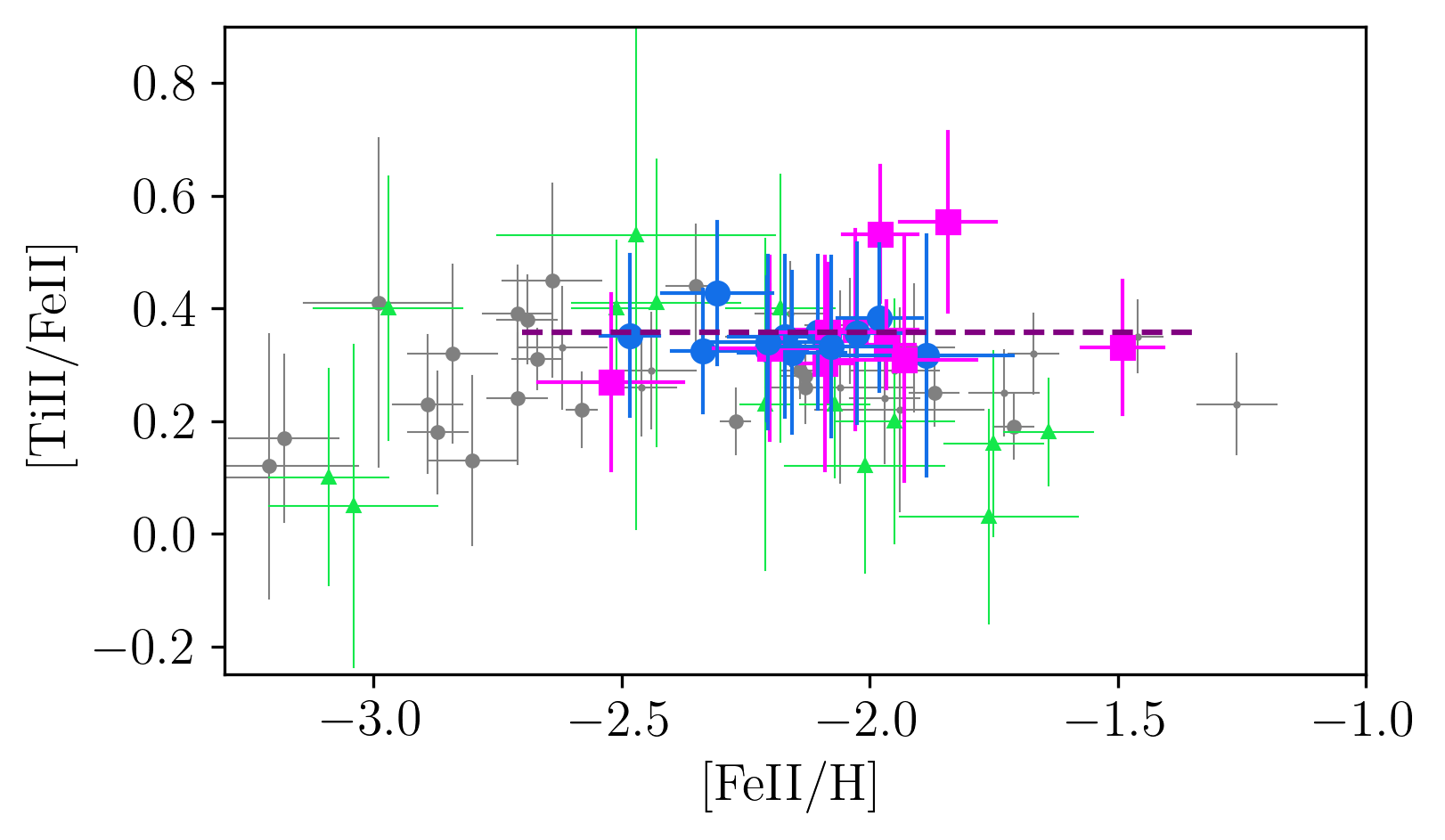} 
   \caption{NLTE abundance ratios for $\alpha$-elements and titanium with respect to iron as a function of [Fe/H] in the Cetus-New (magenta squares) and the Cetus-Palca (blue circles). The dashed line indicates the average element ratios in the Cetus stars. For comparison, we plotted MW halo giants (small grey circles) and dwarfs (grey dots) and UMi dSph stars (green triangles).}
   \label{alpha_fe} 
\end{figure}

The Cetus stars show similar enhancement in magnesium throughout the entire metallicity range, with an average ratio [Mg/Fe] =  0.34 $\pm$  0.08. This value is similar to the average values of the MP star samples in \citet[][hereafter MN19]{2019ARep...63..726M}: [Mg/Fe] = 0.36 $\pm$ 0.13 from halo giants, 0.28 $\pm$ 0.07 from halo dwarfs, and 0.29 $\pm$ 0.07 from thick disk stars. A similar ratio [Mg/Fe] = 0.30 $\pm$ 0.11 is observed in  VMP UMi stars with [Fe/H] < $-2$, while more metal-rich UMi stars show lower [Mg/Fe] = 0 (Fig.~\ref{alpha_fe}), indicating the onset of iron production in type Ia supernovae (SNe~Ia).

Our sample stars show [Si/Fe] = 0.35 $\pm$ 0.14. The scatter in [Si/Fe] is larger than that in [Mg/Fe]. This is because, for most stars, the only two lines available to determine the silicon abundance (Si\ione\ 3905 and 4102 \AA) are located in the blue spectral region, which commonly suffers from low S/N and spectral line blending. If we use the eight stars that have three or more measurable Si\ione\ lines, the scatter reduces drastically to [Si/Fe] = 0.33 $\pm$  0.03. Overall, the derived ratio is close to those found by MN19 in the halo dwarfs (0.31 $\pm$ 0.07) and thick disk stars (0.26 $\pm$ 0.06).

Consistent with the magnesium and silicon, the calcium abundances in the Cetus stars exhibit an enhancement of [Ca/Fe] = 0.37 $\pm$  0.08, which is fairly similar to stars in the MN19 VMP sample: [Ca/Fe] = 0.36 $\pm$ 0.11 (halo giants), 0.35 $\pm$ 0.08 (halo dwarfs), 0.24 $\pm$ 0.07 (thick disk dwarfs). The VMP UMi stars show an enhancement of [Ca/Fe] = 0.24 $\pm$ 0.11 (MJ17b), and the MP stars have less enhanced [Ca/Fe], following a similar trend to that found for [Mg/Fe].  

The source of titanium remains poorly understood, and chemical evolution models underestimate the [Ti/Fe] ratio due to low nucleosynthesis yields \citep[see e.g. ][]{2006ApJ...653.1145K,2020ApJ...900..179K}. Regardless of titanium's origin, observations indicate that [Ti/Fe] exhibits a  similar behaviour to [$\alpha$/Fe]. The stars of our sample  show  [Ti/Fe] = 0.36 $\pm$ 0.07, consistent with the [Ti/Fe] = 0.33 $\pm$ 0.07 found in the MW halo VMP giants (MN19), as well as with the [Mg, Si, Ca/Fe] ratios in the Cetus stars.

In summary, for the studied $\alpha$-elements (Mg, Si, Ca), the Cetus stars consistently exhibit a similar [$\alpha$/Fe] $\simeq$ 0.3 throughout the entire metallicity range. The lack of a decreasing trend in the [$\alpha$/Fe] -- [Fe/H] diagram implies that star formation in the Cetus progenitor stopped before iron production due to SNe~Ia became substantial.

\begin{figure}
   \includegraphics[trim={0 30 30 10},width=80mm]{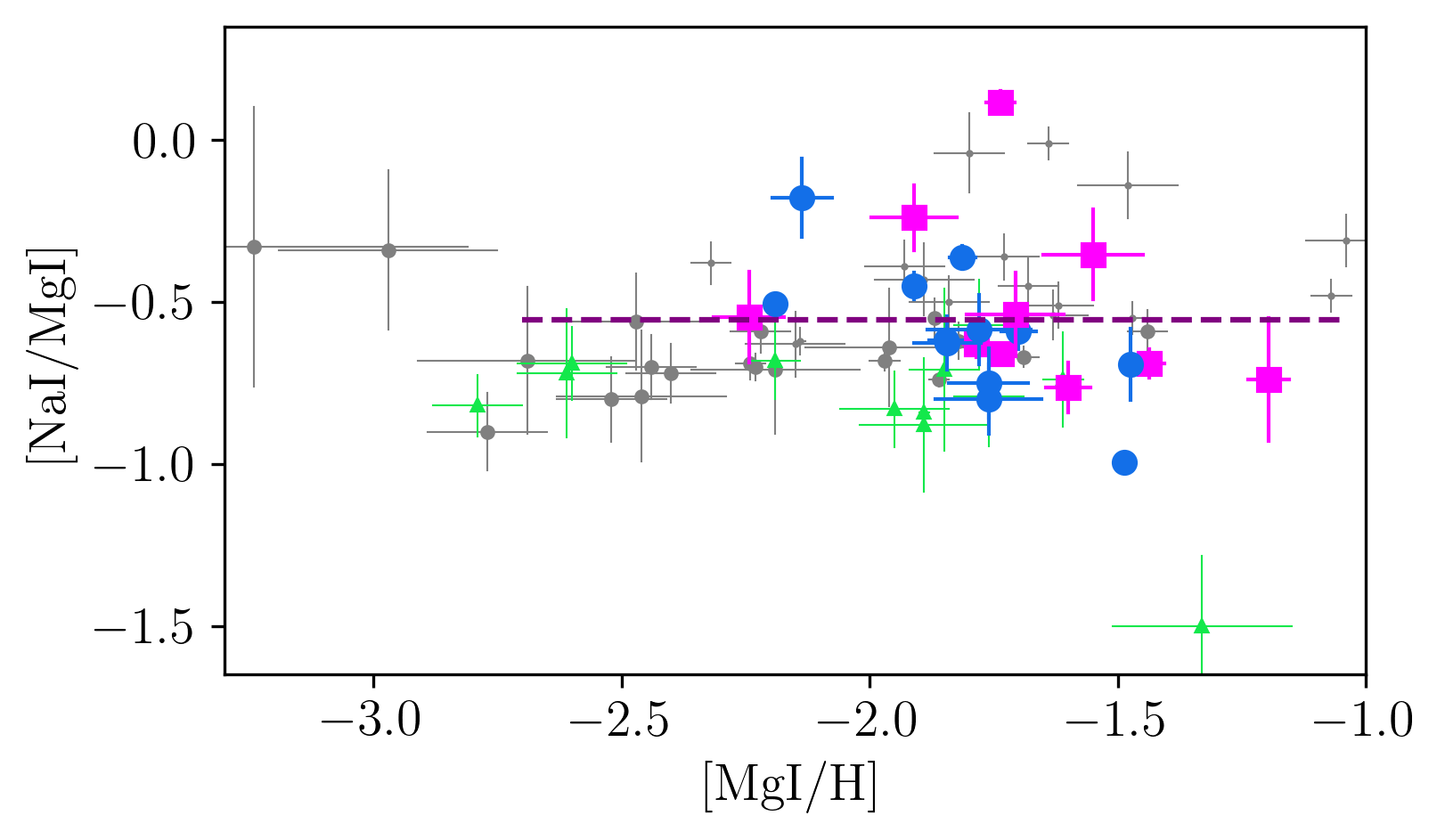}
   \includegraphics[trim={0  0 30 10},width=80mm]{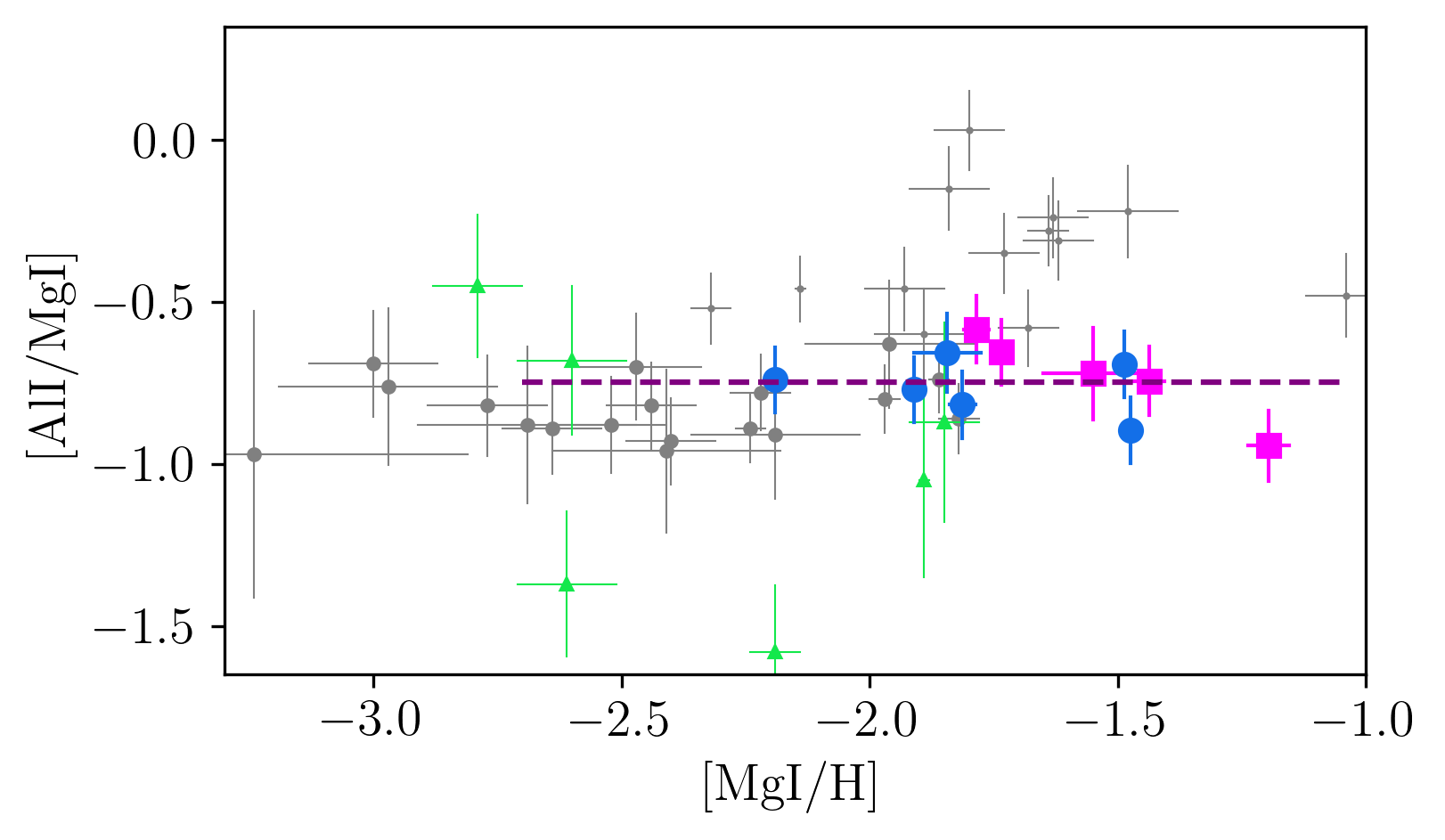}     
   \caption{NLTE abundance ratios as a function of [Mg/H]. The designations are the same as in Fig.~\ref{alpha_fe}.}
   \label{odd_z} 
\end{figure}

\subsubsection{Odd-Z elements}
\begin{figure}
   \includegraphics[trim={0 30 30 0},width=80mm]{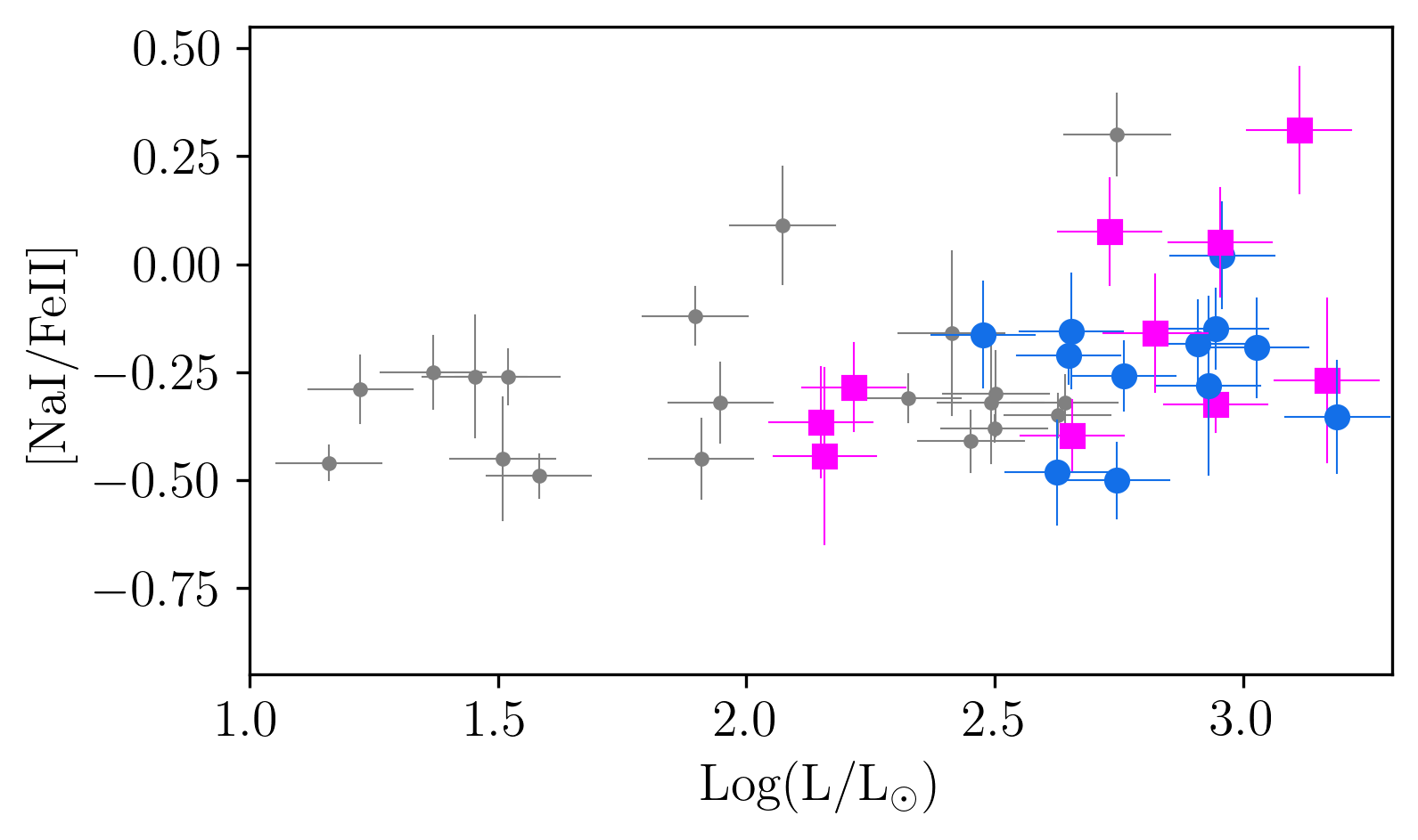}
   \includegraphics[trim={0  0 30 10},width=80mm]{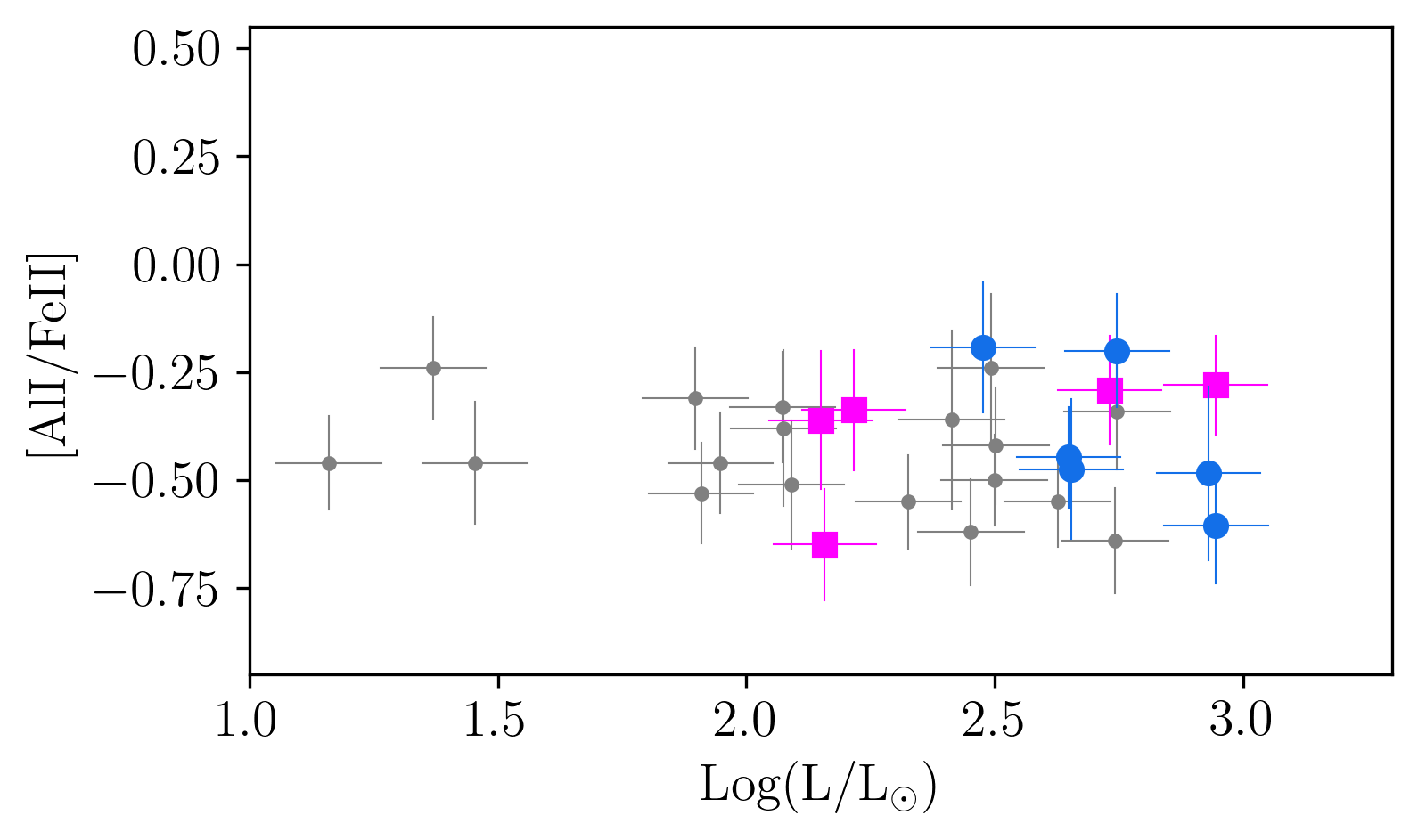} 
   \caption{NLTE abundance ratios as a function of luminosity. The designations are the same as in Fig.~\ref{alpha_fe}.}
   \label{na_al_lumin} 
\end{figure}

Figure~\ref{odd_z} shows [Na/Mg] and [Al/Mg] as a function of [Mg/H]. The [Na/Mg] ratio displays considerable scatter, while the nine stars with measured aluminium abundance show coherent ratios with an average [Al/Mg] = $-0.80$ $\pm$ 0.11. Giants with surface gravities of log~g < 2 may exhibit sodium and aluminium overabundance caused by mixing and stellar evolution effects and the overabundances increase with stellar luminosity \citep{2001ARep...45..301B,2009ARep...53..660P,2013AstL...39...54P,2014AstL...40..406A}. Figure~\ref{na_al_lumin} shows [Na/Fe] and [Al/Fe] in our sample of stars as a function of luminosity. On average, we find [Al/Fe] = $-0.39$ $\pm$ 0.15 and the scatter remains independent of stellar luminosity. The situation is different for sodium: the scatter in [Na/Fe] increases with  luminosity; although stars with normal [Na/Fe] = $-0.4$ are found across the entire luminosity range. 

\begin{figure}
\includegraphics[trim={4 30 45 10},width=80mm]{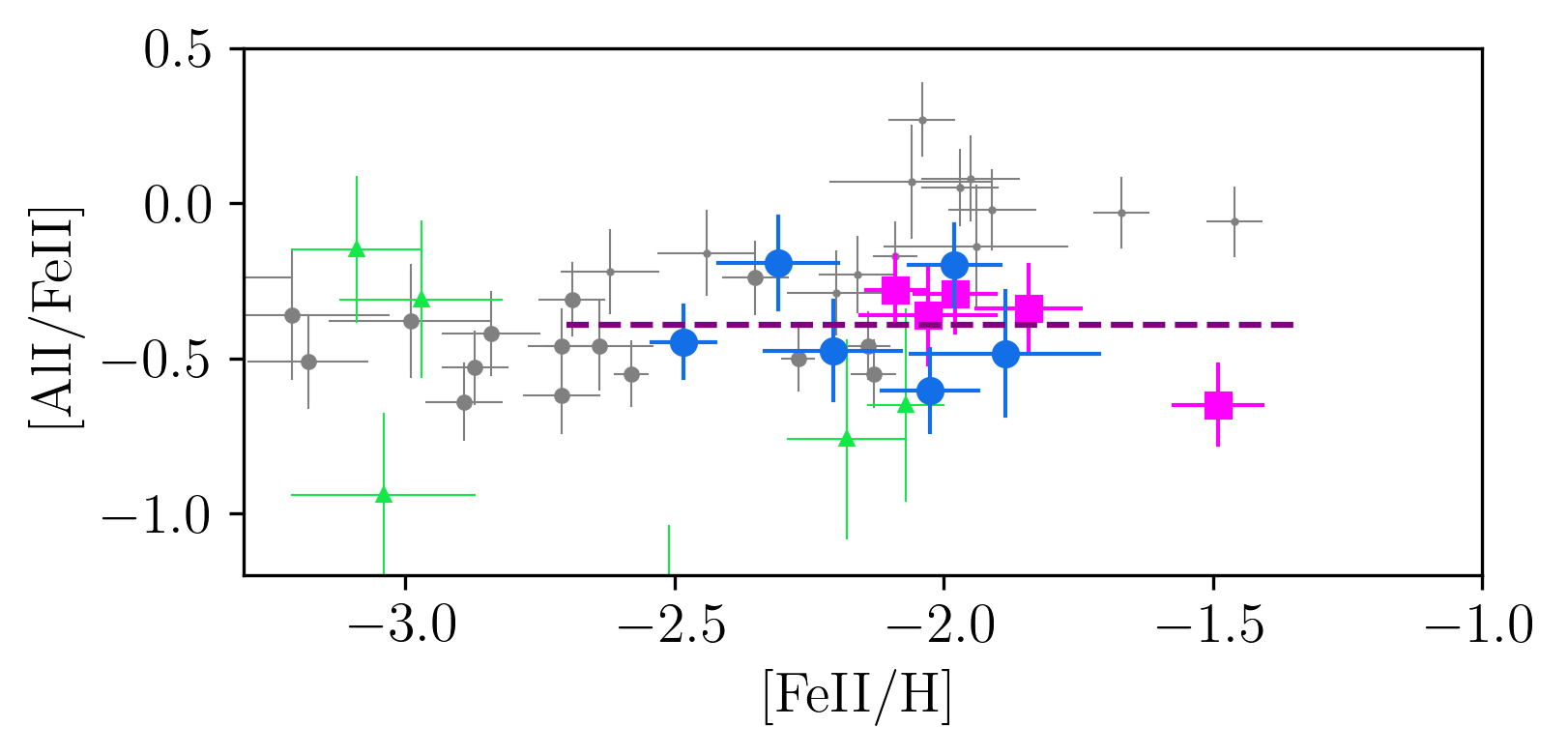}
\includegraphics[trim={10 30 45 10},width=80mm]{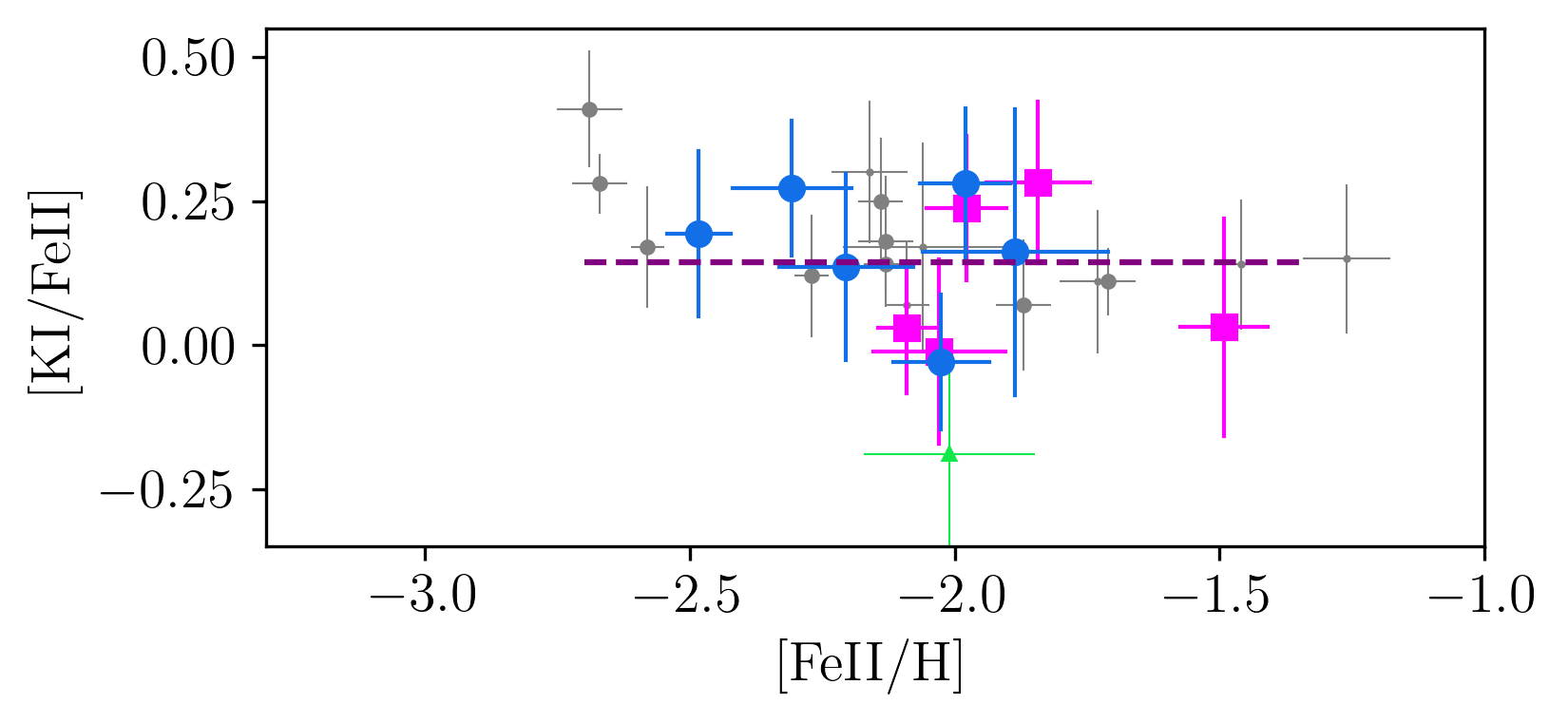} 
\includegraphics[trim={10 30 45 10},width=80mm]{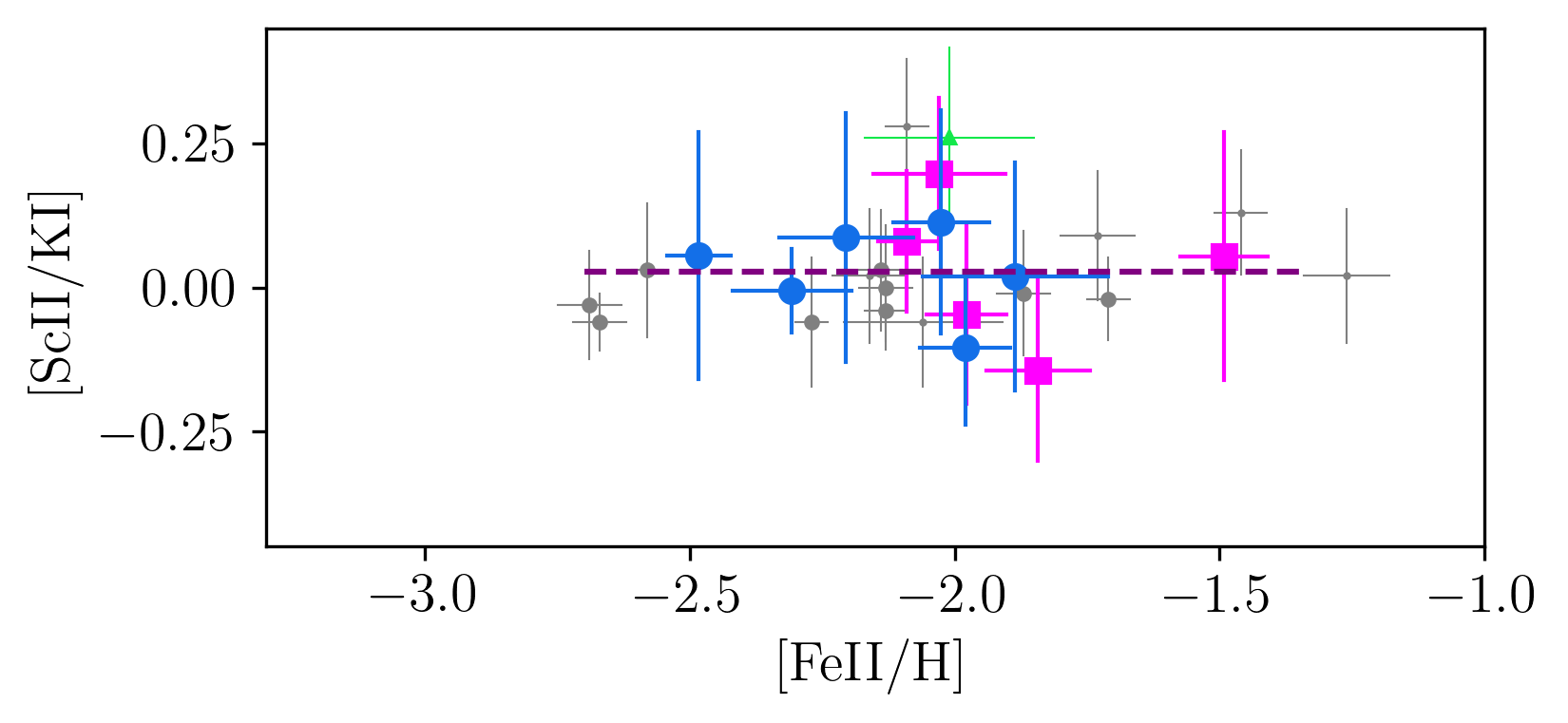}
\includegraphics[trim={10 30 45 10},width=80mm]{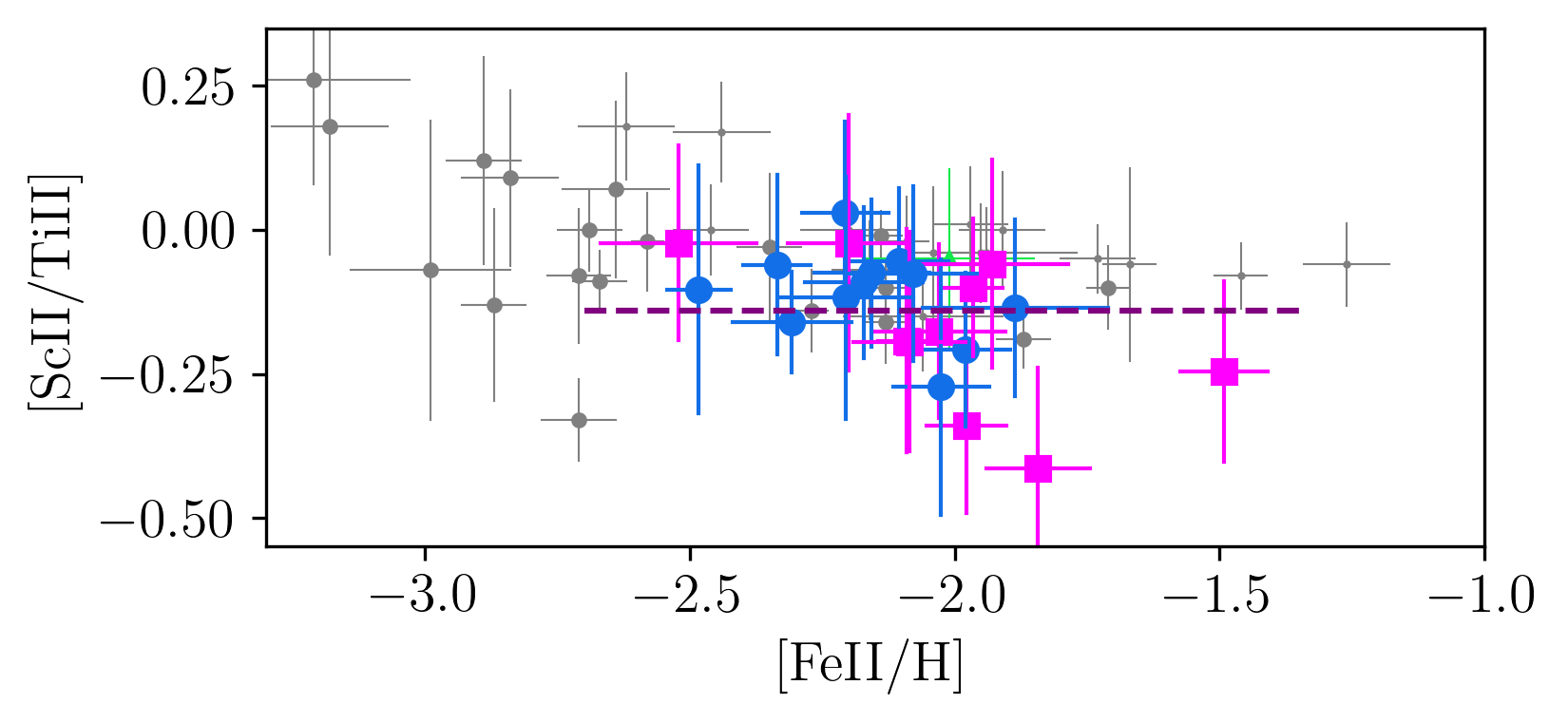}
\includegraphics[trim={4  0 45 10},width=80mm]{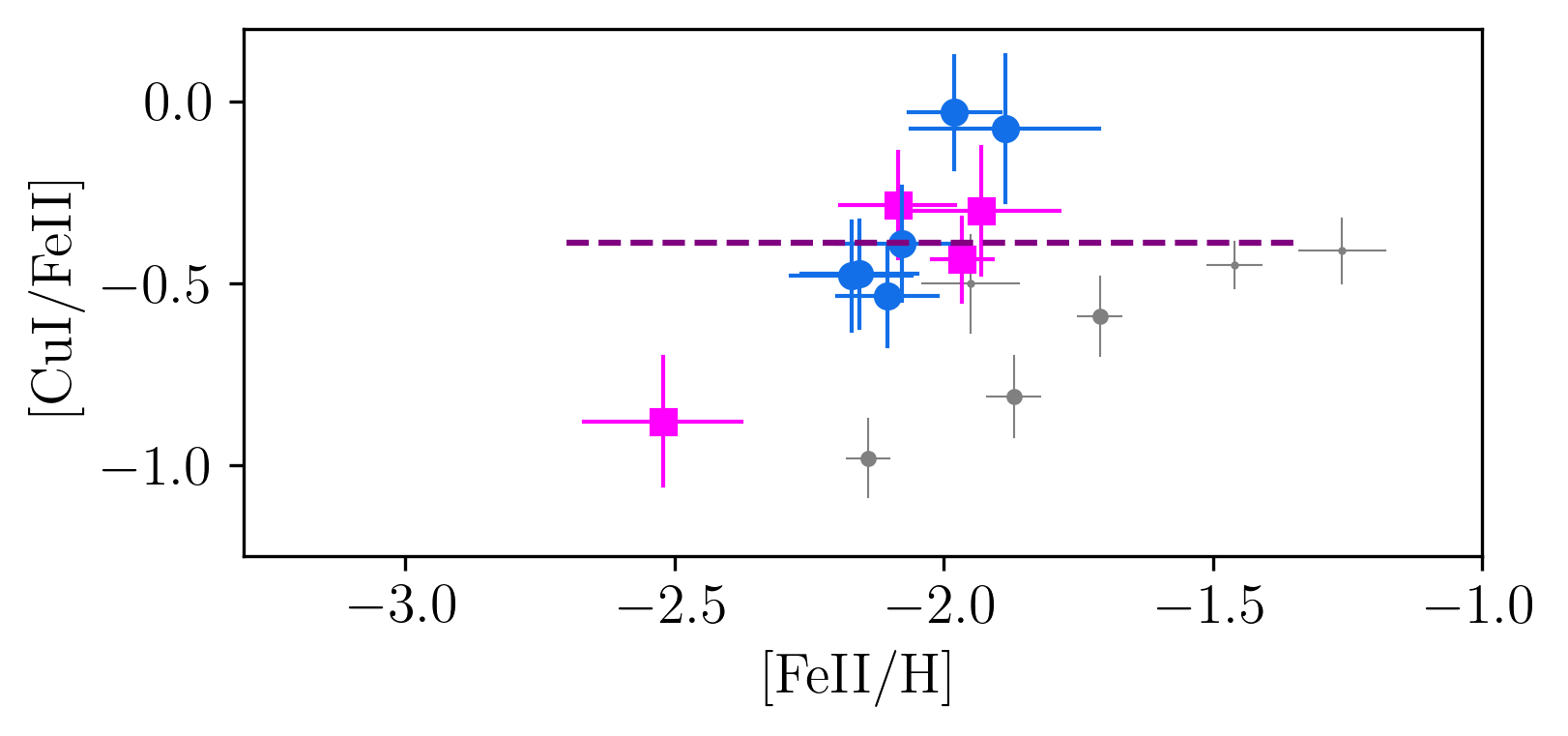}
   \caption{NLTE abundance ratios as a function of [Fe/H]. The designations are the same as in Fig.~\ref{alpha_fe}.}
   \label{ksc} 
\end{figure}

In our sample stars, variations in abundance due to stellar evolution impact sodium but not aluminium, and the latter can be used for galactic chemical evolution analysis. Figure~\ref{ksc} shows [Al/Fe] as a function of [Fe/H]. In the MW halo, [Al/Fe] rises from $-0.6$ to 0 with [Fe/H] increasing from $-3.0$ to $-1.5$. 
Unlike the stars in the MW halo, the Cetus stars do not exhibit an upward trend in [Al/Fe].

Aluminium is produced by massive stars with initial masses ranging from 10 to 40~m$_{\rm \odot}$. The yields of aluminium are strongly dependent on the metallicity and the initial mass of the progenitor star, with higher CNO abundances and greater progenitor mass resulting in higher aluminium abundance \citep{2006ApJ...653.1145K}. Given the rapid stellar evolution of massive stars and the nearly instantaneous enrichment of the interstellar medium, it is more likely that the difference in [Al/Fe] trends arises from a variation in the progenitor mass function rather than a variation in their metallicities. In the Cetus progenitor, the underproduction of aluminium at [Fe/H] > $-2$ could indicate a lack of the most massive stars compared to the MW.
It is worth noting that only half of the  stars of our sample have aluminium abundance measurements, which is due the wavelength coverage of the observed spectra. Increasing the sample size by adding more Cetus stars with aluminium abundance determinations could help us to obtain more accurate observational constraints on the [Al/Fe] trend.

The origin of heavier odd-Z elements (K, Sc, V) remains poorly understood and their abundances are underestimated by chemical evolution models and stellar yields \citep{2006ApJ...653.1145K,2020ApJ...900..179K}. For these elements, our results can serve as observational constraints on chemical evolution models. We also aim to deduce how these element abundances correspond to other chemical elements. In our sample of stars, potassium, scandium, and vanadium abundances follow each other with average ratios of [Sc/K] = 0.03 $\pm$ 0.09 and [Sc/VI\ii ] = 0.08 $\pm$ 0.16. When calculating the [K/Fe] ratio, we find [K/Fe] = 0.11 $\pm$ 0.12 (Fig.~\ref{ksc}). However, the scatter reduces significantly when comparing potassium with respect to magnesium and [K/Mg] = $-0.23$~$\pm$ 0.06. 

Following the example of \citet{2022AstL...48..455M}, we plot [Sc/Ti] as a function of [Fe/H] (Fig.~\ref{ksc}). Our analysis of the Cetus stars reveals [Sc/Ti] = $-0.14 \pm\ 0.11$, which is close to the [Sc/Ti]  found in MW halo stars within the same [Fe/H] range. The two Cetus-New stars exhibit lower [Sc/Ti] $\simeq$ $-0.4$ compared to the other stars. This is due to their higher titanium abundance with [Ti/Fe] = 0.5 (see Fig.~\ref{alpha_fe}). \citet{2003ApJ...598.1163M} found that  Sc and Ti can be produced in  bipolar SNe explosions, and Sc and, to a slightly lesser extent, Ti serve as asphericity indicators of these explosions. Our results concerning the [Sc/Ti] trend can be adopted as observational constraints for bipolar SNe models.

We observe an increase in [Cu/Fe] from $-0.8$ to 0 with [Fe/H] increasing from $-2.5$ to $-1.7$. Our analysis of the MW comparison sample also reveals an increasing trend; however it is slightly shifted towards higher [Fe/H] compared to the Cetus stars. Copper production is significantly influenced by metallicity ---the higher the metallicity, the more copper is produced--- and the MW chemical evolution models predict a rise in [Cu/Fe] from $-1$ to 0 as [Fe/H] increases from $-3$ to $-1$ \citep{1993A&A...272..421M,2006ApJ...653.1145K,2010A&A...522A..32R}. 
Observational constraints based on NLTE studies of the MW stars generally agree on a  rise in [Cu/Fe] from $-0.4 \pm 0.1$ to 0 as [Fe/H] increases from $-2$ to $-1$, while significant discrepancies in [Cu/Fe] from $-0.9$ to $-0.4$ are found at [Fe/H] < $-2$ \citep{2018ApJ...862...71S,2018MNRAS.480..965K,2018MNRAS.473.3377A,2022ApJ...936....4X,2023MNRAS.518.3796C}. These discrepancies may arise from uncertainties in stellar atmosphere parameters or copper and iron abundances, including differences in codes, model atoms, atomic data, namely, in rate coefficients for inelastic collisions with hydrogen atoms \citep{2022ApJ...936....4X,2023MNRAS.518.3796C}.

In summary, some of the most luminous stars exhibit sodium enhancements, which are potentially attributed to evolutionary changes in abundances.
For odd-Z elements (Al, K, Sc), their [X/Fe] trends are found to be flat in the Cetus stars. The limited metallicity range of the Cetus sample does not allow the detection of subtle changes in the [X/Fe] ratios for these elements. An exception is [Cu/Fe], where we found a prominent increase with increasing [Fe/H].

\subsubsection{Iron-peak elements and zinc}
\begin{figure}
\includegraphics[trim={10 30 30 10},width=80mm]{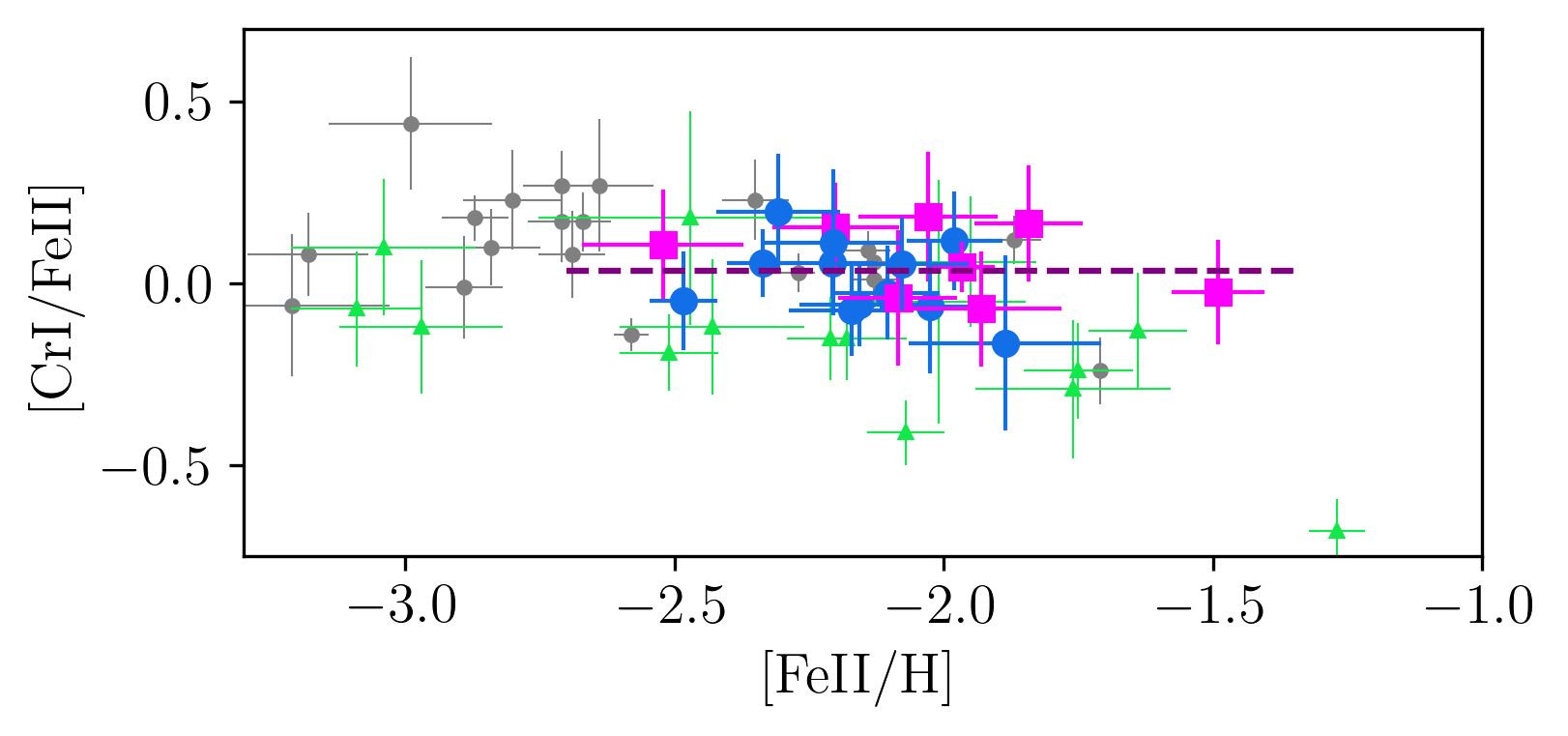}
\includegraphics[trim={10 30 30 10},width=80mm]{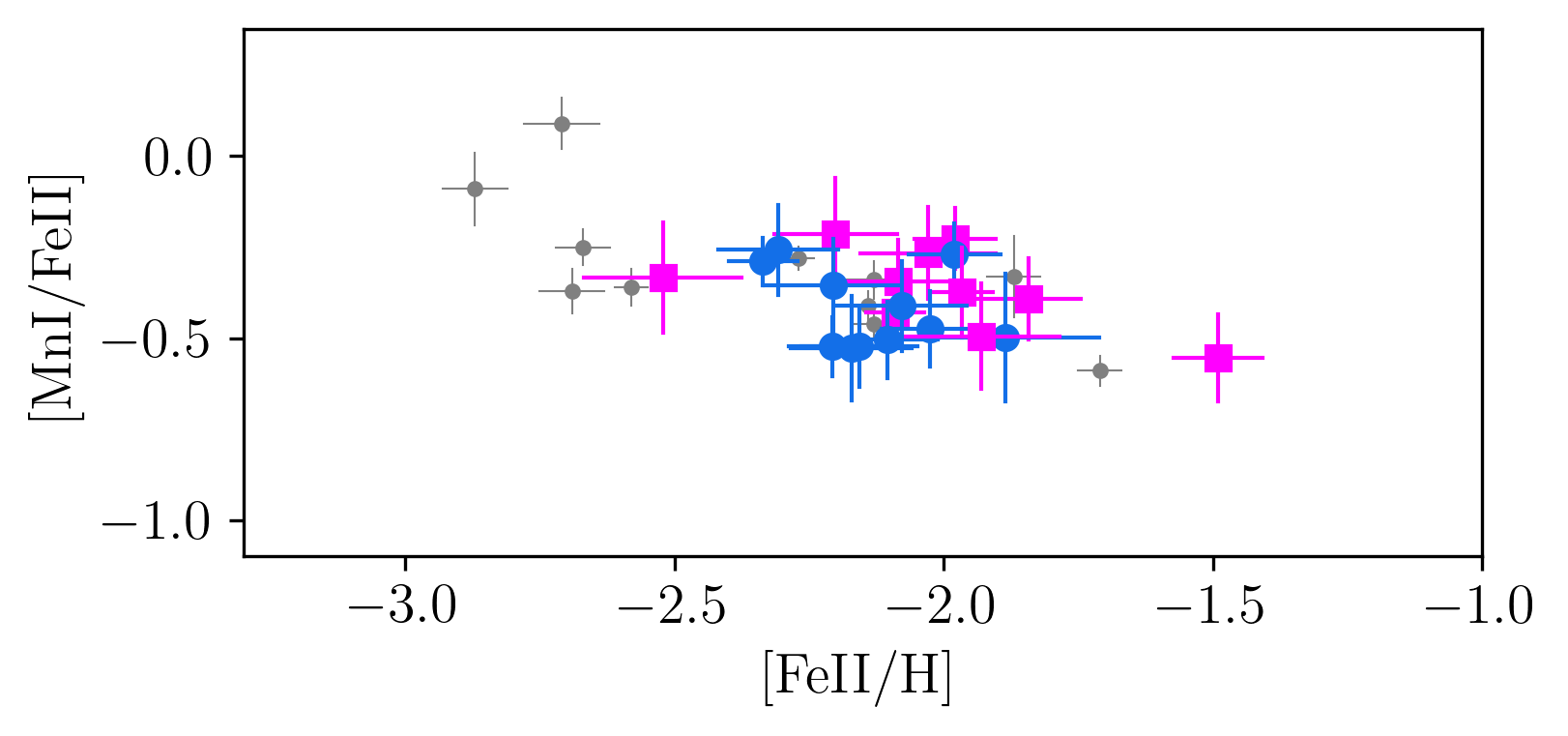}  
\includegraphics[trim={ 0 30 30 10},width=80mm]{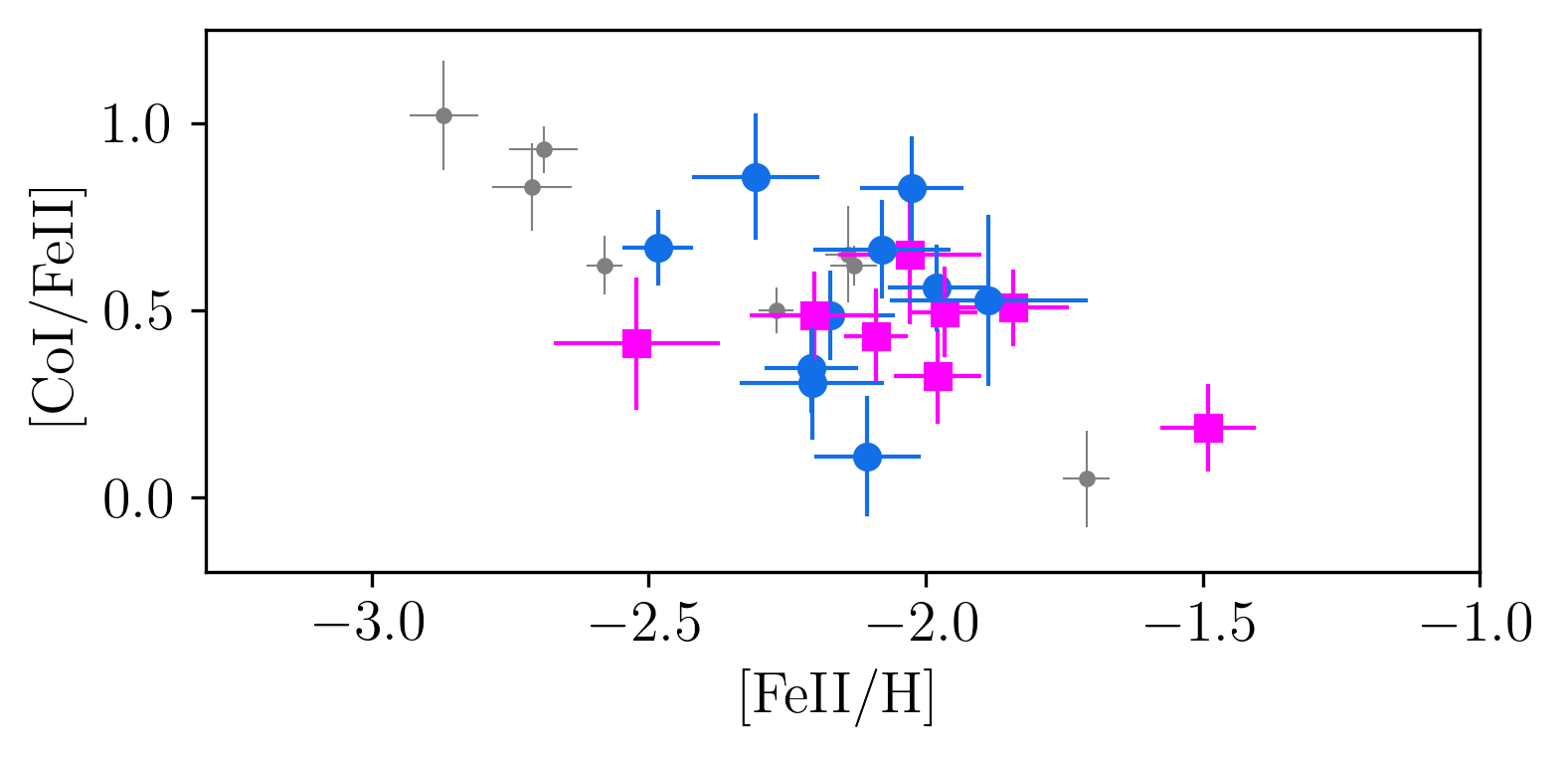}
\includegraphics[trim={10 30 30 10},width=80mm]{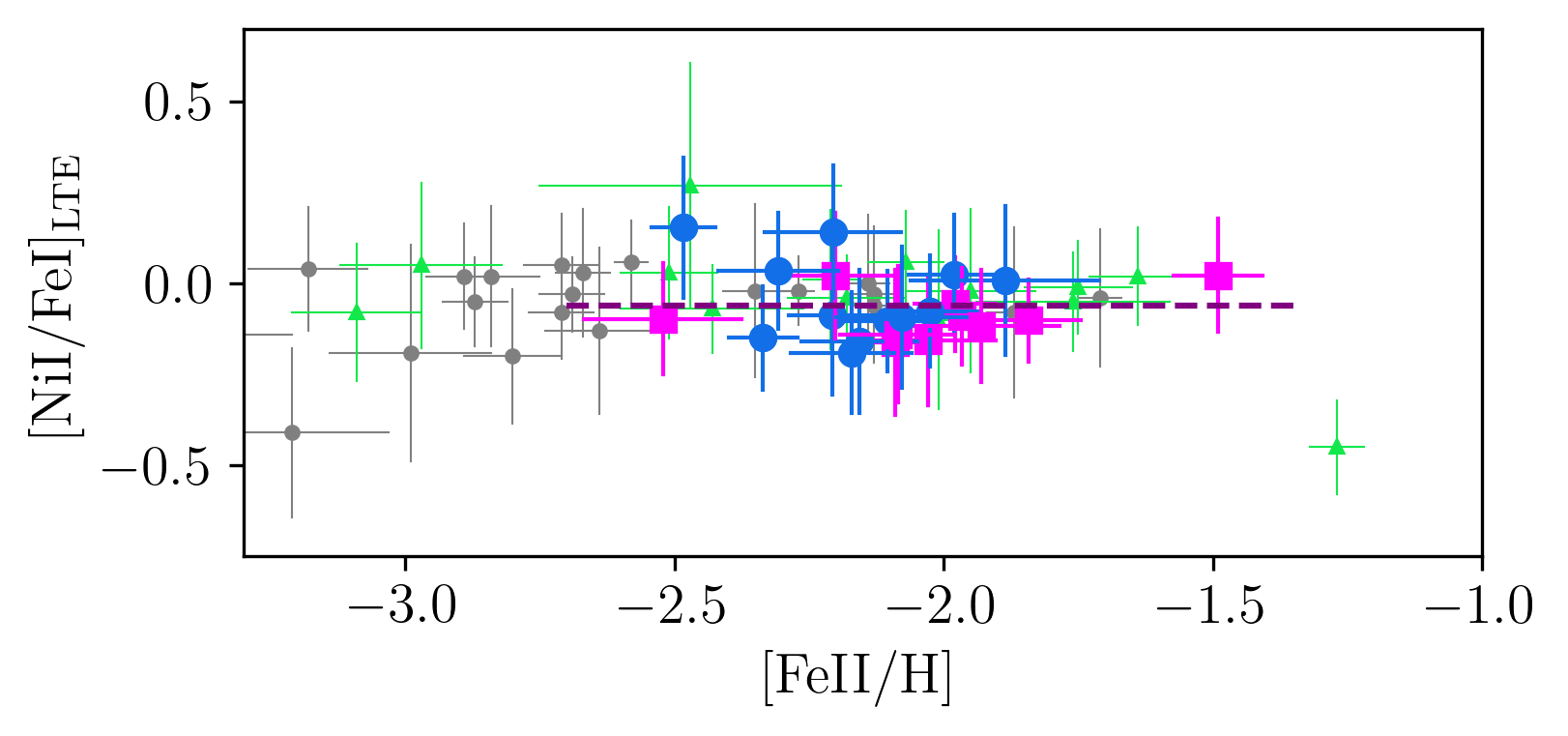}
\includegraphics[trim={10  0 30 10},width=80mm]{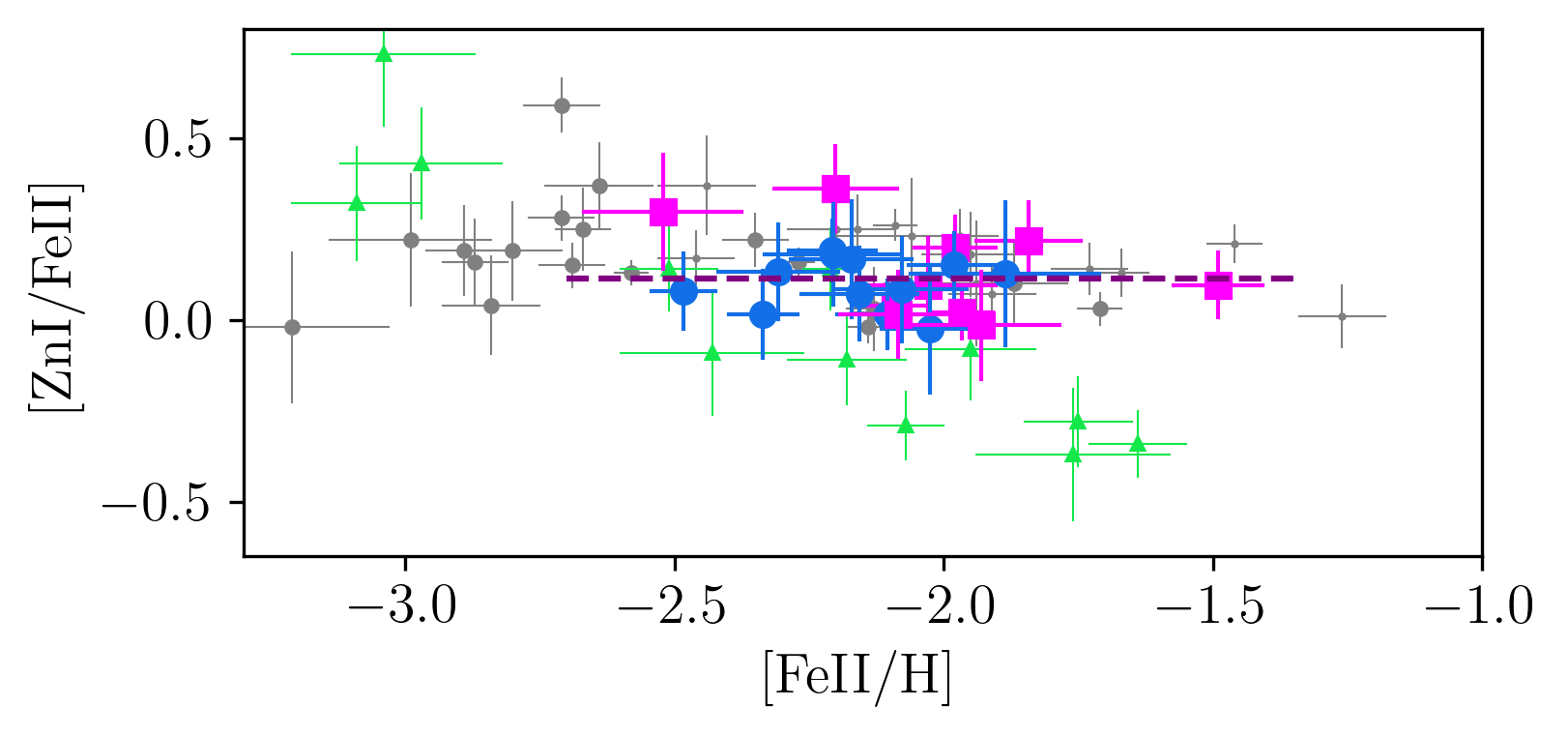}
\caption{NLTE abundance ratios for iron-peak elements. For Ni\ione, LTE abundances are plotted with respect to the LTE abundances from Fe\ione. The designations are the same as in Fig.~\ref{alpha_fe}.}
\label{iron_peak} 
\end{figure}

The [X/Fe] ratios of the iron peak elements (Cr, Mn, Co, Ni) and zinc in the Cetus stars align well with those in the MW comparison sample stars with similar metallicities (Fig.~\ref{iron_peak}.)

In Cetus, chromium follows iron and exhibits an average ratio of [Cr/Fe] = 0.03 $\pm$ 0.10, consistent with the NLTE study of [Cr/Fe] performed by \citet{2010A&A...522A...9B}. For [Mn/Fe], the Cetus stars show a constant value of [Mn/Fe] = $-0.39$ $\pm$ 0.11. Our average ratio in the Cetus and comparison sample stars is 0.2~dex lower than the nearly constant [Mn/Fe] = $-0.2$ reported for the VMP stars in the NLTE study of \citet{2019A&A...631A..80B}.

Regarding [Co/Fe], the MW comparison sample shows a steep decrease of $\sim$1 dex as [Fe/H] increases from $-2.8$ to $-1.7$, consistent with the earlier NLTE study of \citet{2010MNRAS.401.1334B}. The [Co/Fe] trend in Cetus is consistent with that in the MW (Fig.~\ref{iron_peak}). However, the scatter in Cetus is larger  and can be attributed to the uncertainties in the observed spectra: the two lines of Co\ione\  at 4110 and 4121~\AA\ are available for abundance determination and they are displaced in the blue spectral region, where the S/N is lower compared to the red region.

For nickel, our determinations rely on the LTE analysis. We assume that the NLTE effects are nearly the same for Ni\ione\ and Fe\ione\ and analyse the [Ni\ione/Fe\ione]$_{\rm LTE}$ ratio, where both Ni\ione\ and Fe\ione\ abundances are taken in LTE. Using the same approach, M17b found a solar value of [Ni/Fe] for the VMP stars in the MW halo and dwarf galaxies. We find the same result for the Cetus stars with an average [Ni\ione/Fe\ione]$_{\rm LTE}$ = $-0.06 \pm$ 0.09. For Cetus stars, on average, we find [Zn/Fe] = 0.11 $\pm$ 0.10, in agreement with the MW comparison sample stars with similar metallicities. In summary, nucleosynthesis in massive stars appears to be the primary source of iron-peak elements and zinc in the Cetus stars, with no clear evidence of a contribution from SNe~Ia.

\subsubsection{Neutron capture elements}
We adopt a commonly accepted opinion on the origin of neutron capture (n-capture) elements in rapid (r-) and slow (s-) neutron capture processes. Europium is a typical  r-element and is almost totally synthesised via the r-process. Its site and nuclear reactions branching along its path are debated  \citep[e.g. see][]{2021RvMP...93a5002C}. Barium is known as an s-element, as 81~\% of the barium in the  matter of the Solar System originates from the main s-process, which takes place in the low- to intermediate-mass (1--6 m$_\odot$) AGB stars \citep{1999ApJ...525..886A,1999ARA&A..37..239B,2011MNRAS.418..284B,2012ApJ...747....2L}. The r-process also produces a certain amount of barium.

For light n-capture elements (Sr, Y, Zr), observations of VMP stars with supersolar [Sr/Ba] ratios and [Ba/H] < $-2.5$ suggest the existence of an additional production channel besides the r-process and the main s-process \citep{2004ApJ...601..864T}. Its nuclear reactions and astrophysical sites are not yet known, while many hypotheses have been suggested (for their discussion and references, see Sect.~5.4 in MJ17b). It is important to highlight that the suggested sources share a common signature: they efficiently produce the light s-process elements (Sr, Y, Zr), while the production of barium either does not occur, as in the weak s-process \citep{1991ApJ...367..228R}, or occurs to a much lesser degree, as in fast rotating massive stars \citep[][]{2016MNRAS.456.1803F,2018ApJS..237...13L} or  in the intermediate (i-)process occurring in rapidly accreting white dwarfs \citep{2018ApJ...854..105C} and AGB stars with masses ranging from 1 to 4 m$_{\odot}$ \citep{2024A&A...684A.206C}.

\begin{figure}
\includegraphics[trim={10 30 12 10},width=80mm]{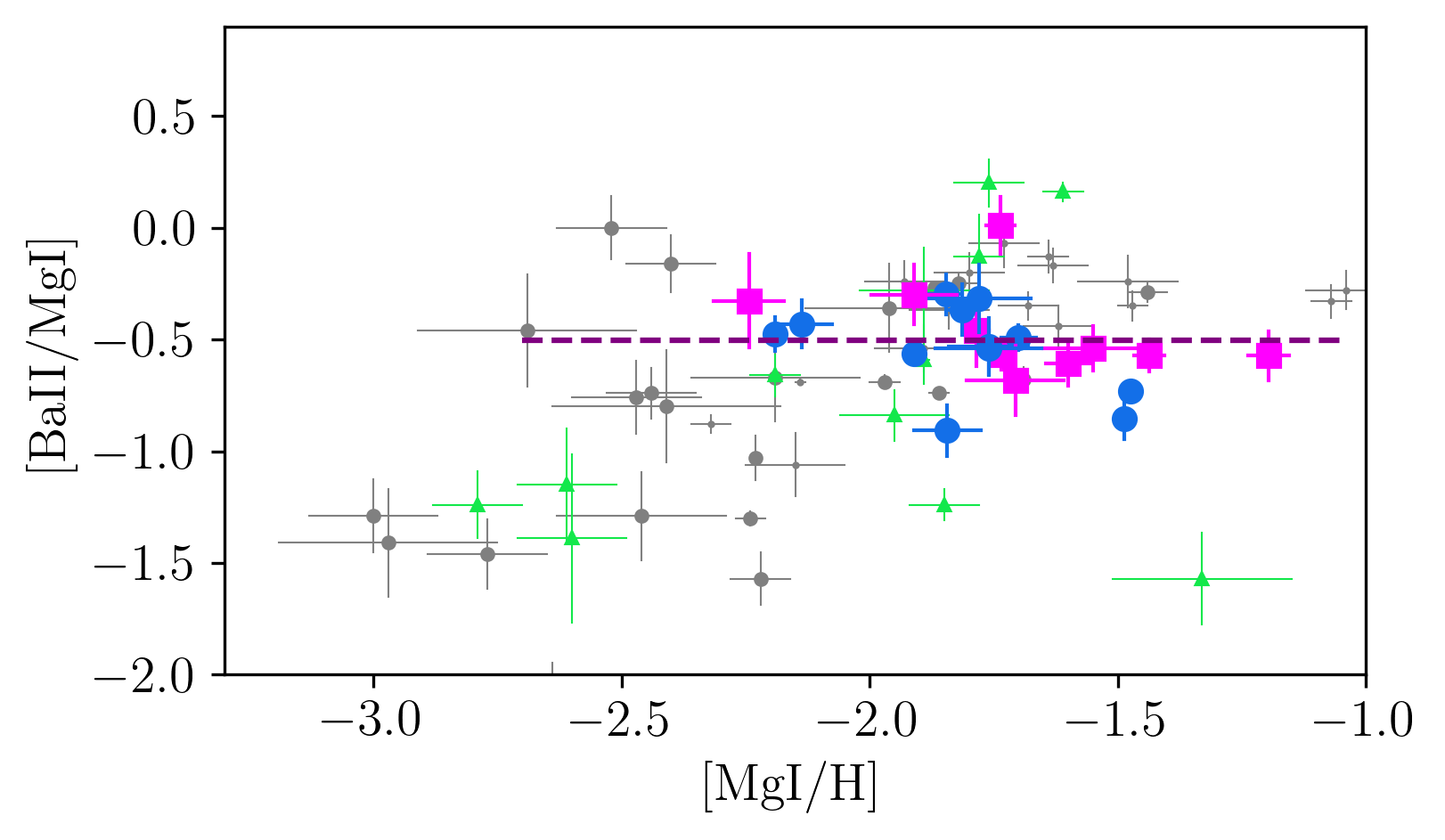}
\includegraphics[trim={10 15 12 10},width=80mm]{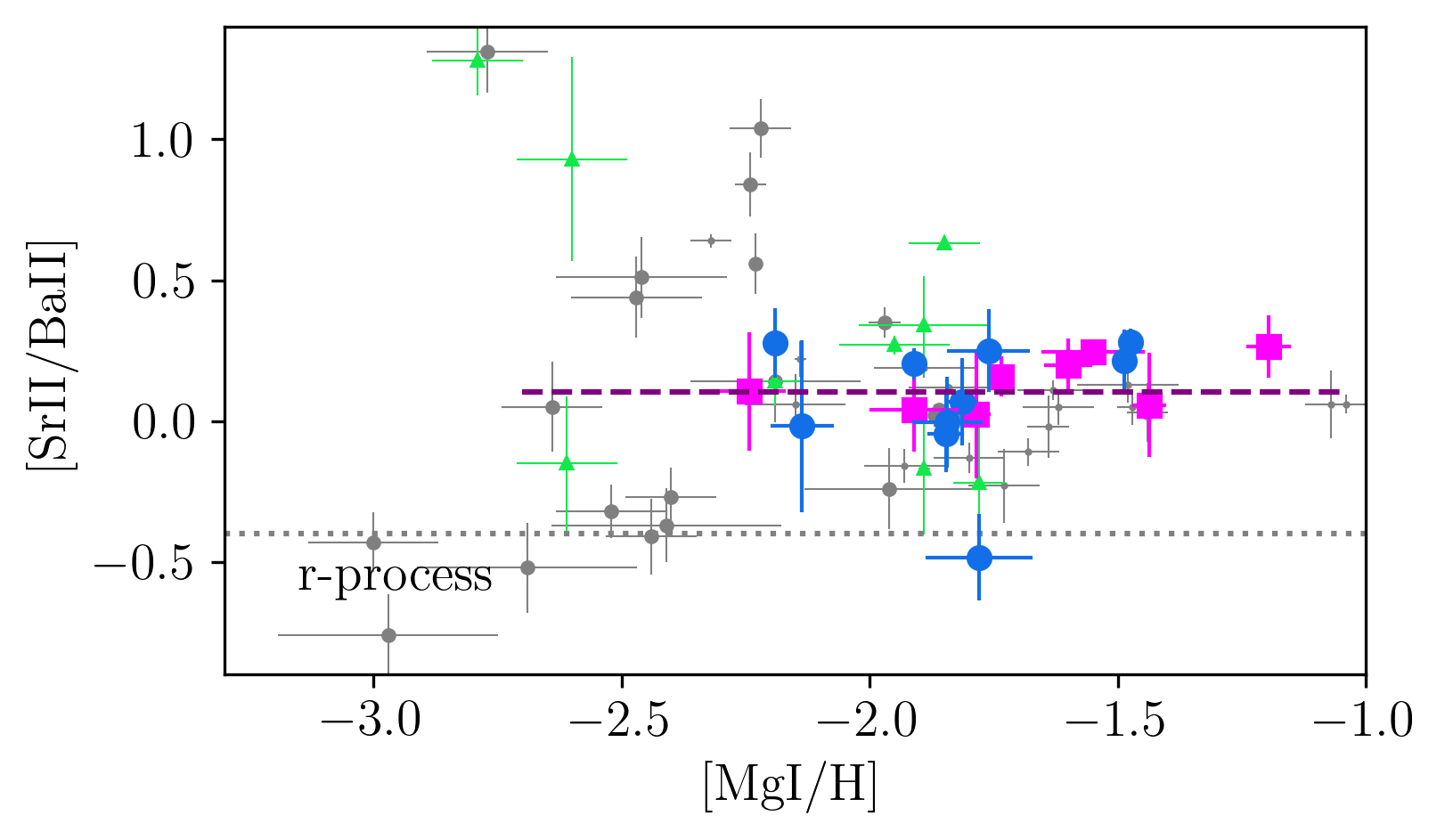}
\caption{Neutron capture element abundance ratios. The dotted line indicates the empirical r-process ratios observed in r-\ii\ type stars. The designations are the same as in Fig.~\ref{alpha_fe}.}
\label{rs_ncapture} 
\end{figure}

The upper panel of Fig.~\ref{rs_ncapture} shows [Ba/Mg] as a function of [Mg/H]. Magnesium, originating primarily from massive stars, can serve as a benchmark element for VMP stars, in contrast to iron, which stems from stars of various masses. In the Appendix, we present conventional diagrams employing iron as the reference element. The Cetus stars have [Mg/H] > $-2.2$ and exhibit a flat [Ba/Mg] trend with an average ratio [Ba/Mg] = $-0.51$ $\pm$ 0.20. This value aligns well with the [Ba/Mg] ratio in the MW halo stars with similar [Mg/H]. For [Mg/H] < $-2.3$, the MW halo stars exhibit an increase in [Ba/Mg] as [Mg/H] rises, although there is substantial scatter of [Ba/Mg], namely from $-1.5$ to 0. Our Cetus sample lacks stars with [Mg/H] < $-2.3$ and does not allow us to track its earliest stages.

To understand the sources responsible for n-capture element production within different regimes, let us take a look at the [Sr/Ba] -- [Mg/H] diagram (Fig.~\ref{rs_ncapture}, bottom panel). The Cetus stars exhibit a nearly solar [Sr/Ba] ratio with an average [Sr/Ba] = 0.10 $\pm$ 0.18, in line with the MW halo stars with similar [Mg/H]. The MW halo stars with [Mg/H] < $-2.2$ show a dispersion of 2 dex in [Sr/Ba], with a minimum value at [Sr/Ba]$_{\rm r}$ = $-0.4$. The latter value is known as an empirical r-process ratio observed in stars strongly enhanced in r-process elements \citep{2003ApJ...591..936S,2005A&A...439..129B,2009A&A...504..511H,2010A&A...516A..46M,2018ApJ...865..129R,2024MNRAS.529.1917S}. The supersolar [Sr/Ba] ratios in stars with [Mg/H] < $-2.2$ are due to the contribution to Sr from an additional source(s) of unknown nature, as discussed earlier.

The Cetus stars do not exhibit substantially supersolar [Sr/Ba] ratios. One possibility explaining this situation could be that the stars were formed after completing that additional source in the Cetus progenitor; another possibility is that the source of unknown nature did not run at all. Further observations and increased statistics of Cetus VMP stars with [Mg/H] < $-2.3$  are needed to study its earliest epochs.

\begin{figure}
\includegraphics[trim={20  0  0 10},width=80mm]{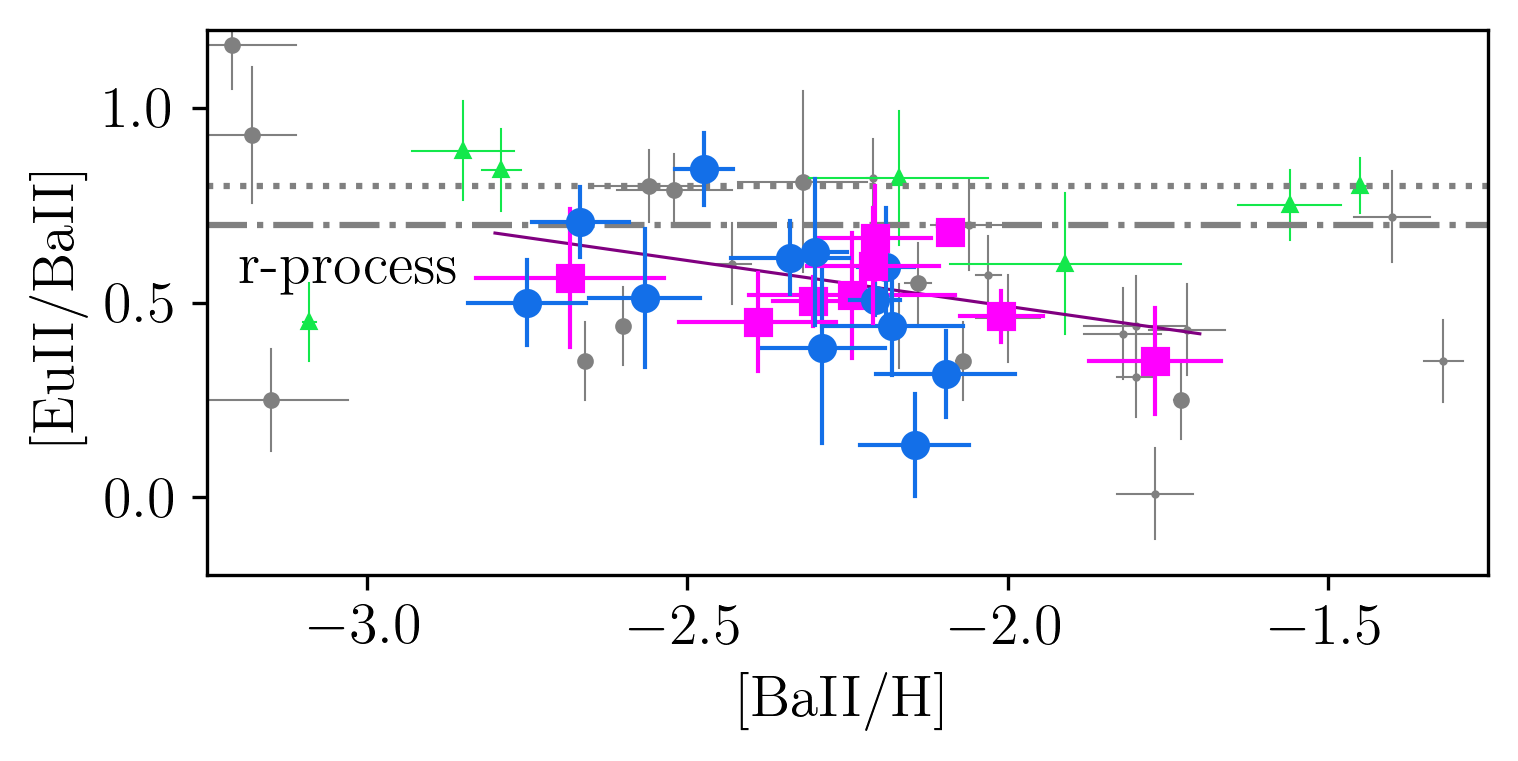}
\includegraphics[trim={20  15 0  7},width=80mm]{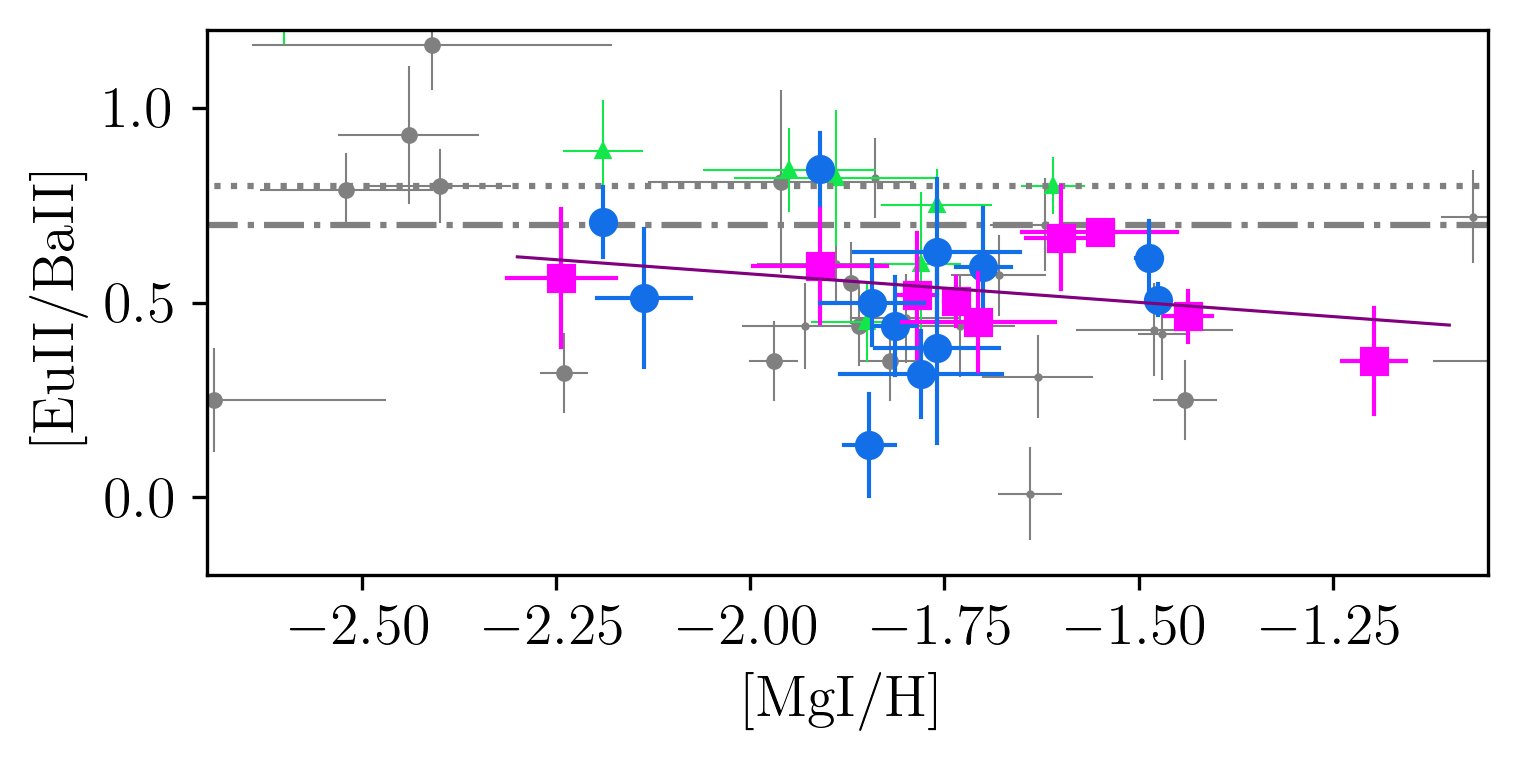}
\caption{Neutron capture element abundance ratios. The predicted r-process element ratios are shown with the dotted line \citep{1999ApJ...525..886A} and dashdotted line \citep{2014ApJ...787...10B}. The solid line represents a linear regression of the trend. The designations are the same as in Fig.~\ref{alpha_fe}.}
\label{ntrend}
\end{figure}

For our sample stars, we estimate the contribution of the r-process to their chemical composition using their [Eu/Ba] ratios. Figure~\ref{ntrend} shows [Eu/Ba] as a function of [Ba/H]. We find an average value [Eu/Ba] = $0.53$ $\pm$ 0.13. This ratio is between the solar value and the r-process [Eu/Ba]$_{\rm r}$ = $0.80$ \citep[][here the r-residual is defined as the difference between solar total and s-process abundance]{2014ApJ...787...10B}. Our average value [Eu/Ba]$_{\rm Cetus}$ $\simeq$ 0.5 indicates that the Eu to Ba number density ratio in r-process material is twice (10$^{0.3}$) that in the Cetus stars. In other words, on average, half of the barium is produced in the r-process. This estimate is close to the s-process contribution of 30\% $\pm$ 30\% to barium in metal-poor thick-disk stars, as found by \citet{2003A&A...397..275M}. 

In different Cetus stars, [Eu/Ba] takes values from $-0.1$ to 0.8, corresponding to r-process contributions ranging from 20\%  to 100\%, respectively. We note a decreasing trend in [Eu/Ba] relative to [Ba/H] (Fig.~\ref{ntrend}). We find this trend to be statistically significant with a slope of $-0.23$ $\pm$ 0.14 and a p-value of 0.06. In other words, assuming a decrease in the [Eu/Ba] -- [Ba/H] diagram, there is only a 6~\% chance that our hypothesis is wrong. 

Figure~\ref{ntrend} (bottom panel) also shows [Eu/Ba] as a function of [Mg/H]. In the latter case, the decreasing trend is less pronounced, with a slope of $-0.16$ $\pm$ 0.11 and a p-value of 0.11. Nevertheless, we can conclude that, as metallicity or [Mg/H] increases, the barium production rate grows relative to that for europium, suggesting an increasing contribution from the s-process in AGB stars. We also note a scatter of [Eu/Ba], indicating that the interstellar medium in the Cetus progenitor was not fully mixed.

In Cetus, we find the trend in  [Eu/Ba]  to be consistent with that in the MW halo indicating a similar relative contribution from AGB stars in these two systems at the [Ba/H] ranging from $-2.7$ to $-1.8$. In our sample of UMi comparison stars, the r-process dominates and stars with [Ba/H] of $-1.5$ still exhibit close to r-process [Eu/Ba] ratios of 0.7. A different situation was found by \citet{2020A&A...634A..84S} from an n-capture element abundance analysis in Sculptor (Scl) dSph, a dwarf galaxy with a stellar mass of 10$^6$m$_{\rm \odot}$, which is close to that of the Cetus and UMi. Stars in Scl with [Ba/H] > $-1.5$ exhibit lower [Y,La,Nd,Eu/Ba] compared to the MW halo stars, suggesting a higher relative contribution from AGB stars at these [Ba/H]. The distinct trends observed in these systems can be explained by their different star formation histories. In the Cetus progenitor, within the studied metallicity range, the star formation history was similar to that in the MW halo. It is worth noting that, in Scl, [Ba/H] spans a wide range up to [Ba/H] = $-0.5$, while in our Cetus sample, [Ba/H] reaches $-1.8$.

\begin{figure}
\includegraphics[trim={16  30 7 12},width=80mm]{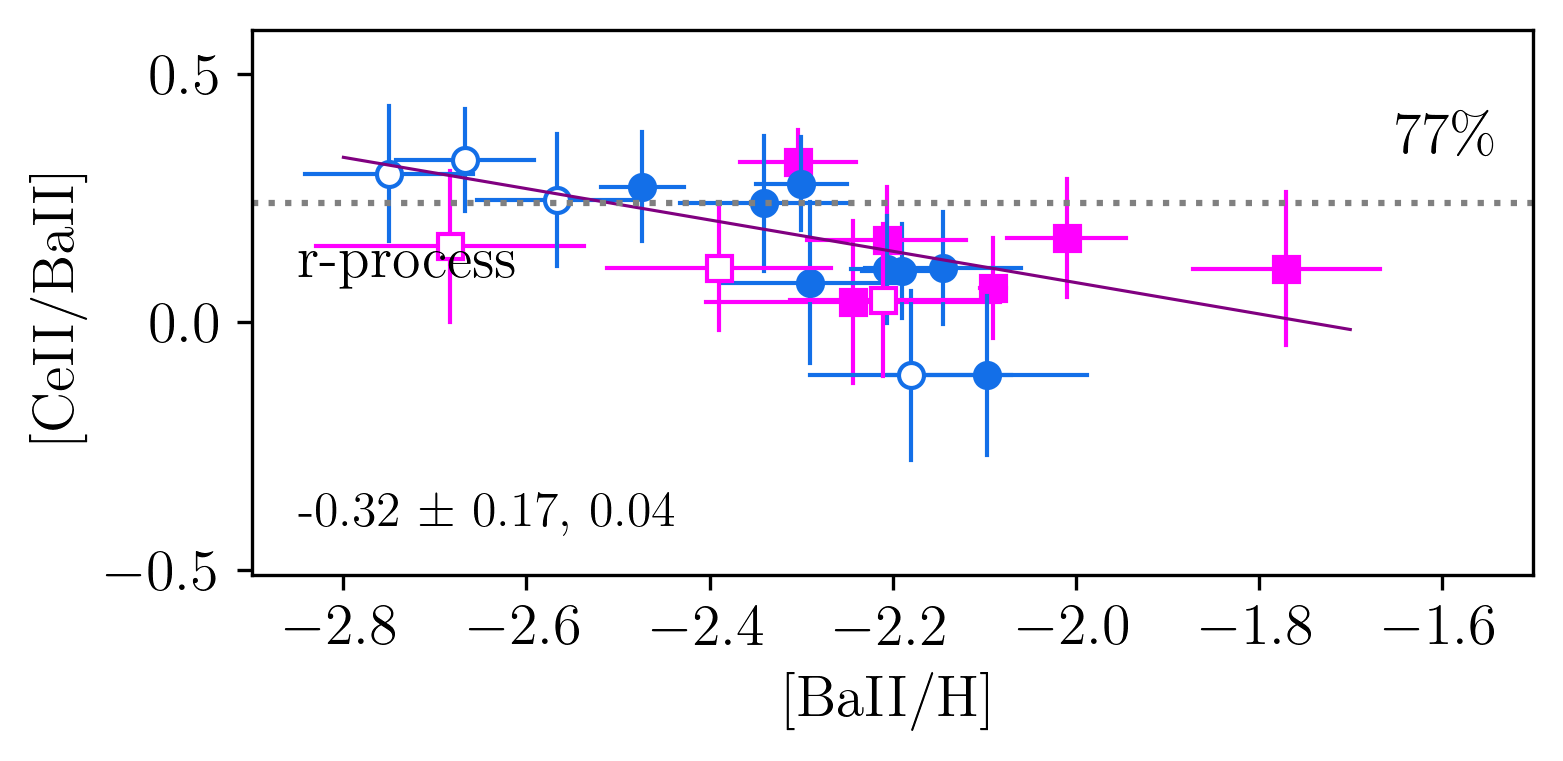}
\includegraphics[trim={16  30 7 12},width=80mm]{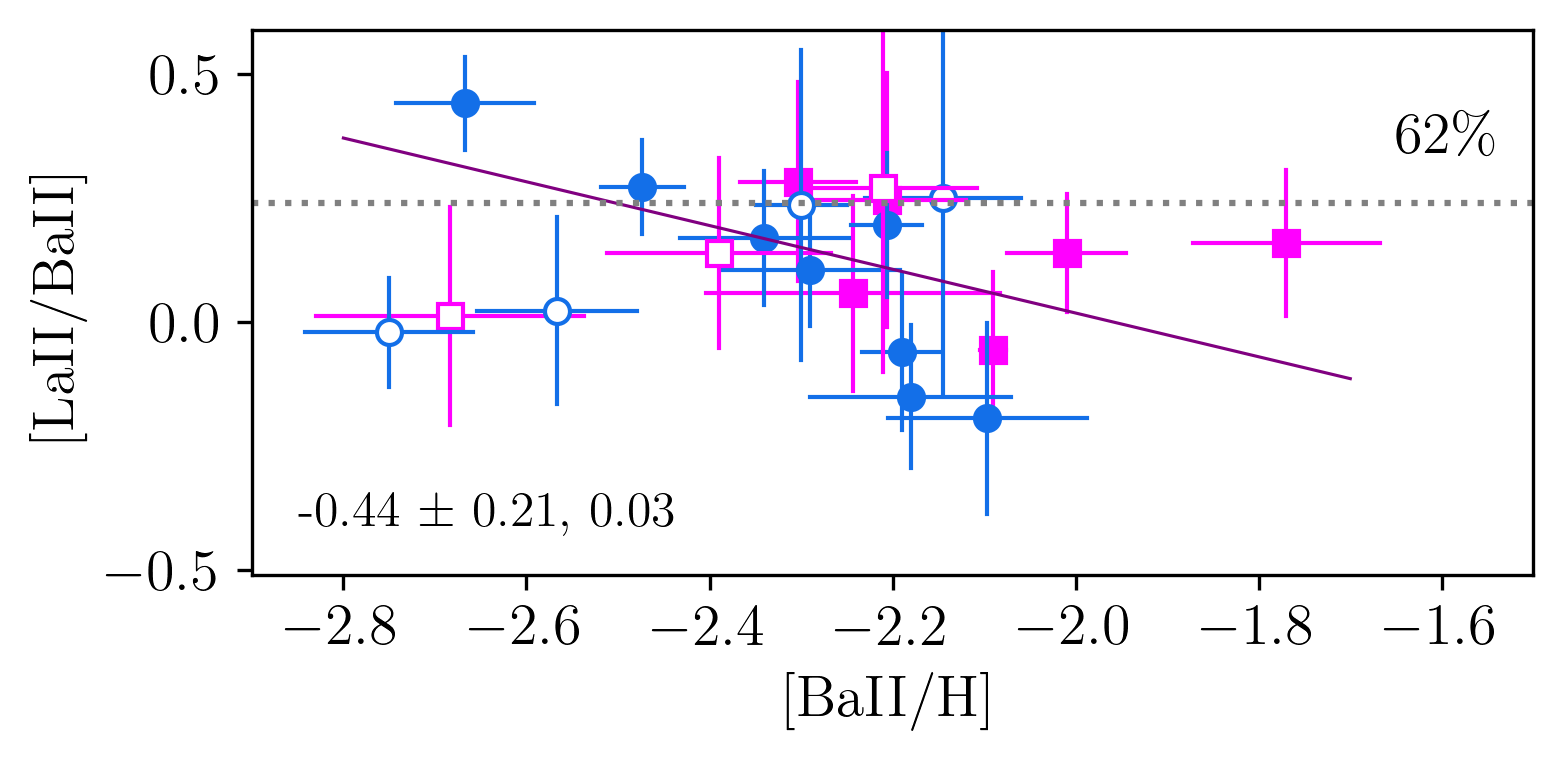}
\includegraphics[trim={5   30 7 12},width=80mm]{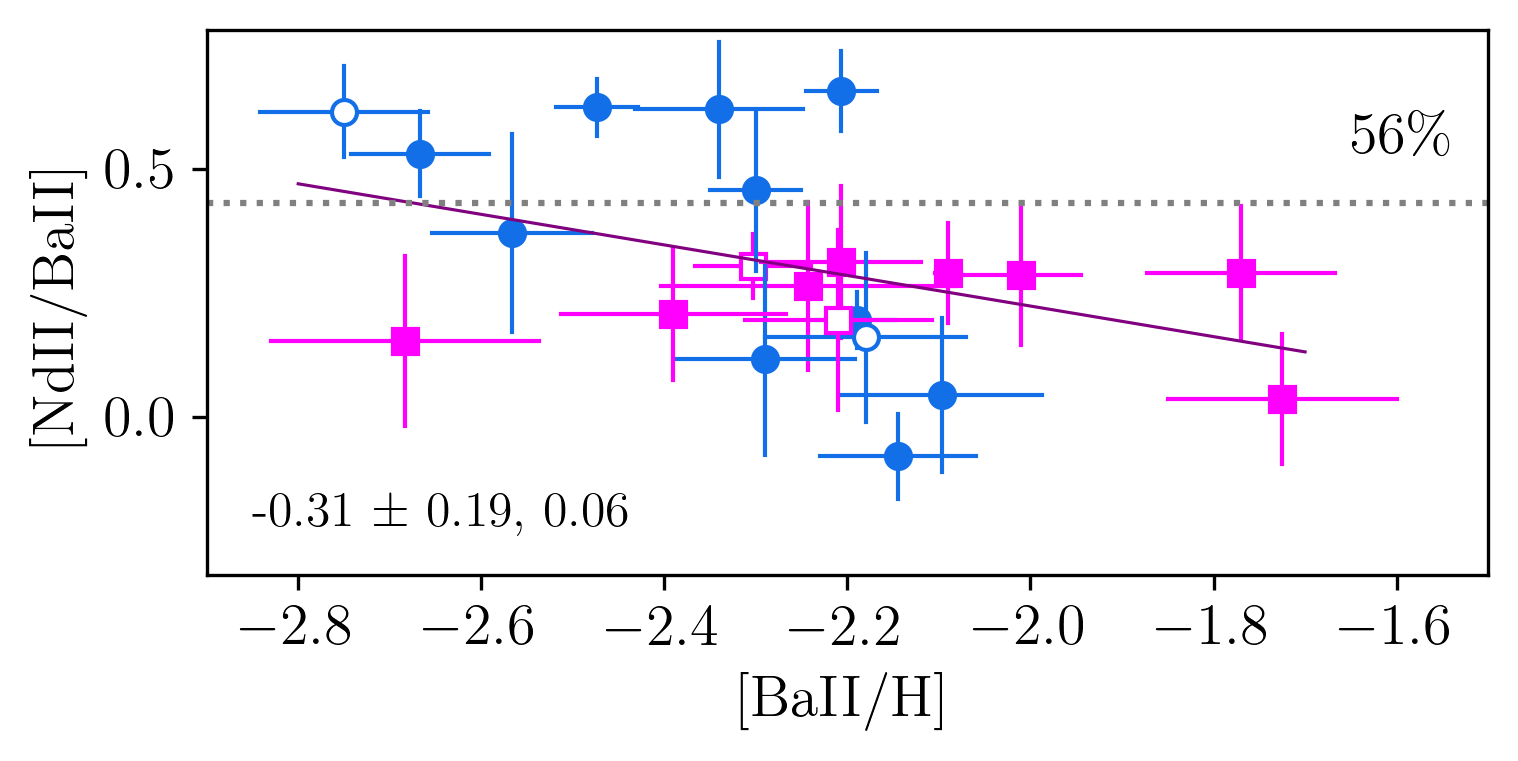}
\includegraphics[trim={5   15 7 12},width=80mm]{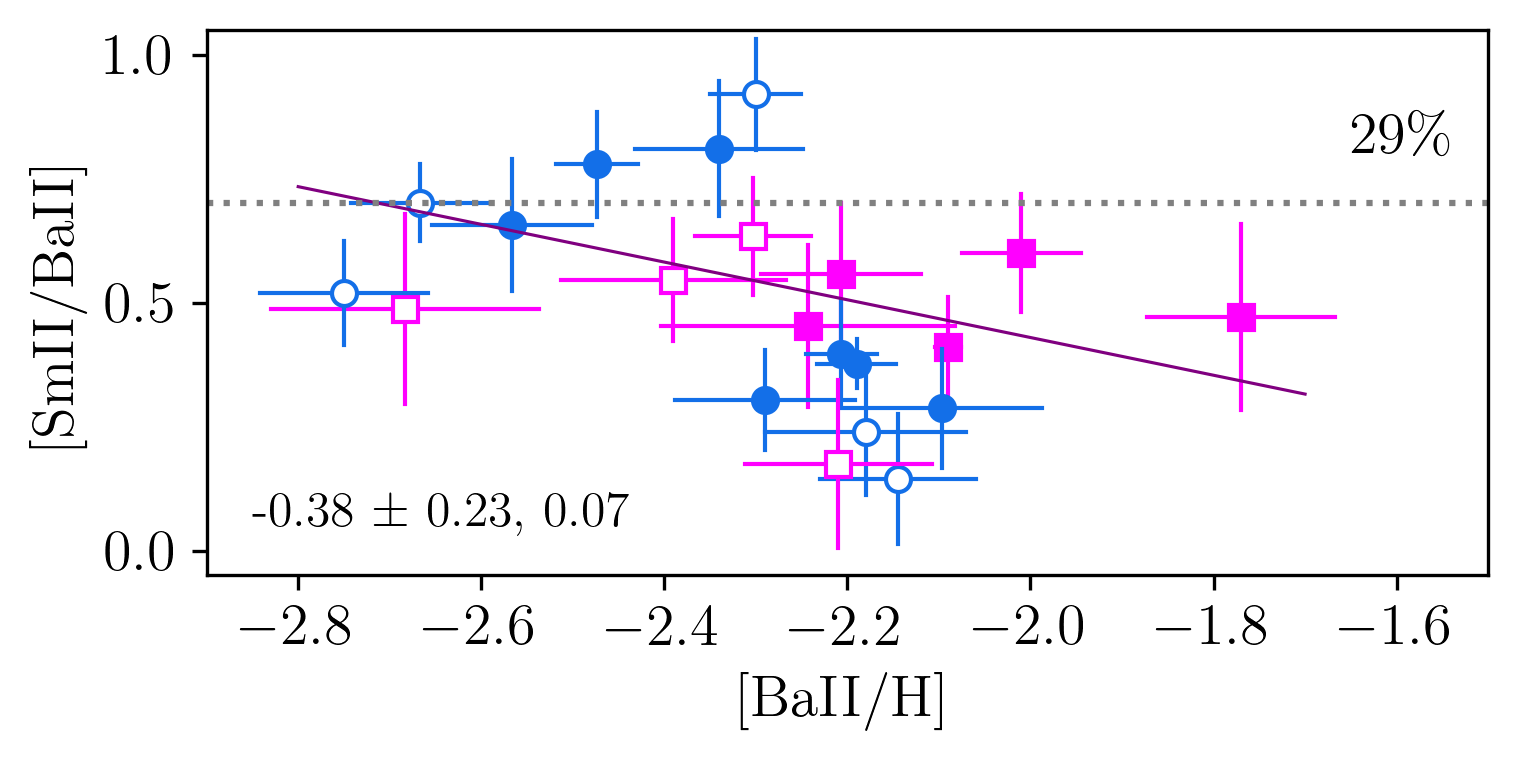}
\caption{Neutron capture element abundance ratios. The solid line represents a linear regression of the trend. For each trend, the slope and the corresponding p-value are indicated. The dotted line indicates the empirical r-process ratios observed in r-\ii\ type stars; see text for details. Open symbols represent stars with less reliable abundance measurements. The fraction of s-nuclei in the Solar System is indicated for each element. The designations are the same as in Fig.~\ref{alpha_fe}.}
\label{mains}
\end{figure}

Besides barium, the main s-process produces effectively second peak s-process elements such as cerium, lanthanum, neodymium, and samarium. The Solar System s-nuclei fractions for these elements are 77 \%, 62 \%, 65 \%, and 29 \%, respectively \citep{1999ApJ...525..886A}. However, their production in the main s-process is less effective compared to barium, which has an s-nuclei fraction of 81 \% in the Solar System. 

Relative contributions of the s- and r-processes to chemical abundances of the Cetus stars can be drawn by plotting the abundance ratios for the n-capture elements with different contributions of the s- and r-processes to their solar abundances with respect to barium. As seen in Fig.~\ref{mains}, each of the four elements (Ce, La, Nd, Sm) shows statistically significant decreasing trends as barium increases, indicating an increasing contribution from the s-process in AGB stars. The slopes and p-values are quoted in the corresponding panels of Fig.~\ref{mains}. For [Ba/H] > $-2.2$, the [Ce, La, Nd, Sm/Ba] ratios decrease below the empirical r-process ratios calculated for the r-\ii\ stars using the data from \citet{2003ApJ...591..936S}, \citet{2005A&A...439..129B}, \citet{2009A&A...504..511H}, \citet{2010A&A...516A..46M}, \citet{2018ApJ...865..129R}, and \citet{2024MNRAS.529.1917S}.

Examining the diagrams more closely, we note a hint of distinct behaviour between the Cetus-New and Cetus-Palca stars. The Cetus-Palca stars show a prominent decrease with a variation of about 0.5~dex in abundance ratios, while the Cetus-New trends remain almost flat. However, this impression mainly relies on three Cetus-New stars -- two with [Ba/H] $\geq -2.0$ and one with [Ba/H] = $-2.7$. Therefore, our current statistics do not allow us to draw solid conclusions about the distinct behaviour of n-capture elements in the two wraps. 

\begin{figure}
\includegraphics[trim={ 0 30 20 10},width=80mm]{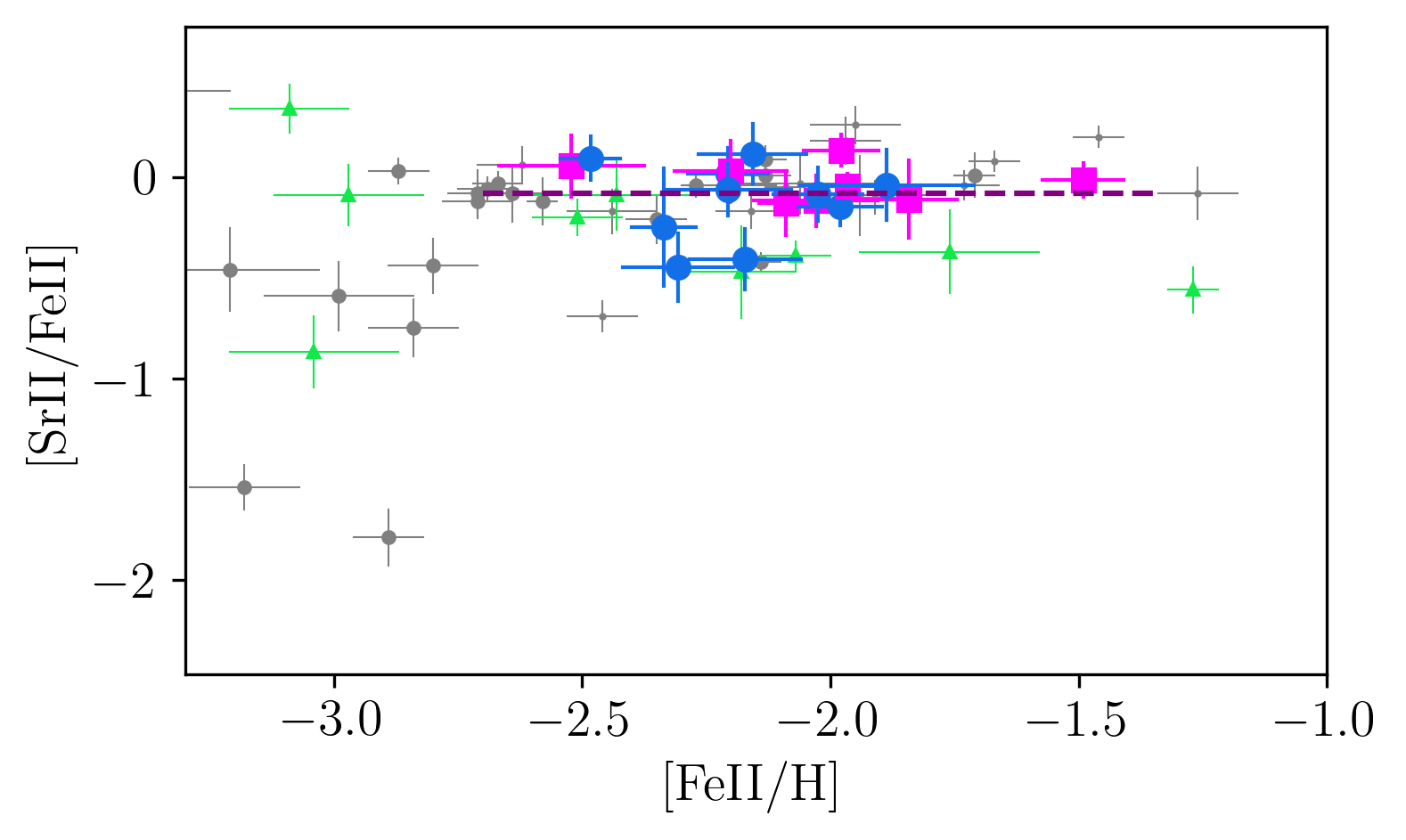}
\includegraphics[trim={10 30 20 10},width=80mm]{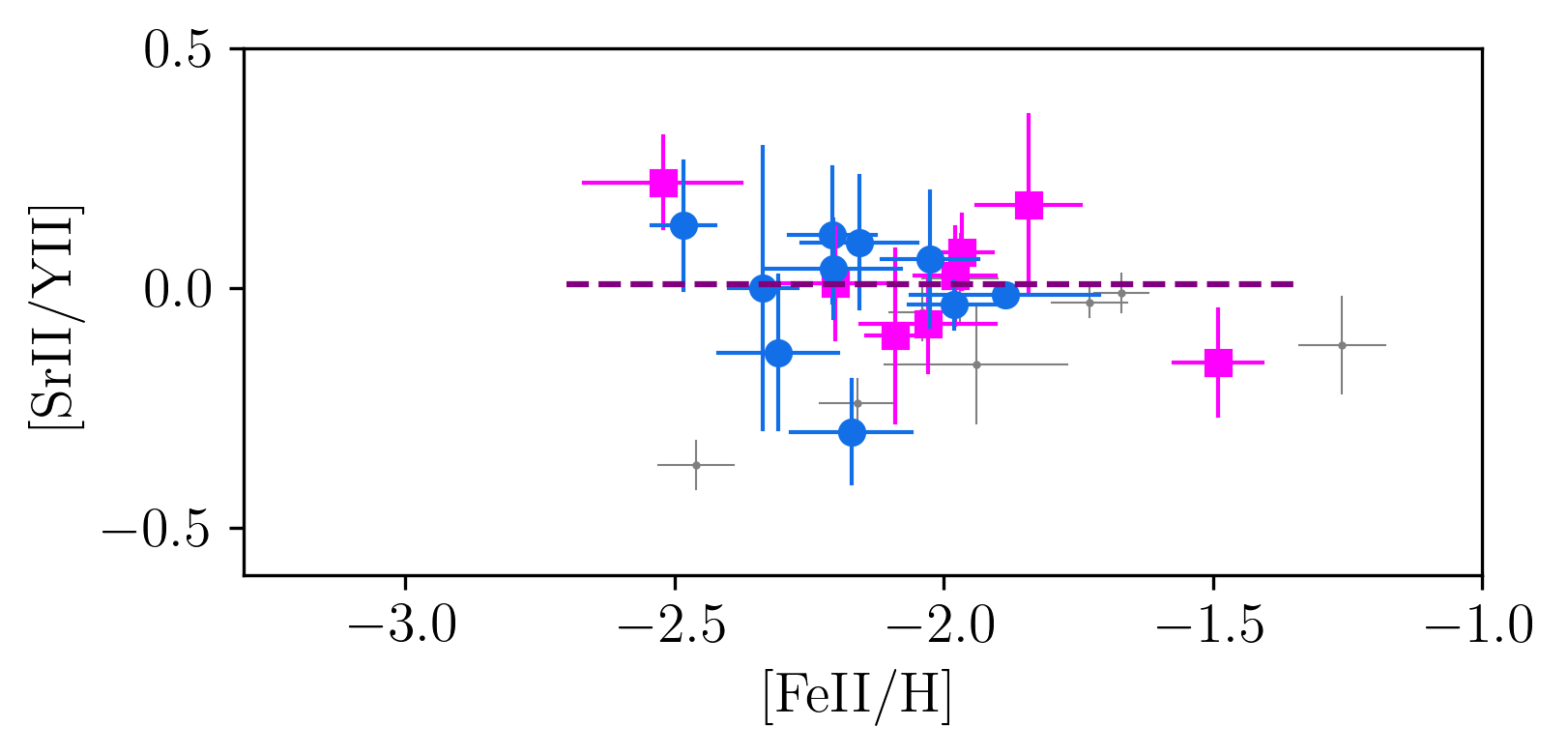}
\includegraphics[trim={ 0  7 20 10},width=80mm]{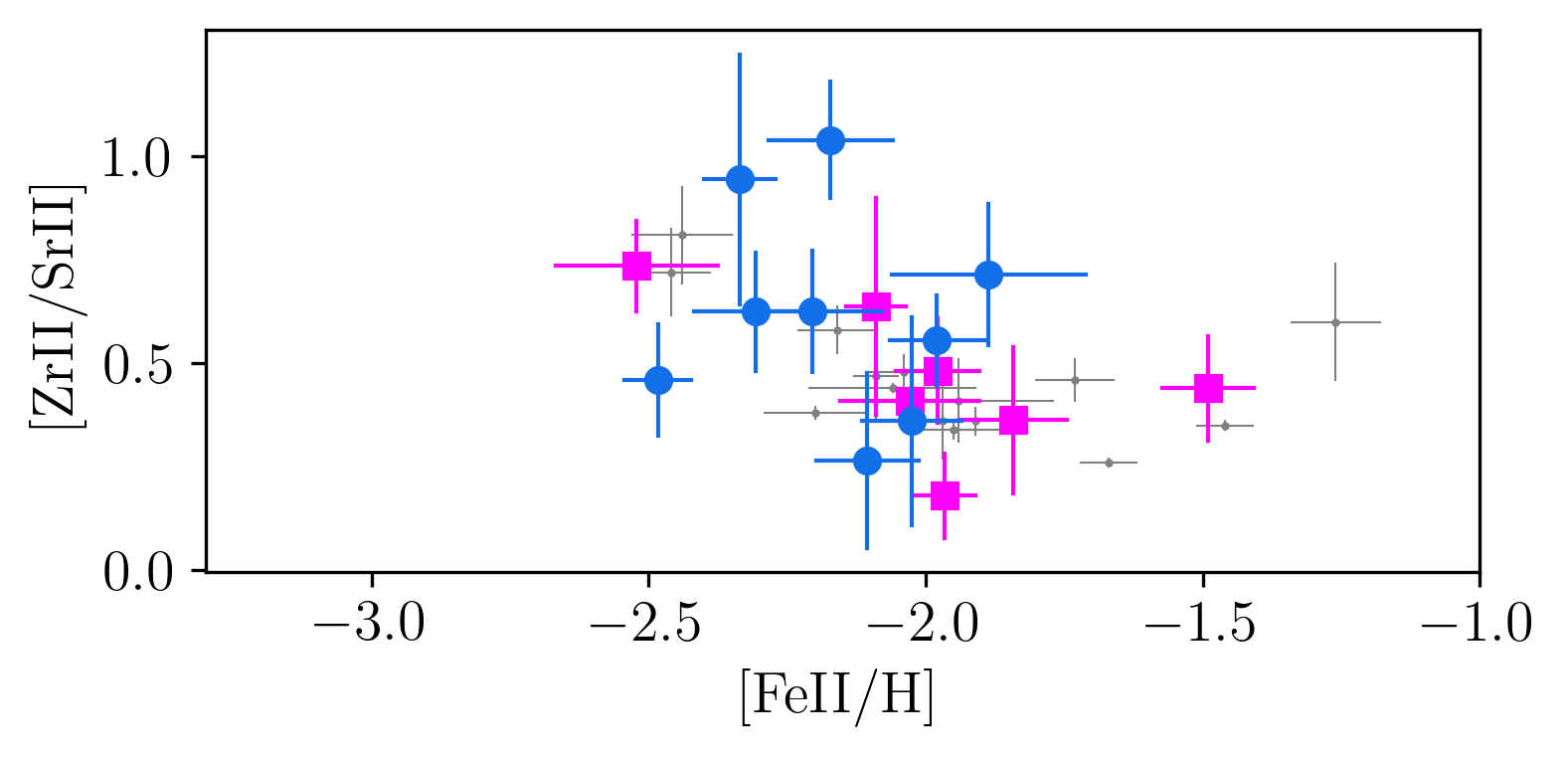}
\includegraphics[trim={5   15 7  0},width=80mm]{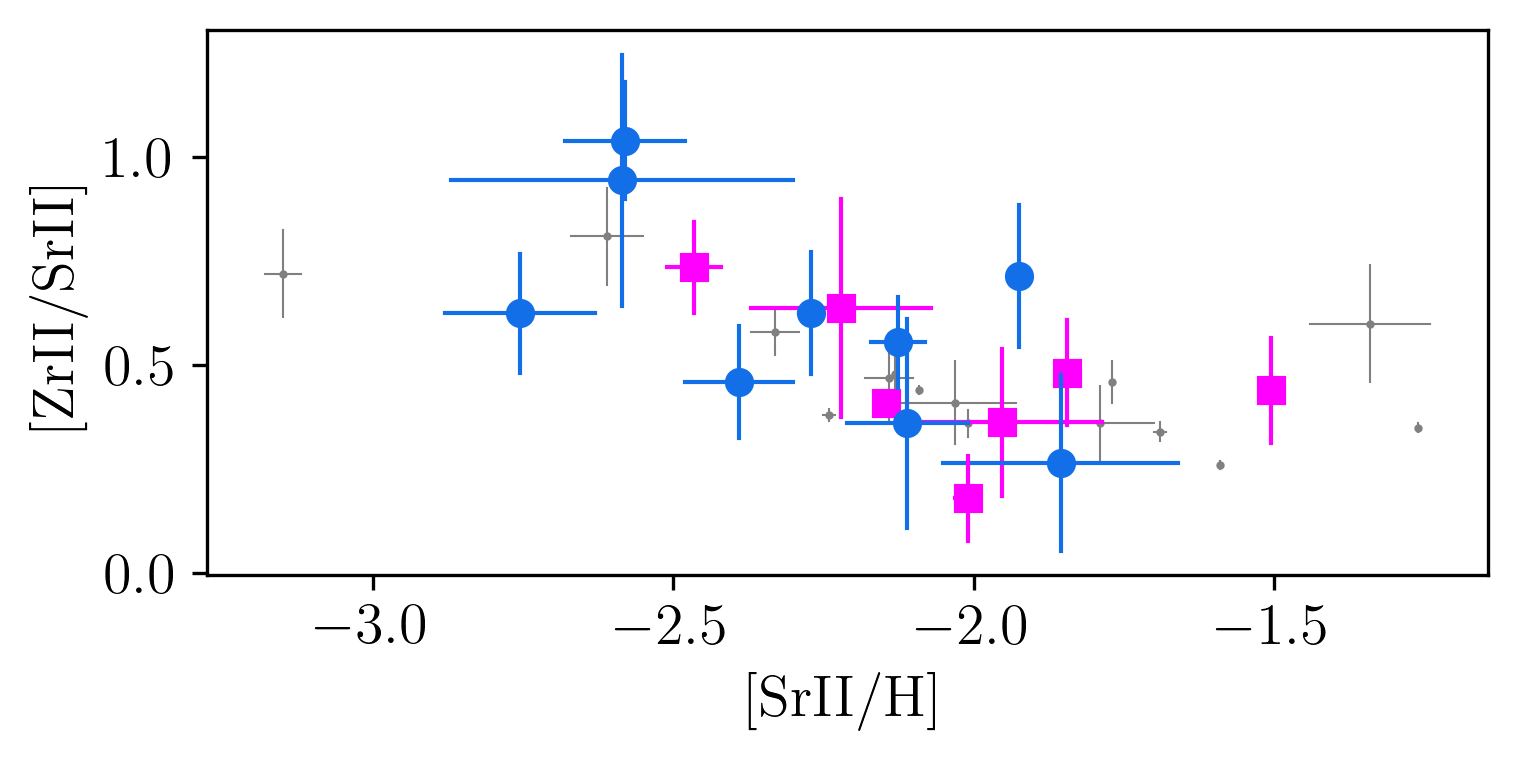}
\caption{Abundance ratios for light neutron-capture elements. The designations are the same as in Fig.~\ref{alpha_fe}.}
\label{lepp_ncapture} 
\end{figure}

In both the Cetus stream and the MW comparison sample, the [Sr/Y] trend is flat (Fig.~\ref{lepp_ncapture}), with values close to the solar ratio, indicating similar production rates for these light n-capture elements. The situation is different for [Zr/Sr], which ranges from about 0.2~dex to 1~dex in the Cetus stars. These ratios may look like a scatter in the diagrams versus metallicity [Fe/H], while they form a clear declining trend in the plot against [Sr/H] (Fig.~\ref{lepp_ncapture}). The latter suggests that the Cetus stars formed in the environment with a higher production rate for strontium and yttrium compared to that for zirconium. This is one more argument for the onset of the main s-process in the Cetus progenitor, because zirconium is less efficiently produced in the main s-process compared to strontium and yttrium. The corresponding contributions of Sr, Y, and Zr to the Solar System amount to 71\%, 69\%, and 65\%, respectively \citep[main s-process,][]{2004ApJ...601..864T}. In the MW halo, an increase in [Zr/Sr] and [Zr/Y] towards lower metallicity was reported by \citet{2016ApJ...833..225Z} and \citet{2023ApJ...957...10A}, respectively, while the [Zr/Sr] plateau was found by \citet{2022A&A...665A..10L} for their MW stellar sample.

From analyses of the [Sr/Ba] and [Sr/Y], we conclude that strontium, yttrium, and barium in the Cetus stars have a common nucleosynthetic origin from the r-process and the main s-process. Zirconium in the Cetus stars behaves distinctly, indicating its less effective production in the main s-process compared to the above elements.

In summary, the n-capture elements in the Cetus stars originate from both the r-process and the main s-process. In the different  stars of our sample, the contribution from these two processes is different, and, for example, for barium, the r/s ratio varies from 100\%/0\% to 20\%/80\% with an average value of 50\%/50\%.

\section{Conclusions}\label{conclusions}

We provide a comprehensive analysis of the chemical composition of 22 stars in the Cetus stream based on their high-resolution spectra. The abundances are derived by taking into account NLTE effects if possible. In total, we determined abundances for up to 28 chemical species from C to Dy and account for the NLTE effects for 20 of them. We summarise  the conclusions we draw from the derived abundances below.

\begin{itemize}
\item[$\circ$] The Cetus stars cover a metallicity [Fe/H]  range from $-2.5$ to $-1.5$. The mean metallicity of the stream is [Fe/H] = $-$2.11 $\pm$ 0.21. 
The chemical compositions of the Cetus-New  and Cetus-Palca wraps suggests that both follow the same chemical evolution path.

\item[$\circ$] All sample stars are $\alpha$ enhanced with [$\alpha$/Fe] = 0.3. The absence of a `knee' in the [$\alpha$/Fe] -- [Fe/H] diagram indicates that star formation stopped before iron production due to SNe~Ia became substantial. 

\item[$\circ$] Neutron capture element abundances suggest that  both  r-process and  main s-process contributed to their origin. The decrease in [Eu/Ba] from  a typical r-process value [Eu/Ba] = 0.7 to 0.3 with increasing [Ba/H] indicates a distinct contribution of the r- and s-processes to the chemical composition of different Cetus stars. For barium, the r-process contribution varies from 100\%  to 20\% in different sample stars, with an average value of 50\%.

\item[$\circ$] Star formation in the Cetus progenitor ceased after the onset of the main s-process in low-intermediate-mass AGB stars, but before SNe~Ia played an important role in the chemical enrichment of the galaxy.

\end{itemize}

When comparing the abundance ratios of the Cetus stream with the MW halo, we observe that, in the same metallicity range, these two samples generally align.
It is essential to bear in mind that MW halo stars exhibit a much broader range of abundance ratios, given the vast and complex origins of the MW halo. Regarding the observed differences, the most notable discrepancy is found in the [Cu/Fe] -- [Fe/H] trends: in Cetus, it is shifted by 0.5~dex towards lower [Fe/H] compared to the MW trend. 

Regarding our comparison with the UMi dSph, we observe significant differences between these two systems. In contrast to Cetus, UMi stars exhibit a `knee' in [$\alpha$/Fe], which we can use as an indicator of the onset of contributions from SNe~Ia. This implies that UMi dSph has a more extended star formation history compared to Cetus. This finding aligns with the estimates of \citet{2022arXiv220409057N}, who found the star formation quenching redshift in Cetus to be larger than that in the UMi dSph. Regarding neutron capture elements, UMi stars display a diversity, including strongly r-process-enhanced stars, stars with overabundances in light n-capture elements, and stars with contributions from the main s-process. The distinct features in elemental abundances between these two systems with similar stellar mass show that the star formation histories in low-mass dwarf galaxies are not as simple as what we would probably expect. Instead, the diversity and uniqueness of each system can be revealed by their detailed chemical compositions.

This work shows that the high-resolution spectra for a stellar stream allow us to reveal the origins of the elements in a low-mass dwarf galaxy and, moreover, quantify their relative contributions from different production channels. Combining the knowledge from nucleosynthesis predictions, we are able to constrain the star formation history of the progenitor dwarf galaxy. This is the first work in the HR-GO series, and the detailed abundances of more stellar streams and substructures will be systematically analysed in the coming papers. Ultimately, we will obtain a sample of dwarf galaxies with diverse evolution histories based on the library of stellar debris in the MW.

\section{Data availability}
We present the Source IDs, distances, as well as the abundances and the equivalent widths for individual lines of the Cetus member stars in the HR-GO program.
The full Tables 2 and 4 are available at the CDS 
 \url{https://cdsarc.cds.unistra.fr/vizier.submit/uws/listfiles?format=html&repository=HRGO_Cetus_2024/}.
The list of sample stars and their kinematical properties is provided by Z. Yuan at \url{https://zenodo.org/records/13334600}.

\begin{acknowledgements}
The authors are indebted to T. Li, A. Bonaca, K. Venn for their assistance in getting the observed spectra.
T.S. is grateful to Yu.~V.~Pakhomov for providing his code for echelle orders merging. The authors thank the referee for carefully reading the manuscript and providing valuable feedback.
L.L. acknowledges the support by the State of Hesse within the Research Cluster ELEMENTS (Project ID 500/10.006).
N.F.M. gratefully acknowledge support from the French National Research Agency (ANR) funded project "Pristine" (ANR-18-CE31-0017) along with funding from the European Research Council (ERC) under the European Unions Horizon 2020 research and innovation programme (grant agreement No. 834148).
This research is based in part on data collected at the Subaru Telescope, which is operated by the National Astronomical Observatory of Japan. We are honoured and grateful for the opportunity of observing the Universe from Maunakea, which has the cultural, historical, and natural significance in Hawaii. 
This work has made use of data from the European Space Agency mission {\it Gaia} (https://www.cosmos.esa.int/gaia), processed by the {\it Gaia} Data Processing and Analysis Consortium (DPAC, https://www.cosmos.esa.int/web/gaia/dpac/consortium).
\end{acknowledgements}

\bibliographystyle{aa} 
\bibliography{cetus}

\begin{appendix}
\section{[X/Fe] -- [Fe/H] diagrams for the abundance ratios not discussed in the main text}

\begin{figure}
\includegraphics[trim={10  30 0 10},width=80mm]{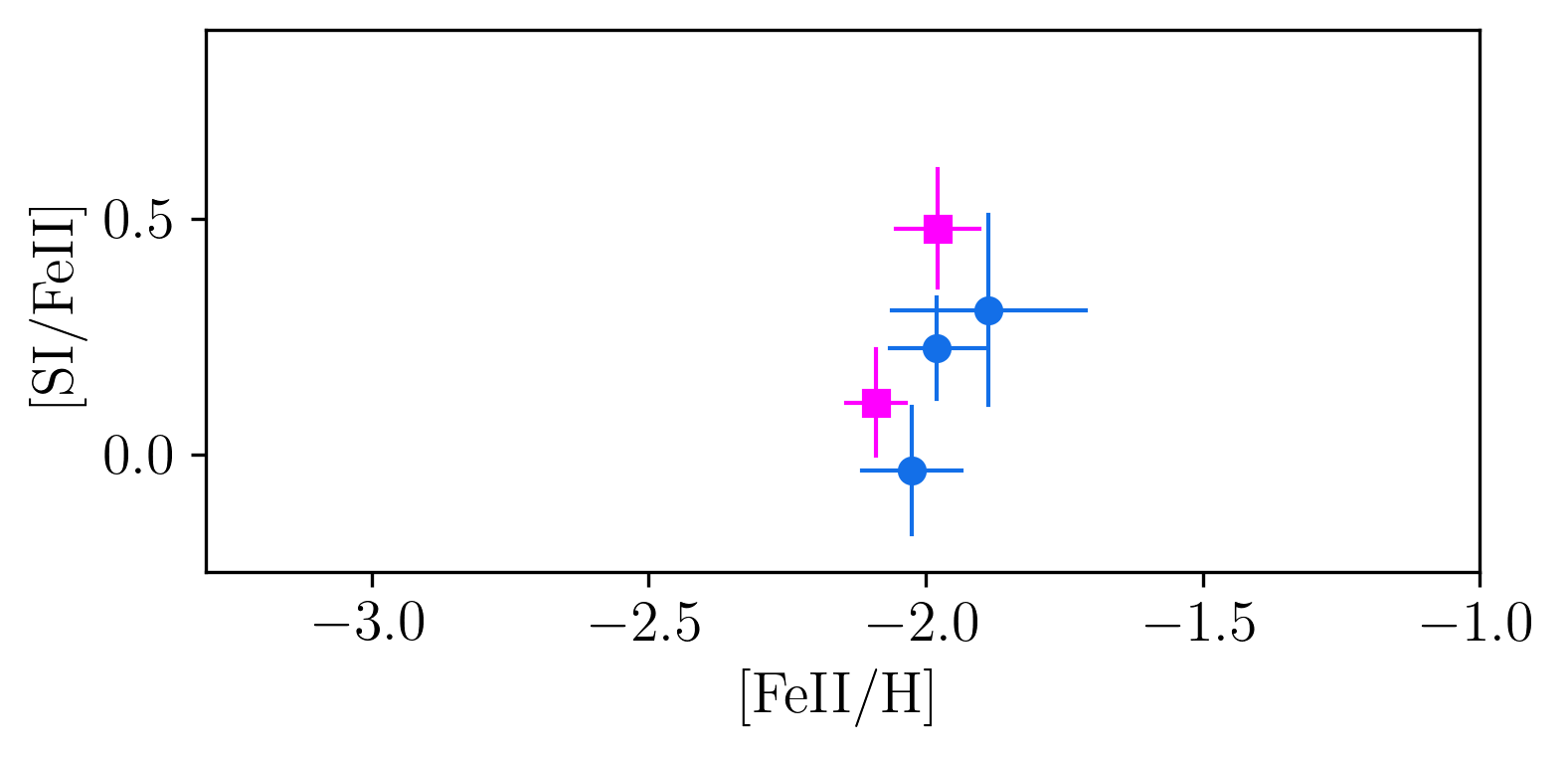}
\includegraphics[trim={0  30 0 10},width=80mm]{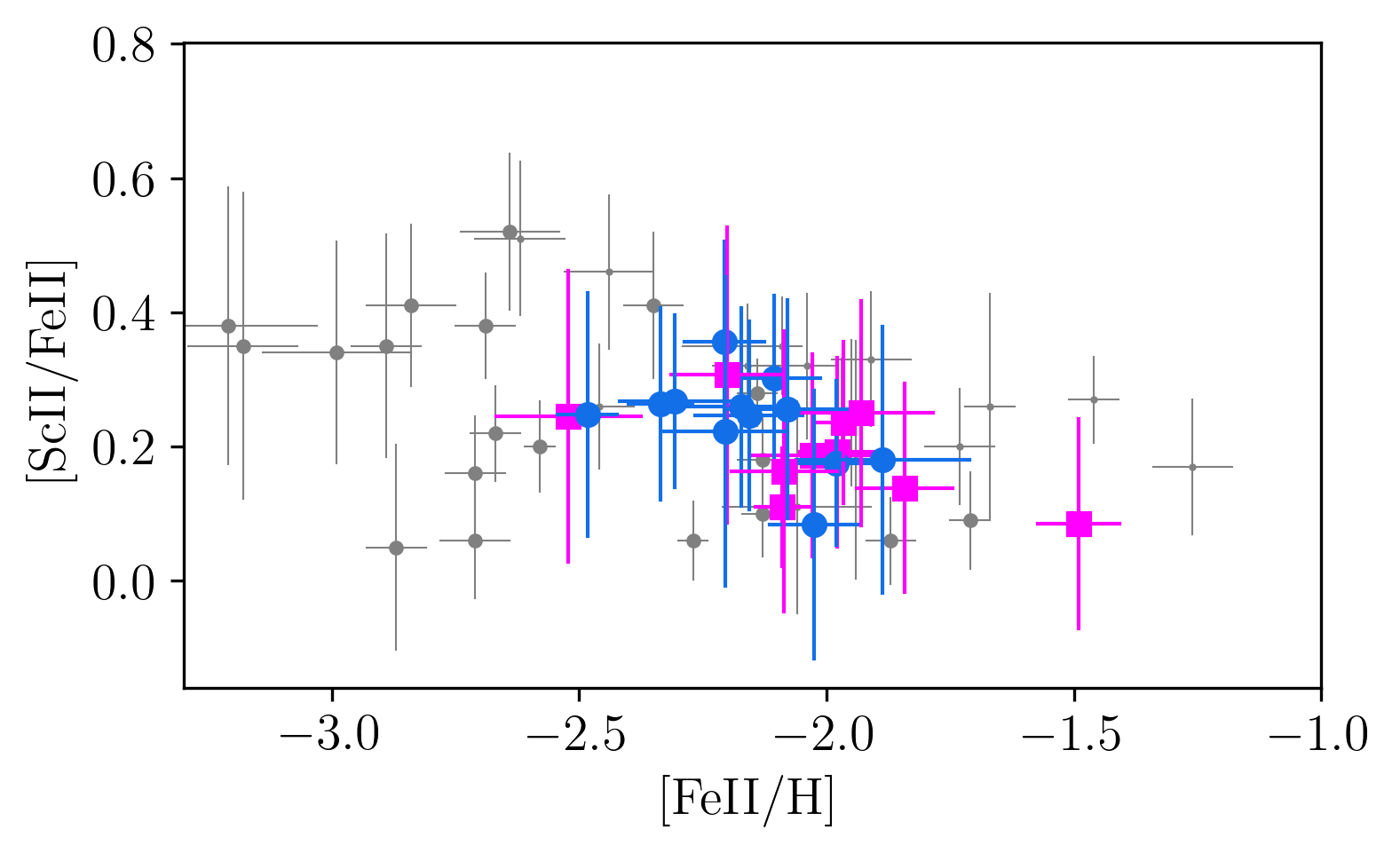}
\includegraphics[trim={10  0 0 10},width=80mm]{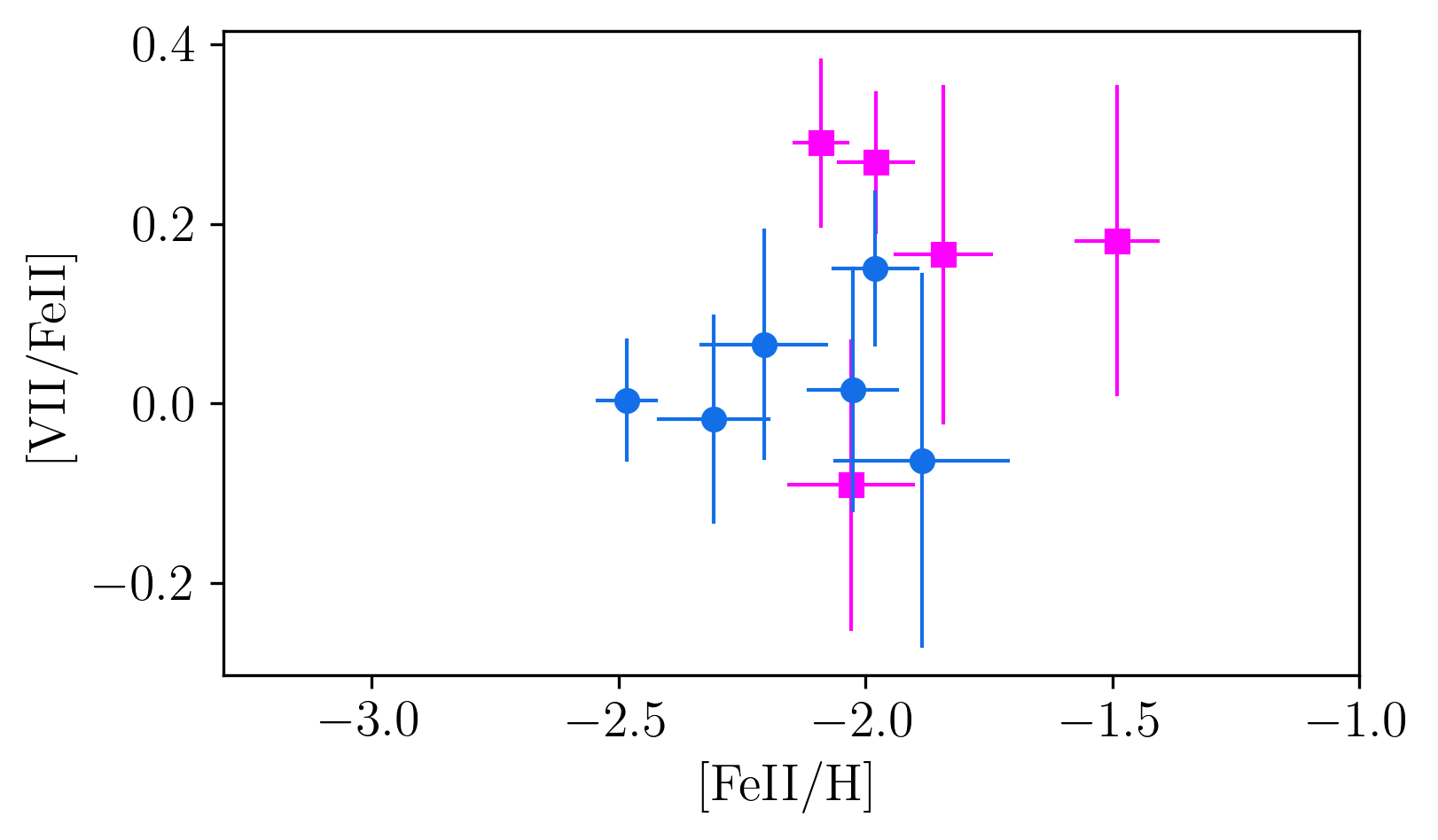}
\caption{Chemical element abundance ratios with respect to iron. The designations are the same as in Fig.~\ref{alpha_fe}.}
\label{elfe_feh_1} 
\end{figure}

\begin{figure}
\includegraphics[trim={10  30 0 10},width=80mm]{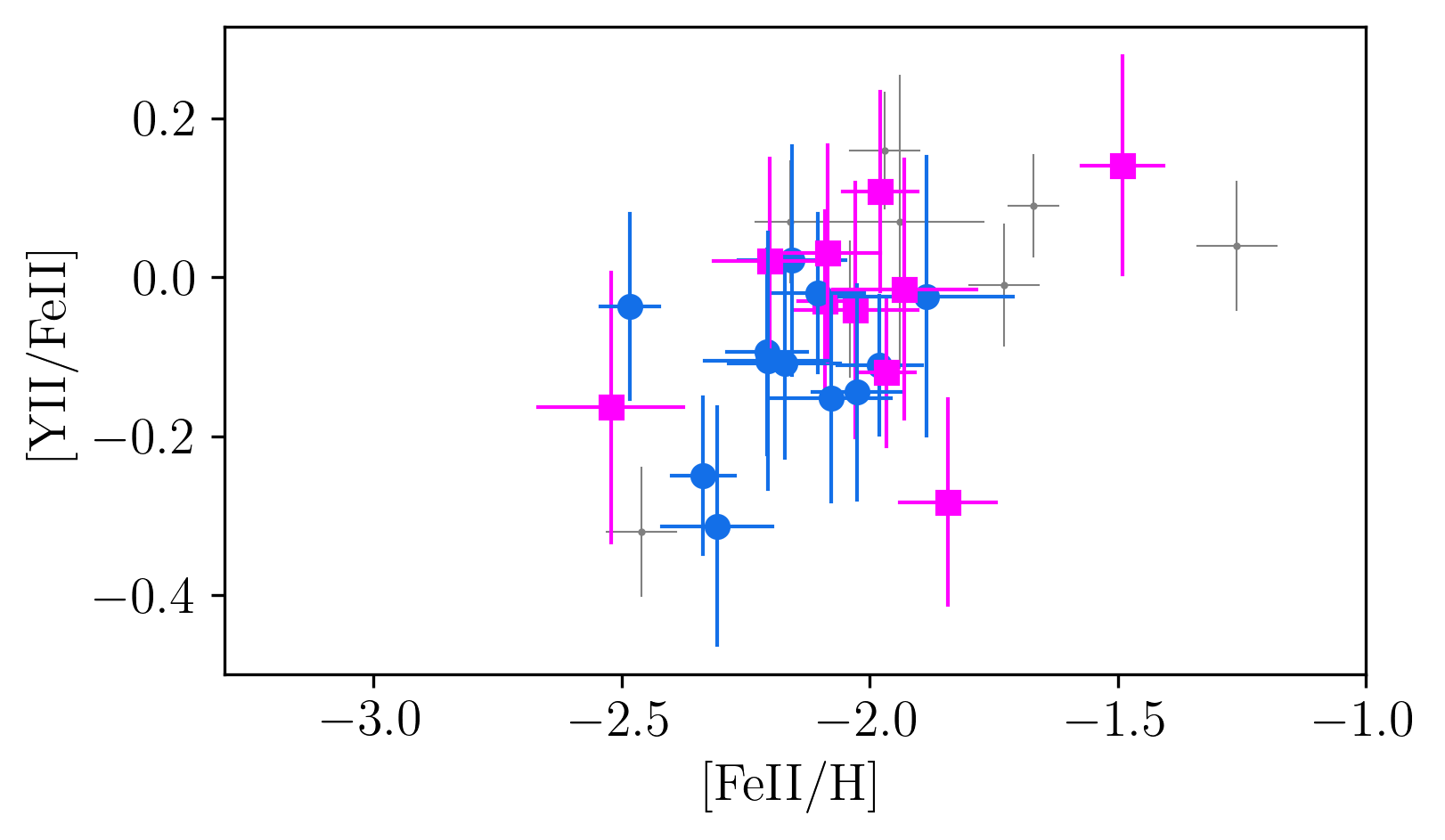}
\includegraphics[trim={0  30 0 10},width=80mm]{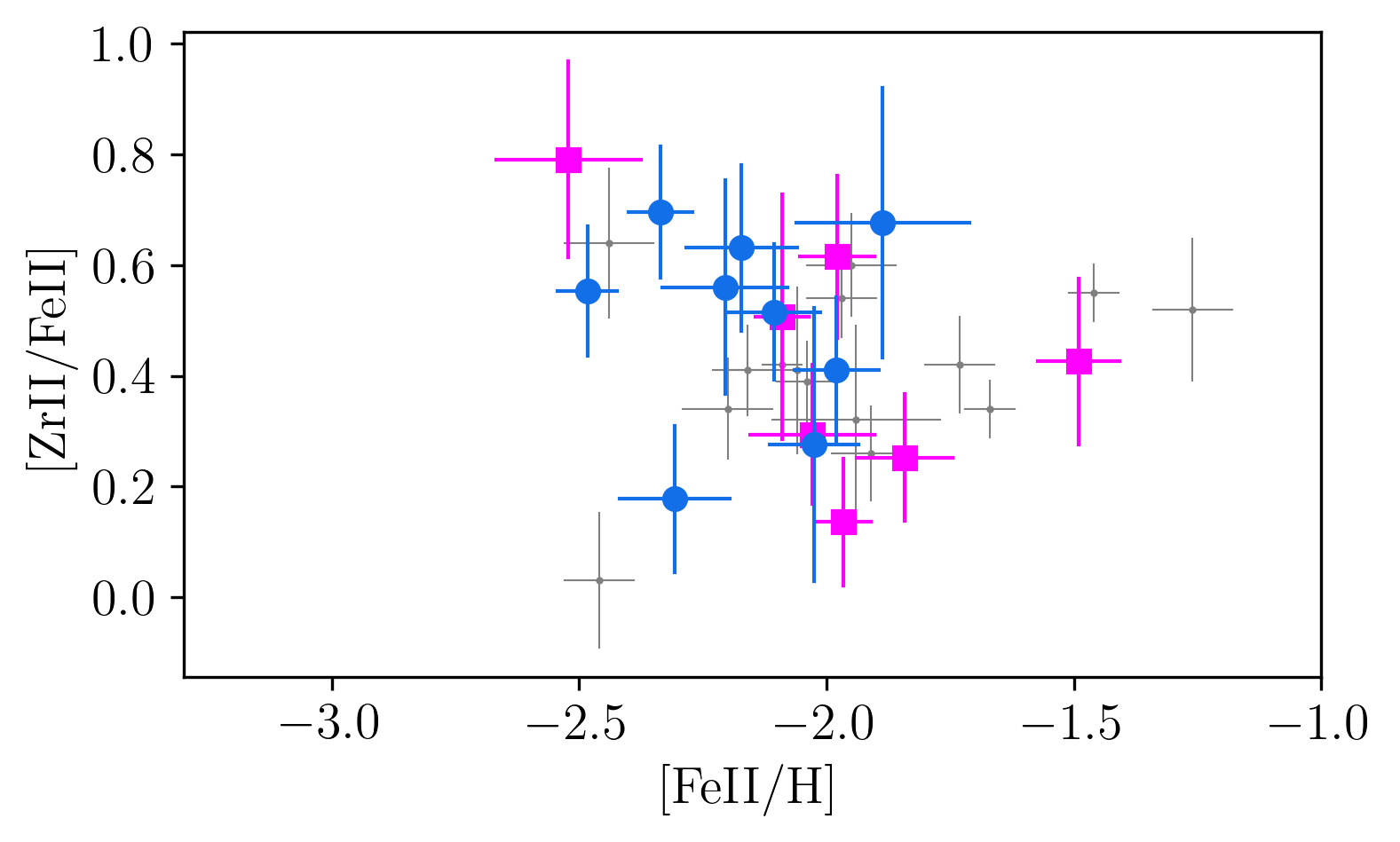}
\includegraphics[trim={10  30 0 10},width=80mm]{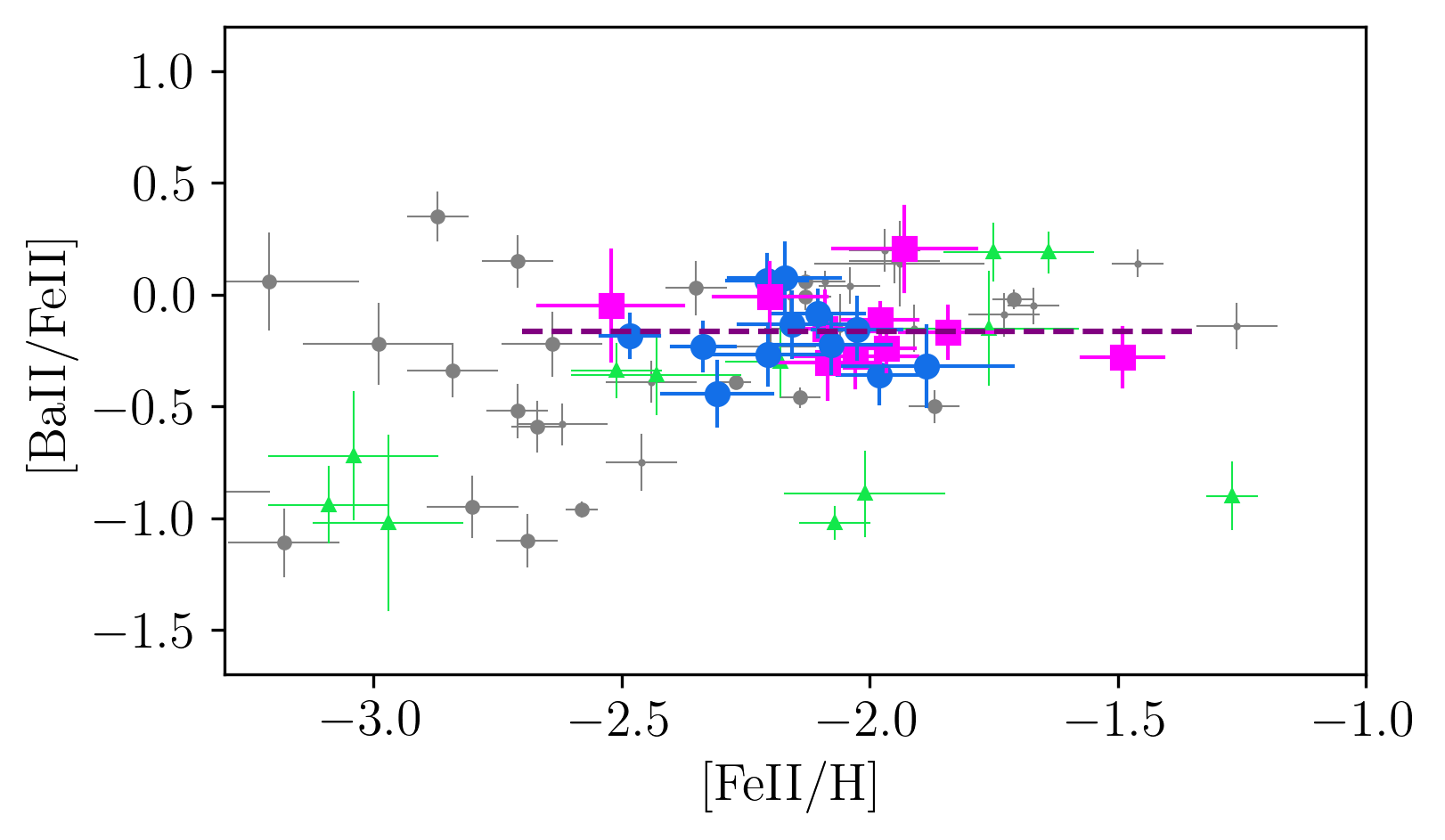}
\includegraphics[trim={5  0 0 10},width=80mm]{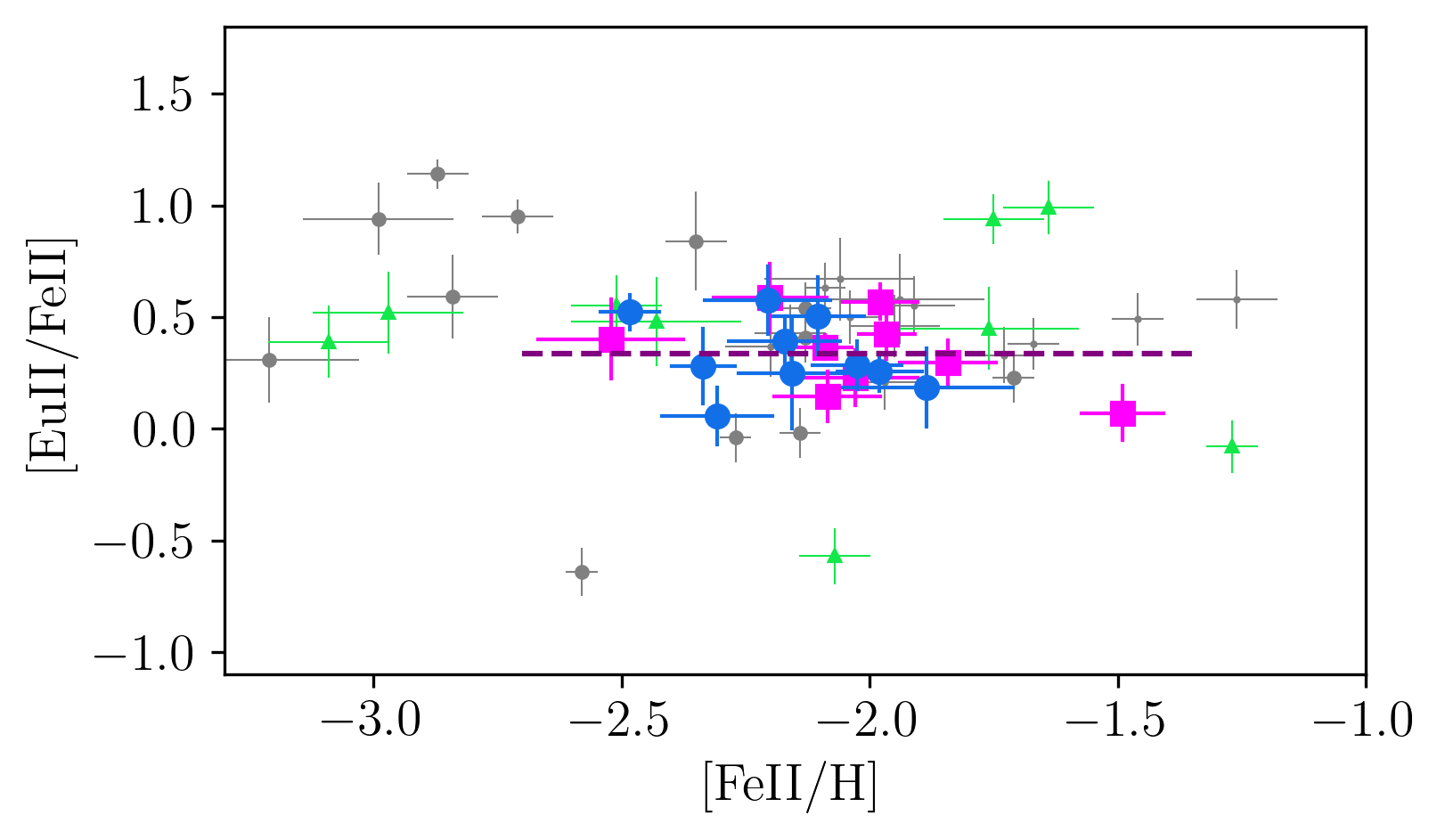}
\caption{Chemical element abundance ratios with respect to iron. The designations are the same as in Fig.~\ref{alpha_fe}.}
\label{elfe_feh_2} 
\end{figure}

\end{appendix}

\end{document}